\pdfoutput=1
\documentclass[12pt,a4paper]{article}
\usepackage{ifthen} % for conditional statements
\newboolean{pdflatex}
\setboolean{pdflatex}{true} % False for eps figures 
\newboolean{articletitles}
\setboolean{articletitles}{true} % False removes titles in references
\newboolean{uprightparticles}
\setboolean{uprightparticles}{false} %True for upright particle symbols
\newboolean{inbibliography}
\setboolean{inbibliography}{false} %True once you enter the bibliography
\usepackage[top=1in, bottom=1.25in, left=1in, right=1in]{geometry}

\usepackage{caption} %these three command get the figure and table captions automatically small

\usepackage{lineno}  % for line numbering during review
\usepackage{listings}
\usepackage{longtable} % only for template; not usually to be used in PAPERs
\usepackage{microtype}
\usepackage{xspace} % To avoid problems with missing or double spaces after
\usepackage{graphicx}  % to include figures (can also use other packages)
\usepackage{color}
\usepackage{colortbl}
\graphicspath{{./figs/}} % Make Latex search fig subdir for figures

\usepackage[percent]{overpic}

\usepackage{amsmath} % Adds a large collection of math symbols
\usepackage{amssymb}
\usepackage{amsfonts}
\usepackage{upgreek} % Adds in support for greek letters in roman typeset

\newcommand*\patchAmsMathEnvironmentForLineno[1]{%
\expandafter\let\csname old#1\expandafter\endcsname\csname #1\endcsname
\expandafter\let\csname oldend#1\expandafter\endcsname\csname
end#1\endcsname
 \renewenvironment{#1}%
   {\linenomath\csname old#1\endcsname}%
   {\csname oldend#1\endcsname\endlinenomath}%
}
\newcommand*\patchBothAmsMathEnvironmentsForLineno[1]{%
  \patchAmsMathEnvironmentForLineno{#1}%
  \patchAmsMathEnvironmentForLineno{#1*}%
}
\AtBeginDocument{%
\patchBothAmsMathEnvironmentsForLineno{equation}%
\patchBothAmsMathEnvironmentsForLineno{align}%
\patchBothAmsMathEnvironmentsForLineno{flalign}%
\patchBothAmsMathEnvironmentsForLineno{alignat}%
\patchBothAmsMathEnvironmentsForLineno{gather}%
\patchBothAmsMathEnvironmentsForLineno{multline}%
\patchBothAmsMathEnvironmentsForLineno{eqnarray}%
}

\usepackage{hyperxmp}
\usepackage[pdftex]{hyperref}
\usepackage[all]{hypcap} % Internal hyperlinks to floats.
\usepackage{bigints}
\usepackage{mathtools}
\usepackage{multirow}

\usepackage{cite} % Allows for ranges in citations
\usepackage{mciteplus}

\usepackage{xspace} 
\usepackage{upgreek}

\def\lhcb {\mbox{LHCb}\xspace}
\def\atlas  {\mbox{ATLAS}\xspace}
\def\cms    {\mbox{CMS}\xspace}

\def\babar  {\mbox{BaBar}\xspace}
\def\belle  {\mbox{Belle}\xspace}

\def\cdf    {\mbox{CDF}\xspace}

\def\lhc    {\mbox{LHC}\xspace}
\def\lep    {\mbox{LEP}\xspace}
\def\tevatron {Tevatron\xspace}

\def\MagUp {\mbox{\em Mag\kern -0.05em Up}\xspace}

\ifthenelse{\boolean{uprightparticles}}%
{

 \def\Pmu         {\ensuremath{\upmu}\xspace}                 
 \def\Pnu         {\ensuremath{\upnu}\xspace}                 
                  
 \def\Ppi         {\ensuremath{\uppi}\xspace}

 \def\Ptau        {\ensuremath{\uptau}\xspace}

 \def\Ppsi        {\ensuremath{\uppsi}\xspace}

 \def\PDelta      {\ensuremath{\Delta}\xspace}                 
 \def\PXi      {\ensuremath{\Xi}\xspace}                 
 \def\PLambda      {\ensuremath{\Lambda}\xspace}                 
 \def\PSigma      {\ensuremath{\Sigma}\xspace}                 
 \def\POmega      {\ensuremath{\Omega}\xspace}                 
 \def\PUpsilon      {\ensuremath{\Upsilon}\xspace}                 
 
 \def\PB      {\ensuremath{\mathrm{B}}\xspace}                 
                  
 \def\PD      {\ensuremath{\mathrm{D}}\xspace}

 \def\PJ      {\ensuremath{\mathrm{J}}\xspace}                 
 \def\PK      {\ensuremath{\mathrm{K}}\xspace}

 \def\PW      {\ensuremath{\mathrm{W}}\xspace}

 \def\PZ      {\ensuremath{\mathrm{Z}}\xspace}                 
                  
 \def\Pb      {\ensuremath{\mathrm{b}}\xspace}                 
 \def\Pc      {\ensuremath{\mathrm{c}}\xspace}                 
 \def\Pd      {\ensuremath{\mathrm{d}}\xspace}                 
 \def\Pe      {\ensuremath{\mathrm{e}}\xspace}

 \def\Pi      {\ensuremath{\mathrm{i}}\xspace}

 \def\Pp      {\ensuremath{\mathrm{p}}\xspace}

 \def\Ps      {\ensuremath{\mathrm{s}}\xspace}                 
 \def\Pt      {\ensuremath{\mathrm{t}}\xspace}                 
 \def\Pu      {\ensuremath{\mathrm{u}}\xspace}

}
{

 \def\Pmu         {\ensuremath{\mu}\xspace}                 
 \def\Pnu         {\ensuremath{\nu}\xspace}                 
                  
 \def\Ppi         {\ensuremath{\pi}\xspace}

 \def\Ptau        {\ensuremath{\tau}\xspace}

 \def\Ppsi        {\ensuremath{\psi}\xspace}                 
                  
 \mathchardef\PDelta="7101
 \mathchardef\PXi="7104
 \mathchardef\PLambda="7103
 \mathchardef\PSigma="7106
 \mathchardef\POmega="710A
 \mathchardef\PUpsilon="7107
                  
 \def\PB      {\ensuremath{B}\xspace}                 
                  
 \def\PD      {\ensuremath{D}\xspace}

 \def\PJ      {\ensuremath{J}\xspace}                 
 \def\PK      {\ensuremath{K}\xspace}

 \def\PW      {\ensuremath{W}\xspace}

 \def\PZ      {\ensuremath{Z}\xspace}                 
                  
 \def\Pb      {\ensuremath{b}\xspace}                 
 \def\Pc      {\ensuremath{c}\xspace}                 
 \def\Pd      {\ensuremath{d}\xspace}                 
 \def\Pe      {\ensuremath{e}\xspace}

 \def\Pi      {\ensuremath{i}\xspace}

 \def\Pp      {\ensuremath{p}\xspace}

 \def\Ps      {\ensuremath{s}\xspace}                 
 \def\Pt      {\ensuremath{t}\xspace}                 
 \def\Pu      {\ensuremath{u}\xspace}

}

\makeatletter
\ifcase \@ptsize \relax% 10pt
  \newcommand{\miniscule}{\@setfontsize\miniscule{4}{5}}% \tiny: 5/6
\or% 11pt
  \newcommand{\miniscule}{\@setfontsize\miniscule{5}{6}}% \tiny: 6/7
\or% 12pt
  \newcommand{\miniscule}{\@setfontsize\miniscule{5}{6}}% \tiny: 6/7
\fi
\makeatother

\DeclareRobustCommand{\optbar}[1]{\shortstack{{\miniscule (\rule[.5ex]{1.25em}{.18mm})}
  \\ [-.7ex] $#1$}}

\def\electron   {{\ensuremath{\Pe}}\xspace}
\def\en         {{\ensuremath{\Pe^-}}\xspace}   % electron negative (\em is taken)
\def\ep         {{\ensuremath{\Pe^+}}\xspace}
 
\def\epem       {{\ensuremath{\Pe^+\Pe^-}}\xspace}

\def\muon       {{\ensuremath{\Pmu}}\xspace}
\def\mup        {{\ensuremath{\Pmu^+}}\xspace}
\def\mun        {{\ensuremath{\Pmu^-}}\xspace} % muon negative (\mum is taken)
\def\mumu       {{\ensuremath{\Pmu^+\Pmu^-}}\xspace}

\def\tauon      {{\ensuremath{\Ptau}}\xspace}
\def\taup       {{\ensuremath{\Ptau^+}}\xspace}
\def\taum       {{\ensuremath{\Ptau^-}}\xspace}

\def\ellm       {{\ensuremath{\ell^-}}\xspace}
\def\ellp       {{\ensuremath{\ell^+}}\xspace}

\def\neu        {{\ensuremath{\Pnu}}\xspace}
\def\neub       {{\ensuremath{\overline{\Pnu}}}\xspace}

\def\neueb      {{\ensuremath{\neub_e}}\xspace}
\def\neum       {{\ensuremath{\neu_\mu}}\xspace}
\def\neumb      {{\ensuremath{\neub_\mu}}\xspace}
\def\neut       {{\ensuremath{\neu_\tau}}\xspace}
\def\neutb      {{\ensuremath{\neub_\tau}}\xspace}

\def\neulb      {{\ensuremath{\neub_\ell}}\xspace}
\def\W      {{\ensuremath{\PW}}\xspace}
\def\Wp     {{\ensuremath{\PW^+}}\xspace}
\def\Wm     {{\ensuremath{\PW^-}}\xspace}
\def\Wpm    {{\ensuremath{\PW^\pm}}\xspace}
\def\Z      {{\ensuremath{\PZ}}\xspace}

\def\uquark    {{\ensuremath{\Pu}}\xspace}

\def\dquark    {{\ensuremath{\Pd}}\xspace}

\def\squark    {{\ensuremath{\Ps}}\xspace}

\def\cquark    {{\ensuremath{\Pc}}\xspace}
\def\cquarkbar {{\ensuremath{\overline \cquark}}\xspace}
\def\ccbar     {{\ensuremath{\cquark\cquarkbar}}\xspace}
\def\bquark    {{\ensuremath{\Pb}}\xspace}
\def\bquarkbar {{\ensuremath{\overline \bquark}}\xspace}
\def\bbbar     {{\ensuremath{\bquark\bquarkbar}}\xspace}
\def\tquark    {{\ensuremath{\Pt}}\xspace}

\def\pion   {{\ensuremath{\Ppi}}\xspace}
\def\piz    {{\ensuremath{\pion^0}}\xspace}

\def\pip    {{\ensuremath{\pion^+}}\xspace}
\def\pim    {{\ensuremath{\pion^-}}\xspace}
\def\pipm   {{\ensuremath{\pion^\pm}}\xspace}

\def\kaon    {{\ensuremath{\PK}}\xspace}
  \def\Kbar    {{\kern 0.2em\overline{\kern -0.2em \PK}{}}\xspace}

\def\KorKbar    {\kern 0.18em\optbar{\kern -0.18em K}{}\xspace}

\def\Kp      {{\ensuremath{\kaon^+}}\xspace}
\def\Km      {{\ensuremath{\kaon^-}}\xspace}
\def\Kpm     {{\ensuremath{\kaon^\pm}}\xspace}

\def\KS      {{\ensuremath{\kaon^0_{\mathrm{ \scriptscriptstyle S}}}}\xspace}

\def\Kstarz  {{\ensuremath{\kaon^{*0}}}\xspace}

\def\Kstar   {{\ensuremath{\kaon^*}}\xspace}

\def\Kstarpm {{\ensuremath{\kaon^{*\pm}}}\xspace}

  \def\Dbar    {{\kern 0.2em\overline{\kern -0.2em \PD}{}}\xspace}
\def\D       {{\ensuremath{\PD}}\xspace}

\def\DorDbar    {\kern 0.18em\optbar{\kern -0.18em D}{}\xspace}
\def\Dz      {{\ensuremath{\D^0}}\xspace}

\def\Dp      {{\ensuremath{\D^+}}\xspace}

\def\Dstar   {{\ensuremath{\D^*}}\xspace}

\def\Dstarz  {{\ensuremath{\D^{*0}}}\xspace}

\def\Dstarp  {{\ensuremath{\D^{*+}}}\xspace}
\def\Dstarm  {{\ensuremath{\D^{*-}}}\xspace}
\def\Dstarpm {{\ensuremath{\D^{*\pm}}}\xspace}

\def\Dsm     {{\ensuremath{\D^-_\squark}}\xspace}

\def\B       {{\ensuremath{\PB}}\xspace}
\def\Bbar    {{\ensuremath{\kern 0.18em\overline{\kern -0.18em \PB}{}}}\xspace}
\def\Bb      {{\ensuremath{\Bbar}}\xspace}
\def\BorBbar    {\kern 0.18em\optbar{\kern -0.18em B}{}\xspace}
\def\Bz      {{\ensuremath{\B^0}}\xspace}
\def\Bzb     {{\ensuremath{\Bbar{}^0}}\xspace}
\def\Bu      {{\ensuremath{\B^+}}\xspace}
\def\Bub     {{\ensuremath{\B^-}}\xspace}
\def\Bp      {{\ensuremath{\Bu}}\xspace}
\def\Bm      {{\ensuremath{\Bub}}\xspace}
\def\Bpm     {{\ensuremath{\B^\pm}}\xspace}

\def\Bd      {{\ensuremath{\B^0}}\xspace}
\def\Bs      {{\ensuremath{\B^0_\squark}}\xspace}
\def\Bsb     {{\ensuremath{\Bbar{}^0_\squark}}\xspace}

\def\Bc      {{\ensuremath{\B_\cquark^+}}\xspace}
\def\Bcp     {{\ensuremath{\B_\cquark^+}}\xspace}
\def\Bcm     {{\ensuremath{\B_\cquark^-}}\xspace}

\def\jpsi     {{\ensuremath{{\PJ\mskip -3mu/\mskip -2mu\Ppsi\mskip 2mu}}}\xspace}
\def\psitwos  {{\ensuremath{\Ppsi{(2S)}}}\xspace}

  \def\Y#1S{\ensuremath{\PUpsilon{(#1S)}}\xspace}% no space before {...}!

\def\TwoS  {{\Y2S}}

\def\FourS {{\Y4S}}

\def\proton      {{\ensuremath{\Pp}}\xspace}
\def\antiproton  {{\ensuremath{\overline \proton}}\xspace}

\def\Lz          {{\ensuremath{\PLambda}}\xspace}
\def\Lbar        {{\ensuremath{\kern 0.1em\overline{\kern -0.1em\PLambda}}}\xspace}
\def\LorLbar    {\kern 0.18em\optbar{\kern -0.18em \PLambda}{}\xspace}

\def\Lb      {{\ensuremath{\Lz^0_\bquark}}\xspace}

\def\BF         {{\ensuremath{\mathcal{B}}}\xspace}

\def\BR         {\BF}
\newcommand{\decay}[2]{\ensuremath{#1\!\to #2}\xspace}         % {\Pa}{\Pb \Pc}

\def\to                 {\ensuremath{\rightarrow}\xspace}

\def\order   {{\ensuremath{\mathcal{O}}}\xspace}

\def\qsq       {{\ensuremath{q^2}}\xspace}

\def\Vcb  {{\ensuremath{V_{\cquark\bquark}}}\xspace}

\def\BdToKstmm    {\decay{\Bd}{\Kstarz\mup\mun}}

\def\BdTopipi     {\decay{\Bd}{\pip\pim}}

\def\AT#1     {\ensuremath{A_{\mathrm{T}}^{#1}}\xspace}           % 2

\def\C#1      {\ensuremath{\mathcal{C}_{#1}}\xspace}                       % 9
\def\Cp#1     {\ensuremath{\mathcal{C}_{#1}^{'}}\xspace}                    % 7
\def\Ceff#1   {\ensuremath{\mathcal{C}_{#1}^{\mathrm{(eff)}}}\xspace}        % 9  
\def\Cpeff#1  {\ensuremath{\mathcal{C}_{#1}^{'\mathrm{(eff)}}}\xspace}       % 7
\def\Ope#1    {\ensuremath{\mathcal{O}_{#1}}\xspace}                       % 2
\def\Opep#1   {\ensuremath{\mathcal{O}_{#1}^{'}}\xspace}                    % 7

\newcommand{\tev}{\ifthenelse{\boolean{inbibliography}}{\ensuremath{~T\kern -0.05em eV}}{\ensuremath{\mathrm{\,Te\kern -0.1em V}}}\xspace}
\newcommand{\gev}{\ensuremath{\mathrm{\,Ge\kern -0.1em V}}\xspace}
\newcommand{\mev}{\ensuremath{\mathrm{\,Me\kern -0.1em V}}\xspace}
\newcommand{\kev}{\ensuremath{\mathrm{\,ke\kern -0.1em V}}\xspace}
\newcommand{\ev}{\ensuremath{\mathrm{\,e\kern -0.1em V}}\xspace}
\newcommand{\gevc}{\ensuremath{{\mathrm{\,Ge\kern -0.1em V\!/}c}}\xspace}
\newcommand{\mevc}{\ensuremath{{\mathrm{\,Me\kern -0.1em V\!/}c}}\xspace}
\newcommand{\gevcc}{\ensuremath{{\mathrm{\,Ge\kern -0.1em V\!/}c^2}}\xspace}
\newcommand{\gevgevcccc}{\ensuremath{{\mathrm{\,Ge\kern -0.1em V^2\!/}c^4}}\xspace}
\newcommand{\mevcc}{\ensuremath{{\mathrm{\,Me\kern -0.1em V\!/}c^2}}\xspace}

\def\cm   {\ensuremath{\mathrm{ \,cm}}\xspace}

\def\mum  {\ensuremath{{\,\upmu\mathrm{m}}}\xspace}

\def\invfb   {\ensuremath{\mbox{\,fb}^{-1}}\xspace}

\def\invab   {\ensuremath{\mbox{\,ab}^{-1}}\xspace}

\def\sec  {\ensuremath{\mathrm{{\,s}}}\xspace}

\def\order{{\ensuremath{\mathcal{O}}}\xspace}

\def\gsim{{~\raise.15em\hbox{$>$}\kern-.85em
          \lower.35em\hbox{$\sim$}~}\xspace}
\def\lsim{{~\raise.15em\hbox{$<$}\kern-.85em
          \lower.35em\hbox{$\sim$}~}\xspace}

\def\mrad{\ensuremath{\mathrm{ \,mrad}}\xspace}

\def\tell1  {TELL1\xspace}
\def\ukl1   {UKL1\xspace}

\newcommand{\eg}{\mbox{\itshape e.g.}\xspace}
\newcommand{\ie}{\mbox{\itshape i.e.}\xspace}

\newcommand{\vs}{\mbox{\itshape vs.}\xspace}

\def\belleTwo{\mbox{Belle-II}\xspace}
\def\qsqmin{\ensuremath{q^2_{\mathrm{min}}}\xspace}
\def\qsqmax{\ensuremath{q^2_{\mathrm{max}}}\xspace}

\def\BmToDzK{\mbox{\decay{\Bm}{\Dz (\rightarrow \Kp \pim) K^{-}}\xspace}}
\def\ZToll{\mbox{\decay{\Z}{\ell^+\ell^-}}\xspace}
\def\ZToee{\mbox{\decay{\Z}{\ep\en}}\xspace}
\def\ZTomm{\mbox{\decay{\Z}{\mup\mun}}\xspace}
\def\ZTott{\mbox{\decay{\Z}{\taup \taum}}\xspace}
\def\JPsToee{\mbox{\decay{\jpsi}{\ep\en}}\xspace}
\def\JPsTomm{\mbox{\decay{\jpsi}{\mup\mun}}\xspace}
\def\WToe{\mbox{\decay{\Wm}{\en \neueb}}\xspace}
\def\WTom{\mbox{\decay{\Wm}{\mun \neumb}}\xspace}
\def\WTot{\mbox{\decay{\Wm}{\taum \neutb}}\xspace}
\def\WTol{\mbox{\decay{\Wm}{\ellm \neulb}}\xspace}
\def\KToe{\mbox{\decay{\Km}{\en \neueb}}\xspace}
\def\KTom{\mbox{\decay{\Km}{\mun \neumb}}\xspace}
\def\PiToe{\mbox{\decay{\pim}{\en \neueb}}\xspace}
\def\PiTom{\mbox{\decay{\pim}{\mun \neumb}}\xspace}

\def\Hprime{\ensuremath{H^{\prime}}\xspace} 
\def\Wprime{\ensuremath{\W^{\prime}}\xspace} 
\def\Zprime{\ensuremath{\Z^{\prime}}\xspace}

\def\pp{\ensuremath{\proton\proton}\xspace}

\def\mpmm{\ensuremath{\mup\mun}\xspace}

\def\ll{\ensuremath{\ellp\ellm}\xspace}
\def\ee{\ensuremath{\ep\en}\xspace} 
\def\tt{\ensuremath{\taup\taum}\xspace} 

\def\mmsq{\ensuremath{m^{2}_{\textrm{miss}}}\xspace}

\def\RD{\ensuremath{R_{D}}\xspace}
\def\RDm{\ensuremath{R_{D^-}}\xspace}
\def\RDz{\ensuremath{R_{\Dz}}\xspace}
\def\RDst{\ensuremath{R_{D^*}}\xspace}
\def\RDstm{\ensuremath{R_{D^{*-}}}\xspace}
\def\RDstz{\ensuremath{R_{D^{*0}}}\xspace}
\def\RDDst{\ensuremath{R_{D^{(*)}}}\xspace}
\def\RJPs{\ensuremath{R_{\jpsi}}\xspace}

\def\bTouln{\mbox{\decay{\bquark}{\uquark \, \ell^{-}\neub_{\ell}}}\xspace}

\def\bTocln{\mbox{\decay{\bquark}{\cquark \, \ell^{-}\neub_{\ell}}}\xspace}
\def\bToctaun{\mbox{\decay{\bquark}{\cquark \, \tau^{-}\neub_{\tau}}}\xspace}
\def\Hb{\ensuremath{H_{b}}\xspace}
\def\Hc{\ensuremath{H_{c}}\xspace}
\def\RHc{\ensuremath{R_{\Hc}}\xspace}
\def\bToctaun{\mbox{\decay{\bquark}{\cquark \, \tau^{-}\neub_{\tau}}}\xspace}

\def\BToDxlpnu{\mbox{\decay{\B}{\D^{(*)} \, \ell'^{-}\neub_{\ell'}}}\xspace}
\def\BToDxmunu{\mbox{\decay{\B}{\D^{(*)} \, \mu^{-}\neub_{\mu}}}\xspace}
\def\BToDxenu{\mbox{\decay{\B}{\D^{(*)} \, e^{-}\neub_{e}}}\xspace}

\def\BzbToDpmunu{\mbox{\decay{\Bzb}{\D^{+} \, \mu^{-}\neub_{\mu}}}\xspace}
\def\BzbToDpenu{\mbox{\decay{\Bzb}{\D^{+} \, e^{-}\neub_{e}}}\xspace}
\def\BmToDzmunu{\mbox{\decay{\Bm}{\Dz \, \mu^{-}\neub_{\mu}}}\xspace}
\def\BmToDzenu{\mbox{\decay{\Bm}{\Dz \, e^{-}\neub_{e}}}\xspace}

\def\HbToHctaunu{\mbox{\decay{\Hb}{\Hc \, \tau^{-}\neub_{\tau}}}\xspace}

\def\HbToHcmunu{\mbox{\decay{\Hb}{\Hc \, \mu^{-}\neub_{\mu}}}\xspace}
\def\HbToHcenu{\mbox{\decay{\Hb}{\Hc \, e^{-}\neub_{e}}}\xspace}
\def\HbToHcmunuX{\mbox{\decay{\Hb}{\Hc \, \mu^{-}\neub_{\mu} (X)}}\xspace}
\def\HbToHcenuX{\mbox{\decay{\Hb}{\Hc \, e^{-}\neub_{e} (X)}}\xspace}
\def\HbToHclnu{\mbox{\decay{\Hb}{\Hc \, \ellm \neub_{\ell}}}\xspace}

\def\tauTolpnunu{\mbox{\decay{\tau^{-}}{\ell'^{-} \, \neub_{\ell'}\neu_{\tau}}}\xspace}
\def\tauTolpnunu{\mbox{\decay{\tau^{-}}{\ell'^{-} \, \neub_{\ell'}\neu_{\tau}}}\xspace}
\def\tauTomununu{\mbox{\decay{\tau^{-}}{\mu^{-} \, \neub_{\mu}\neu_{\tau}}}\xspace}
\def\tauToenunu{\mbox{\decay{\tau^{-}}{e^{-} \, \neub_{e}\neu_{\tau}}}\xspace}
\def\tauTopinu{\mbox{\decay{\tau^{-}}{\pi^{-} \, \neu_{\tau}}}\xspace}
\def\tauTopipiznu{\mbox{\decay{\tau^{-}}{\pi^{-}\pi^0 \, \neu_{\tau}}}\xspace}
\def\tauTopipipinu{\mbox{\decay{\tau^{-}}{\pi^{-}\pi^+\pi^- \, \neu_{\tau}}}\xspace}
\def\tauTopipipixnu{\mbox{\decay{\tau^{-}}{\pi^{-}\pi^+\pi^-(\pi^0) \, \neu_{\tau}}}\xspace}
\def\tauTopipipipiznu{\mbox{\decay{\tau^{-}}{\pi^{-}\pi^+\pi^-\pi^0 \, \neu_{\tau}}}\xspace}
\def\BzbToDstptaunu{\mbox{\decay{\Bzb}{\D^{*+} \, \tau^{-}\neu_{\tau}}}\xspace}
\def\BzbToDstpmunu{\mbox{\decay{\Bzb}{\D^{*+} \, \mu^{-}\neu_{\mu}}}\xspace}
\def\BzbToDptaunu{\mbox{\decay{\Bzb}{\D^{+} \, \tau^{-}\neu_{\tau}}}\xspace}
\def\BmToDstztaunu{\mbox{\decay{\Bm}{\D^{*0} \, \tau^{-}\neu_{\tau}}}\xspace}
\def\BmToDztaunu{\mbox{\decay{\Bm}{\D^{0} \, \tau^{-}\neu_{\tau}}}\xspace}

\def\BsbToDsxlnu{\mbox{\decay{\Bsb}{\D_s^{(*)+} \, \ell^{-}\neub_{\ell}}}\xspace}
\def\LbToLcxlnu{\mbox{\decay{\Lb}{\PLambda_c^{(*)+} \, \ell^{-}\neub_{\ell}}}\xspace}
\def\BcmToJpsilnu{\mbox{\decay{\Bcm}{J/\psi \, \ell^{-}\neub_{\ell}}}\xspace}

\def\ellpr{\ensuremath{\ell^{\prime}}\xspace} 
\def\ellprm{\ensuremath{\ell^{\prime -}}\xspace} 
\def\HbToHclpnu{\mbox{\decay{\Hb}{\Hc \, \ellprm\neub_{\ellpr}}}\xspace}

\def\Hs{\ensuremath{H_{\squark}}\xspace} 

\def\KX{\ensuremath{\kaon^{(*)} }\xspace}

\def\RHs{\ensuremath{R_{\Hs}}\xspace}
\def\RKX{\ensuremath{R_{\KX}}\xspace}
\def\RK{\ensuremath{R_{\kaon}}\xspace}
\def\RKst{\ensuremath{R_{\Kstar}}\xspace}

\def\RPhi{\ensuremath{R_{\phi}}\xspace}
\def\RPi{\ensuremath{R_{\pi}}\xspace}
\def\RpK{\ensuremath{R_{\proton \kaon}}\xspace}

\def\bTodll{\mbox{\decay{\bquark}{\dquark \, \ll}}\xspace}

\def\bTosll{\mbox{\decay{\bquark}{\squark \, \ll}}\xspace}
\def\bTosmm{\mbox{\decay{\bquark}{\squark \, \mumu}}\xspace}
\def\bTosee{\mbox{\decay{\bquark}{\squark \, \ee}}\xspace}
\def\bTostt{\mbox{\decay{\bquark}{\squark \, \tt}}\xspace}

\def\bTodll{\mbox{\decay{\bquark}{\dquark \, \ll}}\xspace}

\def\BToKXll{\mbox{\decay{\B}{\KX \ll}}\xspace}

\def\BToKtaul{\mbox{\decay{\B}{\kaon \tau \ellpr}}\xspace}
\def\BToKKstemu{\mbox{\decay{\B}{\kaon^{(*)} \Pe \Pmu }}\xspace}
\def\BToKKsttaumu{\mbox{\decay{\B}{\kaon^{(*)} \Ptau \Pmu }}\xspace}
\def\BToKXJPsll{\mbox{\decay{\B}{\KX \jpsi(\decay{}{\ll})}}\xspace}
\def\BToKXJPsee{\mbox{\decay{\B}{\KX \jpsi(\decay{}{\ee})}}\xspace}

\def\BToKll{\mbox{\decay{\B}{\kaon \ll}}\xspace}
\def\BToKmm{\mbox{\decay{\B}{\kaon \mumu}}\xspace}
\def\BToKee{\mbox{\decay{\B}{\kaon \ee}}\xspace}

\def\BToKstmm{\mbox{\decay{\B}{\Kstar \mumu}}\xspace}
\def\BToKstee{\mbox{\decay{\B}{\Kstar \ee}}\xspace}

\def\BToKsttt{\mbox{\decay{\B}{\Kstar \tt}}\xspace}

\def\BdToKstmm{\mbox{\decay{\Bd}{\Kstarz \mumu}}\xspace}
\def\BdToKstee{\mbox{\decay{\Bd}{\Kstarz \ee}}\xspace}

\def\BdToKstJPs{\mbox{\decay{\Bd}{\Kstarz \jpsi}}\xspace}
\def\BdToKstKpiJPsll{\mbox{\decay{\Bd}{\Kstarz(\decay{}{\kaon^+\pion^-}) \jpsi(\decay{}{\ll})}}\xspace}

\def\BdToKstJPsmm{\mbox{\decay{\Bd}{\Kstarz \jpsi(\decay{}{\mumu})}}\xspace}

\def\BuToKmm{\mbox{\decay{\Bu}{\Kp \mumu}}\xspace}

\def\BuToKJPsll{\mbox{\decay{\Bu}{\Kp \jpsi(\decay{}{\ll})}}\xspace}
\def\BuToKJPsmm{\mbox{\decay{\Bu}{\Kp \jpsi(\decay{}{\mumu})}}\xspace}
\def\BuToKJPsee{\mbox{\decay{\Bu}{\Kp \jpsi(\decay{}{\ee})}}\xspace}

\def\BsToPhill{\mbox{\decay{\Bs}{\phi\ll}}\xspace}
\def\BsToPhimm{\mbox{\decay{\Bs}{\phi\mumu}}\xspace}

\def\LbTopKll{\mbox{\decay{\Lb}{\proton\kaon\ll}}\xspace}

\begin{document}
% ==============================================================
\renewcommand{\thefootnote}{\fnsymbol{footnote}}
\setcounter{footnote}{0}
%%%%%%%%%%%%%%%%%%%%%%%%%%%%%%%%%%%%
% !TEX root = main.tex
%%%%%%%%%%%%%%%%%%%%%%%%%%%%%%%%%%%%
% ==============================================================
\begin{titlepage}
\pagenumbering{roman}

% ==============================================================
\noindent
\begin{tabular*}{\linewidth}{lc@{\extracolsep{\fill}}r@{\extracolsep{0pt}}}
& & \today \\
\\
%& & Draft 7 \\
\end{tabular*}
% ==============================================================
\vspace*{4.0cm}

% Title --------------------------------------------------
{\normalfont\bfseries\boldmath\huge\begin{center}
%Everything you always wanted to know about Lepton Universality but never dared to ask
Review of Lepton Universality tests in \B decays
\end{center}}

\vspace*{2.0cm}

% Authors -------------------------------------------------
\begin{center}
Simone Bifani\footnote{University of Birmingham, School of Physics and Astronomy Edgbaston, Birmingham B15 2TT, United Kingdom}, S\'ebastien Descotes-Genon\footnote{Laboratoire de Physique Th\'eorique (UMR 8627), CNRS, Universit\'e Paris-Sud, Universit\'e Paris-Saclay, 91405 Orsay, France}, Antonio Romero Vidal\footnote{Instituto Galego de F\'isica de Altas Enerx\'ias (IGFAE), Universidade de Santiago de Compostela, Santiago de Compostela, Spain}, Marie-H\'el\`ene Schune\footnote{Laboratoire de l'Acc\'el\'erateur Lin\'eaire, CNRS/IN2P3, Universit\'e Paris-Sud, Universit\'e Paris-Saclay, 91440 Orsay, France}
\end{center}

\vspace*{2.0cm}

% Abstract -----------------------------------------------
\begin{abstract}
\noindent Several measurements of tree- and loop-level \bquark-hadron decays performed in the recent years hint at a possible violation of Lepton Universality. 
This article presents an experimental and theoretical overview of the current status of the field.
\end{abstract}

\vspace*{2.0cm}

\begin{center}
Published in  Journal of Physics G: Nuclear and Particle Physics, Volume 46, Number 2
\end{center}

\vspace{\fill}

% ==============================================================
\end{titlepage}
% ==============================================================
\renewcommand{\thefootnote}{\arabic{footnote}}
\setcounter{footnote}{0}
% ==============================================================
\tableofcontents
% ==============================================================
\cleardoublepage
\pagestyle{plain} % restore page numbers for the main text
\setcounter{page}{1}
\pagenumbering{arabic}
% ==============================================================
% Comment before a final submission.
%\linenumbers
% ==============================================================
\clearpage
%%%%%%%%%%%%%%%%%%%%%%%%%%%%%%%%%%%%
% !TEX root = main.tex
%%%%%%%%%%%%%%%%%%%%%%%%%%%%%%%%%%%%

%\clearpage
\section{Introduction}
\label{sec:introduction}

The Standard Model (SM) of particle physics organises the twelve elementary fermions (and their antiparticles) known to date into three families (or generations) and successfully describes their strong and electroweak interactions through the exchange of gauge bosons.

\[
\begin{matrix*}[c]
{\rm I} & {\rm II} & {\rm III} \\
\\
\begin{pmatrix} \Pe^- \\ \nu_e \end{pmatrix}  & \begin{pmatrix} \Pmu^- \\ \nu_{\mu}  \end{pmatrix}  & \begin{pmatrix} \Ptau^- \\ \nu_{\tau} \end{pmatrix} \\
\\
\begin{pmatrix} \uquark \\ \dquark \end{pmatrix}  & \begin{pmatrix} \cquark \\ \squark \end{pmatrix} & \begin{pmatrix} \tquark \\ \bquark \end{pmatrix} \\ \\
\end{matrix*}
\]

Despite its tremendous success in describing all present measurements, the SM can only be regarded as the low-energy, effective, incarnation of a more global theory.
For example, the SM cannot account for the matter-antimatter asymmetry currently present in the Universe, it does not provide a Dark Matter candidate and it does not explain its own gauge group structure, the charge assignment of the fermions or the mass hierarchy between the different families.
A more global theory that extends the SM at higher energies and shorter distances could provide an answer to some of these questions, which are at the core of modern particle physics.

Searches for signs of New Physics (NP) existing beyond the SM are performed in two ways.
The first one looks for the direct production of new particles.
The key ingredient for this so-called \textit{``relativistic path''} is the amount of energy available in the collision, which drives the maximum mass range that can be probed.
For this reason, the vast majority of the most stringent limits on the mass of NP particles have been obtained at the Large Hadron Collider (\lhc) by the \atlas and \cms experiments, as shown for example in Ref.~\cite{PDG2017}. 
The second method, the so-called \textit{``quantum path''}, exploits the presence of virtual states in the decays of SM particles.
Due to quantum mechanics, these intermediate states can be much heavier than the initial and final particles, and can affect the rate of specific decay modes as well as their angular distributions.
The most evident example is the $\beta$ decay of the neutron that probes physics at the \W-boson scale.
Flavour physics, which studies the properties of \squark-, \cquark- and \bquark-hadrons, has been extremely successful in providing information on the virtual particles involved in the processes examined. 
A famous example is the observation of CP violation in \kaon decays~\cite{Christenson:1964fg} that could be interpreted as a sign of the existence of three quark families in the SM\cite{Kobayashi:1973fv}, several years before the discovery of the third generation~\cite{Perl:1975bf,Herb:1977ek}.
Similarly, the first observation of the \Bz-\Bzb mixing phenomenon~\cite{Prentice:1987ap} revealed that the \tquark-quark mass was much larger than anticipated, eight years before its direct measurement~\cite{Abe:1995hr, Abachi:1994td}.
The investigation of the effects induced by virtual particles on SM decays requires very large data samples, since (for a given coupling to NP) the larger the NP scale, the smaller the influence on the process. 
While the relativistic path probes the scale of potential NP contributions in a direct way through specific signatures, the quantum path provides constraints
correlating the scale and the coupling to NP.

In the SM, the leptonic parts of the three families of fermions are identical, except for the different masses of the constituent particles.
In particular, the photon, the \W and the \Z bosons couple in exactly the same manner to the three lepton generations.
This very peculiar aspect of the SM, known as Lepton Universality (LU), can be examined to challenge its validity, since any departure from this identity (once the kinematic differences due to the different leptons masses have been corrected for) would be a clear sign that virtual NP particles are contributing to SM decays.
Studies of this accidental property offer a privileged way to access the quantum path, and tests of LU in \bquark-hadron decays represent the main focus of this review.

The remainder of this paper is organised as follows:
Sec.~\ref{sec:SMstructure} describes LU in the SM;
some of the most precise tests of LU not involving decays of \bquark-hadrons are discussed in Sec~.\ref{sec:history};
the theoretical framework relevant for LU tests in \bquark-quark decays is illustrated in Sec.~\ref{sec:frameworkTh};
Sec.~\ref{sec:frameworkExp} outlines the main characteristics of the experiments that have performed tests of LU in \bquark-hadron decays;
LU tests in \bTocln and \bTosll transitions are discussed in Sec.~\ref{sec:tree} and Sec.~\ref{sec:loop}, respectively;
Sec.~\ref{sec:interpretation} presents possible interpretations of the hints of deviations from LU in \bTocln and \bTosll;
prospects for future measurements in \bquark-hadron decays relevant for LU tests are highlighted in Sec.~\ref{sec:future};
and Sec.~\ref{sec:conclusions} concludes the paper.

%%%%%%%%%%%%%%%%%%%%%%%%%%%%%%%%%%%%
% !TEX root = main.tex
%%%%%%%%%%%%%%%%%%%%%%%%%%%%%%%%%%%%

%\clearpage
\section{Lepton Universality in the Standard Model}
\label{sec:SMstructure}

The importance of LU tests is strongly related to the very structure of the SM, which is based on the gauge group $SU_C(3)\times SU(2)_L\times U(1)_Y$ (corresponding to strong and electroweak interactions).
This group is broken down to $SU(3)_C\times U(1)_{em}$ (\ie QCD and QED) through the non-vanishing vacuum expectation value of the Higgs field.

Three main features are relevant for the tests of LU discussed here.
Firstly, the fermion fields are organised in three generations with the same gauge charge assignments leading to the same structure of couplings in all three generations (universality).
Secondly, the Higgs mechanism for the breakdown of the electroweak gauge symmetry does not affect the universality of the gauge couplings (including electromagnetism). Finally,
the only difference between the three families comes from the Yukawa interaction between the Higgs field and the fermion fields. The diagonalisation of the mass matrices yields mixing matrices between weak and mass eigenstates occurring in the coupling of fermions to the weak gauge boson \Wpm, which are the only source of differences between the three generations: the CKM matrix, $V$, for the quarks and the PMNS matrix, $U$, for the leptons.

\begin{figure}[t!]
\centering
\includegraphics[width=0.4\textwidth]{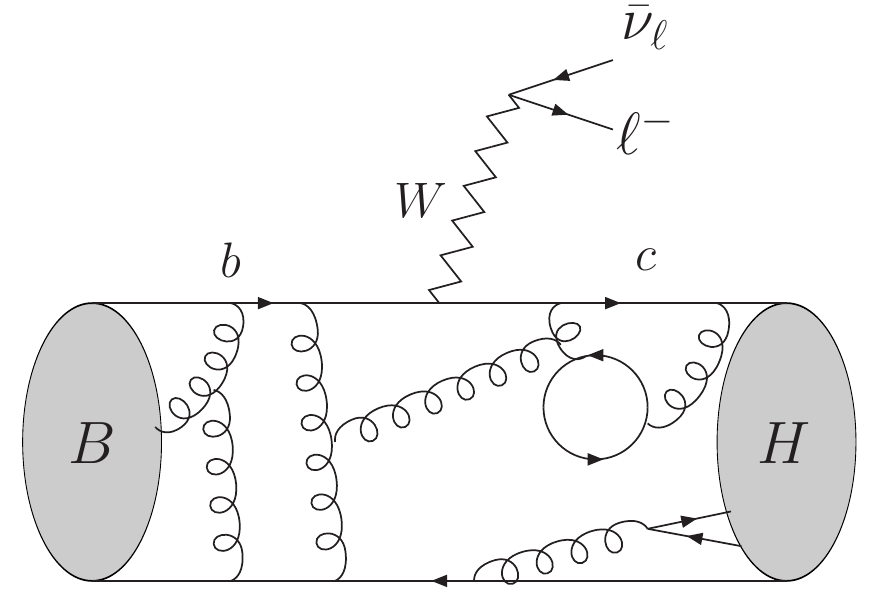}
\caption{Illustration of a \bTocln transition in the SM, as seen at the hadronic level, in the case of a \B meson decaying into an unspecified $H$ meson.}
\label{fig:feynBdecay-btocln}
\end{figure}

\begin{figure}[t!]
\centering
\includegraphics[width=0.4\textwidth]{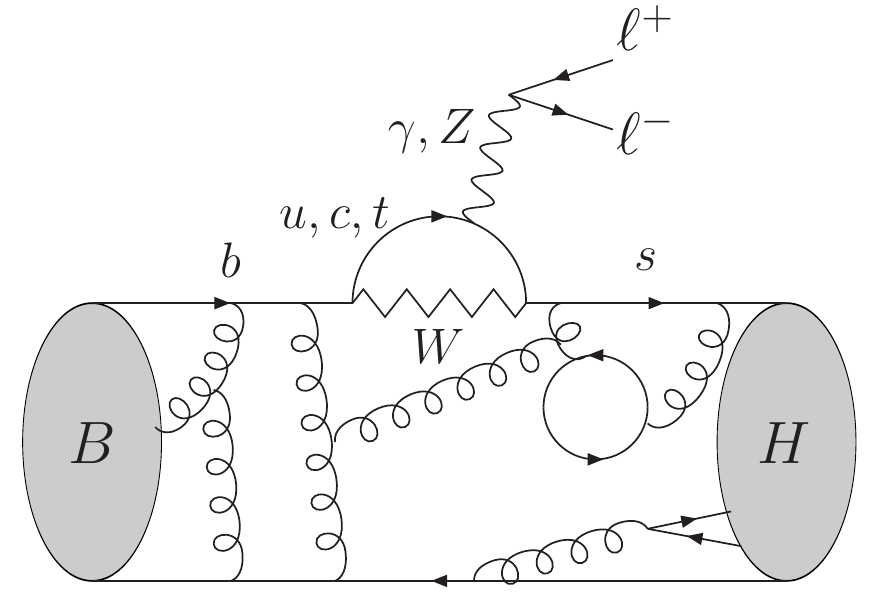}
\hspace{0.5cm}
\includegraphics[width=0.4\textwidth]{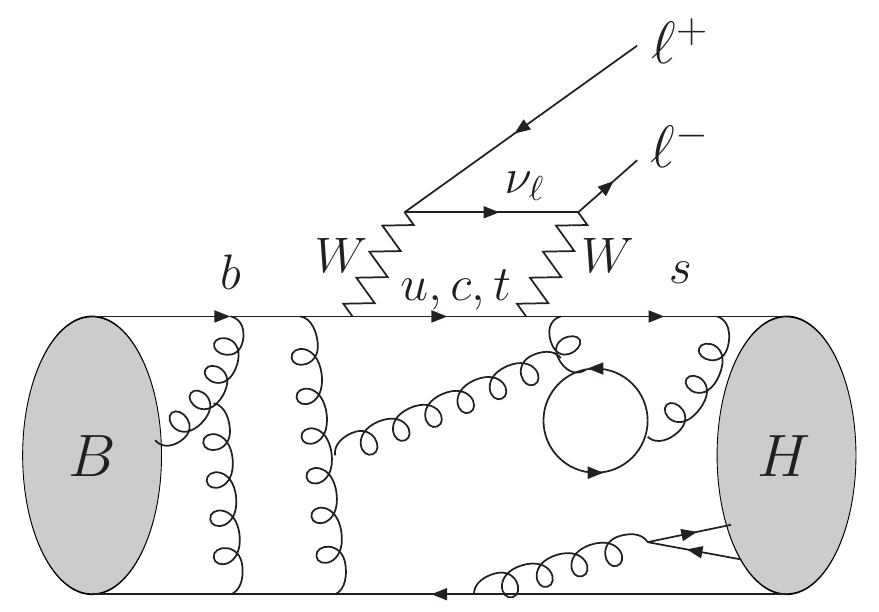}
\caption{Illustration of \bTosll transitions in the SM, as seen at the hadronic level, in the case of a \B meson decaying into an unspecified $H$ meson.}
\label{fig:feynBdecay-btosll}
\end{figure}

This last point is the key element to consider in order to design observables testing LU within the SM.
For a given transition in the SM, quarks and leptons stand on a different footing.
The flavour of the quarks involved in a transition can be determined experimentally (mass and charge of the fermions involved), so that the CKM matrix elements involved can be determined unambiguously for a given process.
In the case of Flavour-Changing Charged-Current (FCCC) transitions like \textit{``tree-level''} \bTocln decays (Fig.~\ref{fig:feynBdecay-btocln}), where $\ell$ represents any of the three charged leptons, a single CKM matrix element is involved ($V_{cb}$).
In the case of Flavour-Changing Neutral-Current (FCNC) processes, like \textit{``loop-level''} \bTosll decays (Fig.~\ref{fig:feynBdecay-btosll}), the CKM matrix elements at play depend on the flavour of the quark running through the loop and have the form $(V_{ib}V_{is}^*)$, where $i=\tquark, \cquark, \uquark$.
The unitarity of the CKM matrix and its hierarchical structure allow one to express all CKM products in terms of a leading term $(V_{tb}V_{ts}^*)$ and a Cabibbo-suppressed contribution $(V_{ub}V_{us}^*)$.
For the lepton part, the charged leptons can be easily distinguished in the same way as quarks, whereas most of the time the neutrino mass eigenstates cannot be distinguished (their mass differences are negligible compared to the other scales and they are not detected in experiments).

For example, considering the decay width of the FCCC transition \bToctaun leading to a final state with a $\tau$ lepton and an unspecified (anti)neutrino mass eigenstate, one has to sum over the amplitudes associated to the production of all the three possible (anti)neutrino mass eigenstates.
The overlap of each mass eigenstate with the produced weak interaction  eigenstate \neutb is given by the PMNS matrix $U$, so that the decay width features a factor of the form $\sum_{i=1,2,3} |U_{\tau i}|^2$, which is equal to 1 due to the unitarity of the PMNS matrix in the SM.
Therefore, the PMNS matrix plays no role in the SM, as well as in extensions featuring only three light neutrinos, and it is ignored in most computations.
These characteristics explain why either purely leptonic or semileptonic processes that involve leptons of different generations, but with the same quark transition, are preferred to test LU, so that there are no PMNS matrix elements and the CKM ones can cancel out in ratios.

%%%%%%%%%%%%%%%%%%%%%%%%%%%%%%%%%%%%
% !TEX root = main.tex
%%%%%%%%%%%%%%%%%%%%%%%%%%%%%%%%%%%%

%\clearpage
\section{Overview of Lepton Universality tests beyond the \B sector}
\label{sec:history}

Before turning to the main focus of the article, \ie tests of LU in semileptonic \B decays, some related tests in other sectors of the SM will be discussed in this section.

\subsection{Electroweak sector} 
\label{sec:WAndZ}

In the SM the couplings of the \W and \Z bosons to all lepton species are identical and a large number of experiments where the electroweak bosons are directly produced have tested this property.
The most precise results were obtained by experiments running at \epem colliders (\lep{1} and SLC, running at the \Z pole, or \lep{2}, where the centre-of-mass energy enabled the direct production of \W boson pairs), at \proton\antiproton (\tevatron) and \proton\proton (\lhc) machines.

The measurements of the \ZToee , \ZTomm and \ZTott  partial widths are in excellent agreement among each other~\cite{PDG2017}. 
Precise tests of the universality of the \Z  couplings to the charged leptons have been performed in Ref.~\cite{Z-Pole} as ratios of the leptonic partial-widths:
\begin{eqnarray}
\frac{\Gamma_{\ZTomm}}{\Gamma_{\ZToee}} &= 1.0009 \pm 0.0028 \, , \\
\frac{\Gamma_{\ZTott}}{\Gamma_{\ZToee}} &= 1.0019 \pm 0.0032 \, .
\end{eqnarray}

Under the assumption that LU holds and in case of massless leptons, all $\Gamma_{\ZToll}$ partial widths are expected to be equal in the SM.
Due to the large mass of the $\tau$ lepton, the prediction for the \ZTott partial width differs from the value computed in the massless limit by $-0.23\%$. 
In addition to the results obtained at the \Z pole, there are also measurements performed at the \lhc that support LU in \Z decays, the most precise being~\cite{Aaboud:2016btc}:
\begin{equation}
\frac{\Gamma_{\ZTomm}}{\Gamma_{\ZToee}} = 0.9974  \pm 0.0050\, .
\end{equation}

\begin{figure}[t!]
\centering
\includegraphics[width=0.8\textwidth]{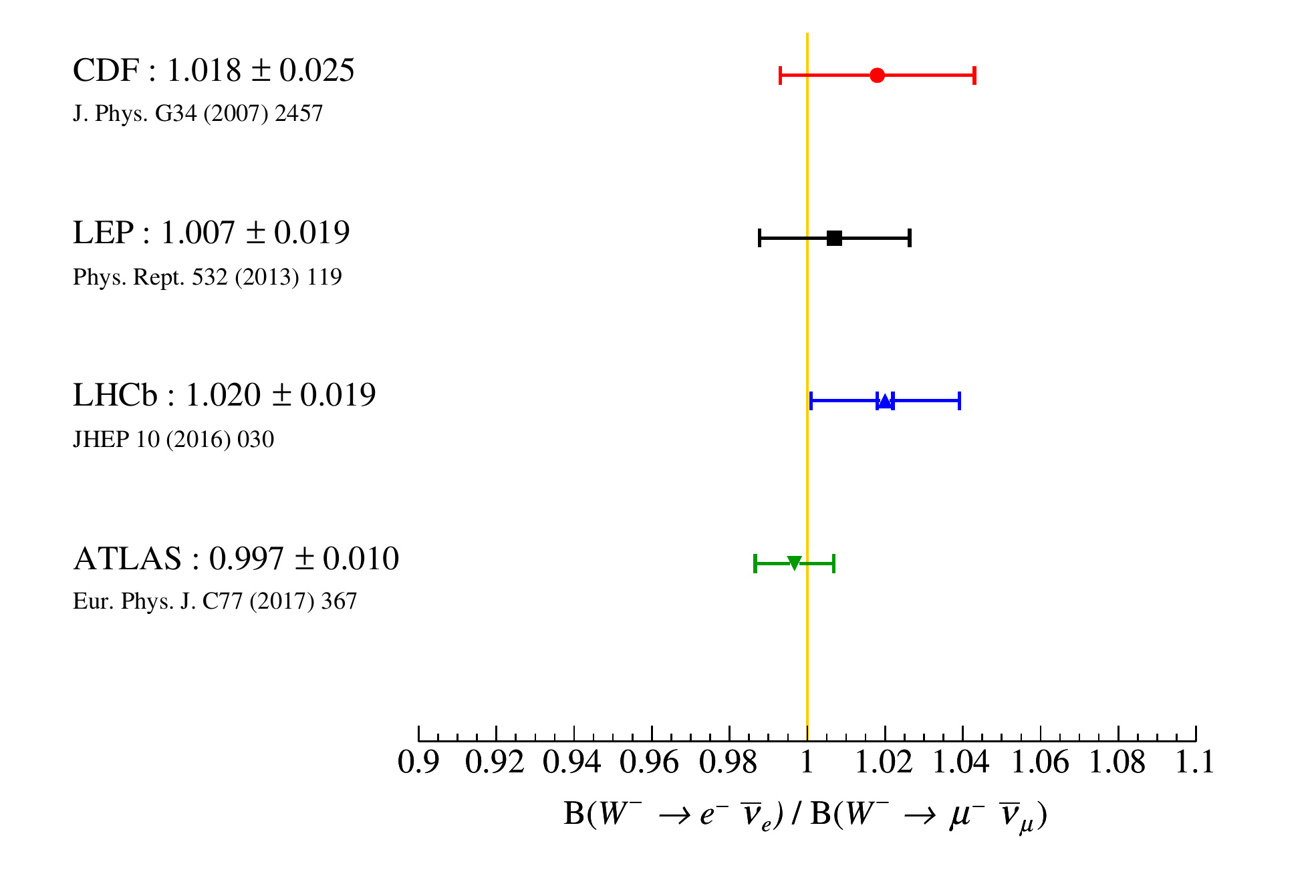}
\caption{Comparison of the ratio of branching fractions for the \WToe\ and \WTom\ decays measured by the \cdf~\cite{Abulencia:2005ix}, \lep~\cite{Schael:2013ita}, \lhcb~\cite{LHCb-PAPER-2016-024} and \atlas~\cite{Aaboud:2016btc} experiments. The SM expectation is shown as the (yellow) vertical line.}
\label{fig:WtoeOverWtom}
\end{figure}

The \lep, \tevatron and \lhc experiments have also performed precise measurements using \W boson decays that can be interpreted as tests of LU.
Following the analysis of Ref.~\cite{Pich:2013lsa}, one can introduce
the strength of the \WTol~coupling, $g_{\ell}$, which is the same in the SM for all the three lepton families.
All experimental results are in good agreement with LU, but the precision is about one order of magnitude worse than those from \Z boson decays. 
Ratios comparing \WToe~and \WTom~decays, which depend on $(g_{\electron} / g_{\muon})^2$, are dominated by the \lep combination and the \lhc results, as shown in Fig.~\ref{fig:WtoeOverWtom}.
A naive average determined by assuming that all uncertainties are fully uncorrelated results in a value of:
\begin{equation}
\frac{\BR(\WToe)}{\BR(\WTom)} = 1.004 \pm 0.008\,.
\end{equation}

Tests involving the third lepton family are less precise due to the more challenging reconstruction of final states with $\tau$ leptons.
The world averages are completely dominated by the combination of the \lep experiments~\cite{Schael:2013ita}:
\begin{eqnarray}
\frac{\Gamma_{\WTot}}{\Gamma_{\WToe}} &= 1.063 \pm 0.027 \, , \\\frac{\Gamma_{\WTot}}{\Gamma_{\WTom}} &= 1.070 \pm 0.026 \, .
\end{eqnarray}

The discussed results impliy that LU between the first two lepton families holds with a precision of better than 0.3\% in \Z boson decays and of about 0.8\% in \W boson decays. 
The constraints on LU between the third and the other two families are of similar precision in \Z boson decays (0.3\%), but of about one order of magnitude worse for \W boson decays (3\%) and somewhat in tension with the SM prediction. Assuming that LU holds between the first and the second families, a slightly more precise test can be performed: 
\begin{equation}
\frac{2 \Gamma_{\WTot}}{\Gamma_{\WToe}+\Gamma_{\WTom}} = 1.066 \pm 0.025 \, ,
\end{equation}
which manifests a tension with the SM expectation at the level of 2.6$\,\sigma$~\cite{Schael:2013ita}. The moderate increase in precision for this ratio with respect to the individual quantities indicates that this test is  systematically limited.  

\subsection{Decays of pseudoscalar mesons} 

Leptonic decays of pseudoscalar mesons also enable powerful tests of LU. The most stringent constraints derive from the study of leptonic decays of charged pions or kaons, which are helicity suppressed in the SM. 

The ratio of the partial decay widths of \KToe~and \KTom, for which hadronic uncertainties cancel, can be precisely computed in the SM~\cite{Cirigliano:2007xi}:
\begin{eqnarray}
\left(\frac{\Gamma_{\KToe}}{\Gamma_{\KTom}} \right)^{\mathrm{SM}}= \left(\frac{M_e}{M_{\mu}}\right)^2 \left( \frac{M_{\kaon}^2 - M_e^2}{M_{\kaon}^2 - M_{\mu}^2}\right) \left( 1 + \delta_{QED} \right) = (2.477 \pm 0.001) \times 10^{-5} \, ,
\end{eqnarray}
where $\delta_{QED} = (-3.78 \pm 0.04)\%$ is the correction for the \decay{\Kpm}{\ell^{\pm}\nu_{\ell}\gamma} inner bremsstrahlung contribution.
Several experimental efforts have been conducted since the 1970s to measure this ratio, which is nowadays precisely known, although its uncertainty is still an order of magnitude larger than the SM prediction.
The world average, dominated by the measurement of the NA62 experiment~\cite{Lazzeroni:2012cx}
and in good agreement with the SM expectation, is measured to be~\cite{PDG2017}: 
\begin{eqnarray}
\frac{\Gamma_{\KToe}}{\Gamma_{\KTom}} = (2.488 \pm 0.009) \times 10^{-5}\,. 
\end{eqnarray}   

The equivalent ratio for charged pion decays has also been measured~\cite{PDG2017,Aguilar-Arevalo:2015cdf}: 
\begin{eqnarray}
\frac{\Gamma_{\PiToe}}{\Gamma_{\PiTom}} = (1.230 \pm 0.004) \times 10^{-4} \, ,
\end{eqnarray}
which is more than one order of magnitude less precise than the SM prediction~\cite{Cirigliano:2007xi}:
\begin{eqnarray}
\left( \frac{\Gamma_{\PiToe}}{\Gamma_{\PiTom}} \right)^{\mathrm{SM}} = (1.2352 \pm 0.0001) \times 10^{-4}\,.
\end{eqnarray}

These measurements test the coupling of the \W boson to the first two families of leptons, $(g_{\electron}/g_{\muon})^2$.
In this sector LU holds at the $0.2\%$ level, which is significantly more precise than the constraints from direct measurements reported in Sec.~\ref{sec:WAndZ}. 

Analogous tests can be performed in the charmed-meson sector.
Only the muonic $\decay{D^-}{\mu^-\overline{\nu}_{\mu}}$ decay has been measured so far, whereas
in the case of the \Dsm meson both the $\muon$ and $\tau$ modes are known~\cite{Amhis:2016xyh}:
\begin{eqnarray}
\frac{\Gamma_{\decay{D_s^-}{\tau^- \neutb}}}{\Gamma_{\decay{D_s^-}{\mun \neumb}}} = 9.95 \pm 0.61 \, ,
\end{eqnarray}
which is consistent with, but less precise than, the SM prediction
\begin{eqnarray}
\left( \frac{\Gamma_{\decay{D_s^-}{\tau^- \neutb}}}{\Gamma_{\decay{D_s^-}{\mun \neumb}}} \right)^{\mathrm{SM}}  = 9.76\pm 0.10
\end{eqnarray} 
even assuming a conservative uncertainty of 1~\% due to the $\decay{D_s^-}{\mu^- \neumb \gamma}$ contribution~\cite{Dobrescu:2008er,Burdman:1994ip}.
LU between the second and third families thus holds at the level of 6\% from $D_s$ meson decays.

Additional tests of LU in FCCC can be performed by comparing semileptonic transitions \decay{H}{H'\ellm\neulb}, such as \decay{K}{\pi\ellm\neulb} and \decay{D}{K\ellm\neulb}, with different lepton flavours.
However, these tests require knowledge of the ratio of the scalar and vector form factors, $f_0/f_+$, with a very high level of accuracy in order to be competitive with the leptonic decays where the main hadronic input (meson decay constants) drops out of the LU ratios.
Concerning FCNC, tests of LU through semileptonic decays involving light quarks require studying transitions such as \decay{K}{\pi\ellp\ellm}, \decay{D}{\pi(\rho)\ellp\ellm} or \decay{D_s^-}{ K^-(K^{*-})\ellp\ellm}.
However, these decays are dominated by long-distance hadronic contributions that are very difficult to estimate theoretically~\cite{DAmbrosio:1998gur,Mescia:2006jd,Fajfer:2001sa,Cappiello:2012vg}, preventing accurate tests of LU with these modes.

\subsection{Purely leptonic decays} 

The LU of FCCC transitions can also be tested using pure leptonic decays of the \tauon lepton. 
In the SM, the only expected difference between the \decay{\taum}{\en\neueb\neut} and \decay{\taum}{\mun\neumb\neut} decays is due to the masses of the final-state charged-leptons.
Following the analysis of Ref.~\cite{Pich:2013lsa} (which introduces the lepton-dependent FCCC couplings $g_\ell$), and exploiting data from Ref.~\cite{PDG2017}, tight constraints can be obtained on the universality of the charged-current couplings to leptons:
\begin{equation}
g_{\muon}/g_{\electron} = 1.0018 \pm 0.0014 \, .
\end{equation}
This ratio implies that the charged-current couplings to the first and second families are universal at the $0.14\%$ level, which is one of the most stringent experimental tests at present.

The combination of the precise measurements of the \decay{\taum}{\en \neueb \neut} branching fraction and of the $\tau$ and $\mu$ lifetimes allows one to compare the FCCC couplings to the second and third lepton families (since $\decay{\mun}{\en \neueb \neum}$ is the only transition contributing to the muon decay width).
The ratio of these couplings is found to be~\cite{Pich:2013lsa}:
\begin{equation}
g_\tau/g_\mu = 1.0011\pm 0.0015 \, .
\end{equation}
Similarly, the ratio of the FCCC couplings to the third and first family can be obtained from the combination of the measurements of the $\decay{\taum}{\mun \neumb \neut}$ and of the $\tau$ and $\mu$ lifetimes~\cite{Pich:2013lsa}: 
\begin{equation}
g_\tau/g_e = 1.0030\pm 0.0015 \, .
\end{equation}
These represent the most stringent experimental tests available today for LU tests involving the coupling of the first or second family to the third one. 

\subsection{Quarkonia decays}
\label{sec:Jpsi}

Leptonic decays of quarkonia resonances can also be used to probe LU. 
The most precise test is obtained from the ratio of the \JPsToee and \JPsTomm partial widths \cite{PDG2017}:
\begin{equation}
\frac{\Gamma_{\JPsToee}}{\Gamma_{\JPsTomm}} = 1.0016 \pm 0.0031 \, ,
\end{equation}
\noindent which is in good agreement with LU with a precision of $0.31\%$. 
The fact that LU holds in \jpsi decays is exploited in the measurements of the \RKX ratios performed by the \lhcb collaboration, which are described in Sec.~\ref{sec:loop}.
Measurements of other leptonic decays of quarkonia (\eg ~\psitwos and \TwoS) lead to constraints that are weaker by an order of magnitude.

\subsection{Summary}

In summary, there are no indications of LU violation in all the processes presented in this section.
The most stringent constraint on LU between the first and the second families are obtained from indirect measurements resulting from the study of $\tau$ leptonic decays and decays of light pseudoscalar mesons.
When LU is tested for the third family, very precise results are obtained from the direct measurements of \Z boson partial widths.
The determinations of the \W couplings involving the third family are significantly less accurate and manifest a tension with the SM expectation at the level of 2.6$\,\sigma$. 

%%%%%%%%%%%%%%%%%%%%%%%%%%%%%%%%%%%%
% !TEX root = main.tex
%%%%%%%%%%%%%%%%%%%%%%%%%%%%%%%%%%%%

%\clearpage
\section{Theoretical treatment of \bquark-quark decays probing Lepton Universality}
\label{sec:frameworkTh}

The theoretical challenges faced and the tools used when examining LU observables in semileptonic \B decays will be addressed in this section.

\subsection{Effective Hamiltonian}

Considering the large variety of scales involved between the electroweak regime and the fermion masses (top quark excluded), it is natural to try and separate the effects coming from different scales through an effective field theory approach~\cite{Buchalla:1995vs,Buras:1998raa}. 
The idea is akin to building the Fermi theory, which is valid for $\beta$ decays, starting from the electroweak theory: even though the latter has desirable features (\eg renormalisability and ultraviolet completion), the former is adapted for low-energy transitions involving only light particles.
The Fermi theory is obtained by neglecting the propagation of the \W gauge boson, leading to a point-like four-fermion interaction.
More generally, the effective theory valid at low energies, \ie large distances, is constructed by ignoring the short-distance propagation of the massive (and/or energetic) degrees of freedom and conserving only the light (and/or soft) particles as dynamical degrees of freedom, propagating over long distances.
The effects of the massive degrees of freedom are absorbed into short-distance coefficients multiplying the operators built from light fields.

\begin{figure}[t!]
\centering
\includegraphics[width=0.4\textwidth]{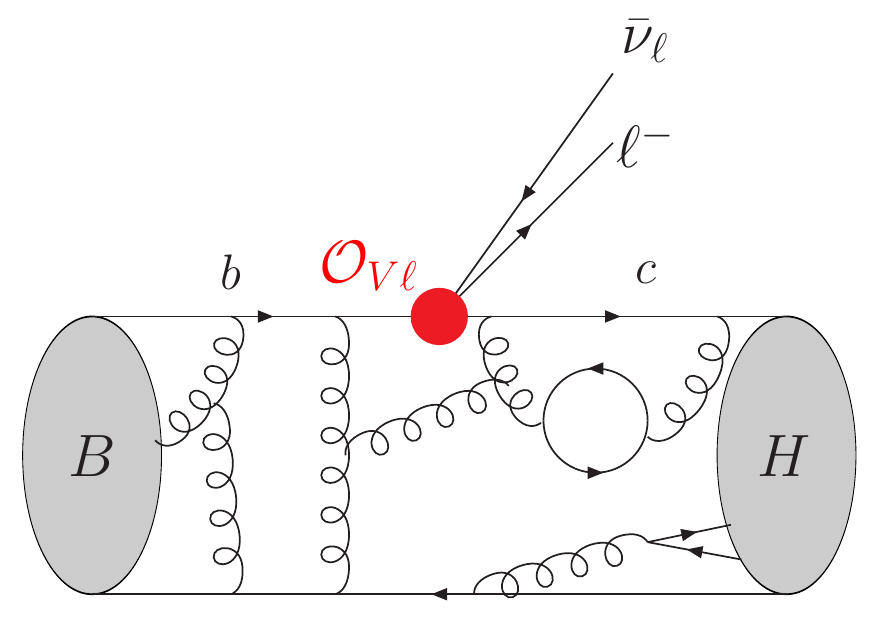}
\caption{Illustration of a \bTocln transition in the effective Hamiltonian approach. The (red) dot corresponds to a local two-quark two-lepton operator. \label{fig:heff-btocln}}
%\end{figure}
%\begin{figure}[t!]
\centering
\includegraphics[width=0.4\textwidth]{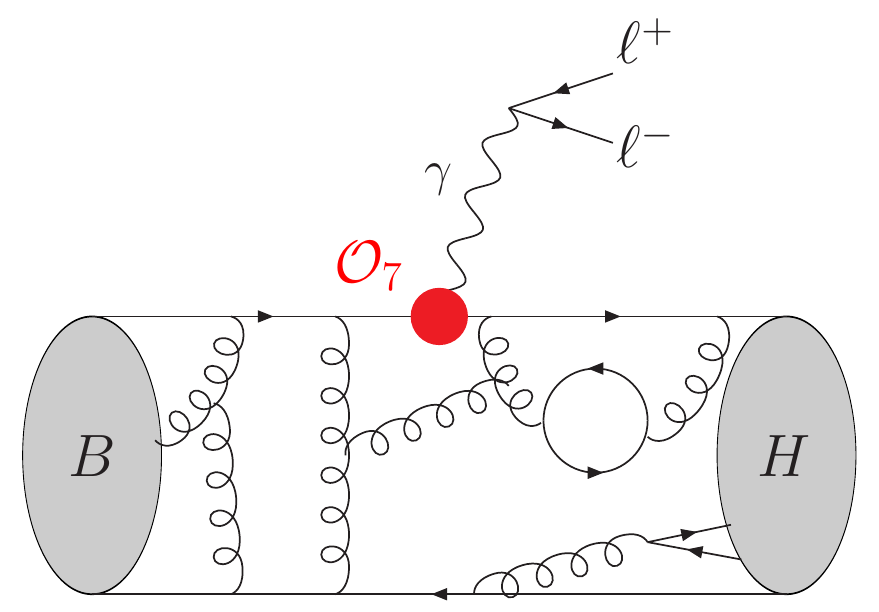}
\hspace{0.5cm}
\includegraphics[width=0.4\textwidth]{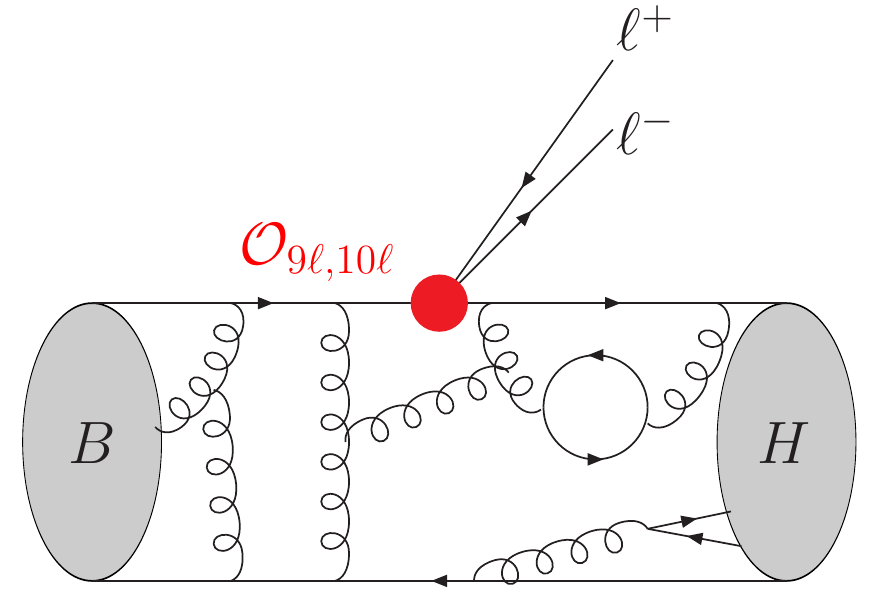}
\caption{Illustration of a \bTosll transition in the effective Hamiltonian approach. The (red) dot corresponds to a local operator. \label{fig:heff-btosll}}
\end{figure}

The effective Hamiltonian approach can be applied to both FCCC and FCNC, which are illustrated in the two cases of interest in the following.
In the case of FCCC decays, such as \bTocln, this idea yields the effective Hamiltonian:
\begin{equation}
{\cal H}_{\rm eff}(\bTocln)=\frac{4G_F}{\sqrt{2}} V_{cb}\sum_i \C{i} {\cal O}_i \, ,
\label{eq:heffbclnu}
\end{equation}
\noindent where the index $i$ runs over the various 4-fermion operators, ${\cal O}_i$, that contain the (modes of the) fields propagating over distances larger than the one associated with the factorisation scale $\mu$. Energetic modes and massive fields that probe short distances (typically smaller than $1/\mu$) are encoded in the Wilson coefficients, ${\cal C}_i$. This separation of scale is illustrated in Fig.~\ref{fig:heff-btocln}, to be compared with Fig.~\ref{fig:feynBdecay-btocln}.
The Wilson coefficients can be computed perturbatively and involve the masses and the couplings of the heavy degrees of freedom (\W, \Z, $H$ and \tquark in the SM).
Due to the universality of lepton couplings for the three generations, the SM Wilson coefficients have the same value for all three lepton generations.
The sum in Eq.~(\ref{eq:heffbclnu}) contains the dominant operator in the SM:
\begin{equation}
{\mathcal{O}}_{V\ell}=(\bar{c}\gamma_\mu P_L b)(\bar{\ell}\gamma^\mu P_L \nu_\ell) \, ,
\end{equation}
\noindent where $P_{L,R}=(1 \mp \gamma_5)/2$ are the projectors on left- (right-) handed chirality ($P_L$ corresponds to the $V$--$A$ structure of the weak interaction in the SM). The index $i=V\ell$ indicates the vector nature of the four-fermion operator, whereas $\ell$ is a reminder of the lepton flavour involved. The normalisation of Eq.~(\ref{eq:heffbclnu}) is chosen to have ${\cal C}_{V\ell}$ of order 1 in the SM. The complete basis allowed by Lorentz invariance and used to study NP contributions involve 
other four-fermion operators of the form $(\bar{c} \Gamma_q b)(\bar{\ell} \Gamma_\ell \nu_\ell)$ such as
scalar/pseudoscalar operators (\eg $\{\Gamma_q,\Gamma_\ell\}=\{1,1\}$ or $\{\gamma_5,\gamma_5\}$),
chirality-flipped operators (\eg \mbox{$\{\Gamma_q,\Gamma_\ell\}=\{\gamma_\mu P_R,\gamma^\mu P_L\}$})
as well as tensor contributions
(\eg $\{\Gamma_q,\Gamma_\ell\}=\{\sigma_{\mu\nu}P_L,\sigma^{\mu\nu}P_L\}$). In most analyses, only operators involving left-handed neutrinos are considered.

For FCNC transitions, such as \bTosll, the effective Hamiltonian takes the form:
\begin{equation}
{\cal H}_{\rm eff}(\bTosll)=-\frac{4G_F}{\sqrt{2}} V_{tb}V_{ts}^\ast\sum_i \C{i}  {\cal O}_i \, ,
\label{eq:heffbsll}
\end{equation}
\noindent up to small corrections proportional to $V_{ub}V_{us}^*$. As indicated in Sec.~\ref{sec:SMstructure}, the CKM unitarity is used to re-express the contributions proportional to $V_{cb}V_{cs}^*$ in terms of the two other products of CKM matrix elements, leading to four-quark operators of the form $(\bar{s}\Gamma_1 c)(\bar{c}\Gamma_2 b)$ among the operators ${\cal O}_i$ in Eq.~(\ref{eq:heffbsll}). The dominant operators in the SM relevant to \bTosll decays are:
\begin{eqnarray}
{\mathcal{O}}_{7} &=& \frac{e}{16 \pi^2} m_b
(\bar{s} \sigma_{\mu \nu} P_R b) F^{\mu \nu} \, ,\\
{\mathcal{O}}_{9\ell} &=& \frac{e^2}{16 \pi^2} 
(\bar{s} \gamma_{\mu} P_L b)(\bar{\ell} \gamma^\mu \ell) \, ,\\
{\mathcal{O}}_{10\ell} &=&\frac{e^2}{16 \pi^2}
(\bar{s}  \gamma_{\mu} P_L b)(  \bar{\ell} \gamma^\mu \gamma_5 \ell) \, ,
\end{eqnarray}
\noindent where $\sigma_{\mu\nu}=i/2[\gamma_\mu,\gamma_\nu]$, $e$ is the QED coupling constant and $F^{\mu\nu}=\partial^\mu A^\nu-\partial^\nu A^\mu$ is the electromagnetic tensor. The overall normalisation of Eq.~(\ref{eq:heffbsll}) is chosen to yield a Wilson coefficient of order 1 for the four-quark operator $(\bar{s}\gamma^\mu(1-\gamma_5)c)(\bar{c}\gamma_\mu(1-\gamma_5)b)$, corresponding to a $W$-boson exchange. This is illustrated in Fig.~\ref{fig:heff-btosll}, to be compared with Fig.~\ref{fig:feynBdecay-btosll}.
The $V$--$A$ structure of the weak interaction is reflected by the chirality of the quarks involved in the operators. Once again, due to the universality of lepton couplings for the three generations, the SM Wilson coefficients have the same value for all three lepton generations.

Analysing the processes within this approach offers several advantages. The determination of the Wilson coefficients requires a matching to the high-energy theory (here, the SM) at the electroweak scale $O(m_\tquark,M_\W)$ and a running from this scale down to the low-energy scale $\mu_\bquark=O(m_\bquark)$. The renormalisation group equations used then provide a re-summation of hard gluon exchanges generating large logarithms, which are then included in the Wilson coefficients~\cite{Buchalla:1995vs,Buras:1998raa,Aebischer:2017gaw} leading to an accurate determination
of the Wilson coefficients at the low-energy scale. At the level of accuracy reached by such computations, electroweak corrections both in the SM and in NP scenarios can also be included and may prove important~\cite{Aebischer:2017gaw,Feruglio:2017rjo}.

Moreover, this framework can be easily extended to include NP effects. These will lead either to shifts in the value of the short-distance Wilson coefficients or the enhancement of additional operators, absent or strongly suppressed due to the symmetries of the SM.
The complete basis allowed by Lorentz invariance and used to study NP contributions involve 
other four-fermion operators of the form $(\bar{s} \Gamma_q b)(\bar{\ell} \Gamma_\ell \ell)$ such as
scalar/pseudoscalar operators (\eg $\{\Gamma_q,\Gamma_\ell\}=\{P_{L,R},1\}$ or $\{P_{L,R},\gamma_5\}$),
chirality-flipped operators (\eg $\{\Gamma_q,\Gamma_\ell\}=\{\gamma_\mu P_R,\gamma^\mu\}$ or $\{\gamma_\mu P_R,\gamma^\mu\gamma_5\}$)
as well as tensor contributions
(\eg $\{\Gamma_q,\Gamma_\ell\}=\{\sigma_{\mu\nu}P_R,\sigma^{\mu\nu}\}$). Like for $\bTocln$ decays, in most analyses, only operators involving left-handed neutrinos are considered.

%The complete basis of operators would involve chirality-flipped operators (corresponding to $V$+$A$ couplings), scalar and pseudoscalar operators (corresponding to exchanges of a scalar boson), or tensor operators. 

Since the extension of the SM may provide an explanation to the existence of three families with a similar structure but very different masses, it can be easily imagined that NP could provide contributions that would not respect the symmetries of the SM regarding flavour, and in particular LU. The NP contributions could thus be lepton-flavour dependent and therefore generate LU-violating effects, leading to different Wilson coefficients for the operators with the same structure but different lepton flavours.

\subsection{Hadronic quantities}
\label{sec:ccbar}

As discussed in the previous section, the contributions to the decay amplitudes starting from a \B hadron and ending with a given final hadron $H$ will be expressed typically as ${\mathcal C}_i \times \langle H | {\mathcal O}_i | \B\rangle$ within the effective field theory approach (additional relevant contributions can arise from higher orders in perturbation theory, \eg for \bTosll).
The contributions from hard gluons are re-summed in the Wilson coefficient ${\mathcal C}_i$, but one must also take into account the contribution from soft gluons corresponding to long-distance QCD dynamics and responsible for the hadronisation of the quarks into the initial and final hadrons.

In the case of FCCC transitions, like \bTocln, the production of the lepton pair goes necessarily via a (heavy) \W exchange in the SM, and the insensitivity of leptons to the strong interaction allows one to factorise lepton and quark contributions (up to tiny higher-order electroweak corrections).
This can be illustrated by considering the simple $B\to D\ell^-\bar\nu_{\ell}$ case:
\begin{equation}
\mathcal{A}(B\to D\ell^-\bar\nu_{\ell})
  =\frac{4G_F}{\sqrt{2}}V_{cb} {\mathcal C}_{V\ell}\ 
       \langle D|\bar{c}\gamma_\mu P_L b|B\rangle
          L^\mu, \qquad
L^\mu=\bar{u}_\ell \gamma^\mu P_L v_\nu ,         
\end{equation}
\noindent where $L^\mu$ is a lepton tensor easily expressed in terms of solutions of the Dirac equation $u_\ell,v_\nu$. The input required to describe this transition at the hadronic level is thus limited to form factors obtained by describing the Lorentz structure of:
\begin{equation}
2\langle D | \bar{c}\gamma_\mu P_L b | B\rangle
 =f_+(q^2)(p_B+p_D)_\mu+[f_0(q^2)-f_+(q^2)]\frac{M_B^2-M_D^2}{q^2}q_\mu
\end{equation}
\noindent with $q^\mu=(p_B-p_D)^\mu$,
in the most simple case of pseudoscalar $B$ and $D$ mesons, leading to two form factors $f_0$ and $f_+$, and $M_B$ and $M_D$ are the masses of the \B and $D$ mesons, respectively.
These scalar quantities depend only on Lorentz-invariant products of kinematic variables (here, the only non-trivial product is \qsq).
This approach can be extended to NP operators, leading to products of different lepton tensors $L$ and hadron matrix elements involving other quark bilinears.
The number of the form factors involved
depends on the spin of the external states considered and on the kind of quark bilinears from the effective Hamiltonian operators (see Tab.~\ref{tab:ffcounting}). The resulting decay amplitudes can be conveniently analysed in the helicity amplitude formalism, leading to a set of angular observables describing the differential cross section for the decay~\cite{Jacob:1959at, Haber:1994pe}.

\begin{table}[t!]
\centering
\renewcommand\arraystretch{1.2}
\begin{tabular}{c|c|c}
\boldmath{$J^P(H)$} & \boldmath{$\Gamma$} & \textbf{Form factors} \\ \hline
$0^-$ & $\gamma_\mu$ & $f_0,f_+$ \\
$0^-$ & $\sigma_{\mu\nu}$  & $f_T$\\
$1^-$ & $\gamma_\mu$ & $A_0,A_1,A_2$ \\
$1^-$ & $\gamma_\mu\gamma_5$ & $V$ \\
$1^-$ & $\sigma_{\mu\nu}$  & $T_2,T_3$\\
$1^-$ & $\sigma_{\mu\nu}\gamma_5$  & $T_1$
\end{tabular}
\caption{Form factors of interest for $B\to H$ transitions for pseudoscalar or vector $H$ mesons for a given bilinear
$\bar{q}\Gamma b$. All form factors are function of \qsq, where $q=p_B-p_H$. The form factors for $J^P(H)=1^-$ also involve the polarisation of the vector meson $H$.}
\label{tab:ffcounting}
\end{table}

\begin{figure}[t!]
\centering
\includegraphics[width=0.4\textwidth]{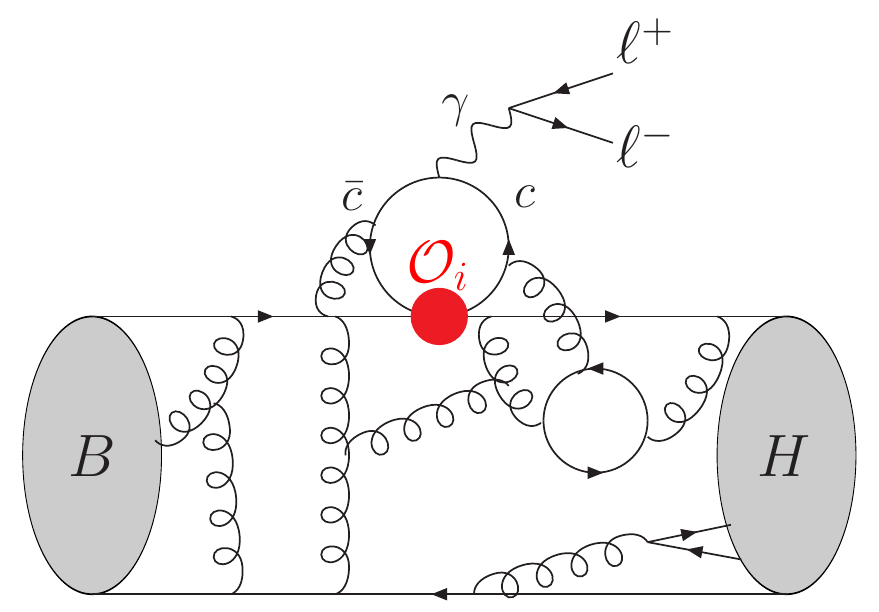}
\caption{Illustration of a \bTosll transition in the effective Hamiltonian approach, with a contribution from charm loops. The (red) dot represents a local four-quark operator corresponding to a \W exchange.}
\label{fig:charmloop}
\end{figure}

For FCNC decays, like \bTosll, the situation is less simple. There are contributions from operators that involve a dilepton pair explicitly (like ${\cal O}_{9\ell}$ or ${\cal O}_{10\ell}$). A decomposition similar to the \bTocln case can be performed, leading to (computable) lepton tensors and hadron matrix elements with quark bilinears that can be projected onto form factors. 
However, it is also possible to create the lepton pair $\ell^+\ell^-$ from 4-quark operators like $b\to sc\bar{c}$ (often denoted ${\cal O}_2$ and corresponding to a \W exchange), leading to a $c\bar{c}$ pair that gets annihilated into a $\ell^+\ell^-$ pair through electromagnetic interactions, as illustrated in Fig.~\ref{fig:charmloop} (the same can be also achieved with $u\bar{u}$). These contributions will have the form:
\begin{equation}
-\frac{16\pi}{\sqrt{2}}G_F\alpha V_{tb}V_{ts}^*\frac{1}{q^2}\ \bar{v}_\ell\gamma^\mu u_\ell\ {\cal H}_\mu
\end{equation}
with the non-local matrix element:
\begin{equation}
{\cal H}_\mu=i\int d^4x\  e^{i(q\cdot x)}
 \langle K^*|T\{j_\mu(x), \sum_{j} {\cal C}_j {\cal O}_j(0)\}| B\rangle
\end{equation}
\noindent where the index $j$ runs over all the operators of the effective Hamiltonian, and $j_\mu=\sum_q Q_q \bar{q}\gamma_\mu q$ is the electromagnetic current (excluding the top quark) with $Q_q$ being the quark charge.
These long-distance contributions are much more difficult to handle theoretically than form factors, due to their non-local nature (\ie, involving two operators at different space-time positions), even though theoretical estimates are available for some of these decays (see Sec.~\ref{sec:loop-possible-interpretations} for more detail). In particular,
these contributions describe the decay through intermediate charmonium resonances such as $B\to K^* \psi(\to \ell^+\ell^-)$. The lightest ones, \jpsi and $\psi(2S)$, provide very large peaked contributions, so that hadronic SM physics dominates \qsq regions close to the charmonium resonance masses. The corresponding regions (bins) of the dilepton invariant mass squared, \qsq, are thus generally not included in the analyses, unless one wants so learn information on the long-distance dynamics of these resonances. The higher charmonium states contribute less significantly, and their contributions appear as wiggles that tend to cancel each other when integrated over large \qsq bins, so that they are expected to yield an average close to the perturbative QCD result~\cite{Shifman:2000jv}. According to specific resonances models~\cite{Beylich:2011aq}, this quark-hadron duality is expected to hold at a few percent level for the branching ratio of \BToKll decays, whereas the extension to other decay modes and observables remain to be investigated. It should be stressed that these additional non-local contributions occur for FCNC, since electromagnetic interactions can convert intermediate light $c\bar{c}$ or $u\bar{u}$ pairs into a final pair of charged leptons, but they do not appear in the case of FCCC, where only form-factor contributions should be present and there are no resonant contributions (which could yield strong variations with \qsq).
Once again, these decay amplitudes can be analysed in the helicity amplitude formalism, leading to a set of angular observables describing the differential cross section for the decay~\cite{Jacob:1959at, Haber:1994pe,Gratrex:2015hna}.

The form factors describe the hadronisation of quarks and gluons: these involve QCD in the non-perturbative regime and are a significant source of theoretical uncertainties. Depending on the energy \qsq released to form the lepton pair, the form factors can be estimated using different theoretical approaches: Light-Cone Sum Rules (LCSR) can be used for small \qsq (\ie large recoil of the final-state hadron in the \bquark-hadron rest frame)~\cite{Balitsky:1989ry, Chernyak:1990ag, Ball:2004ye, Ball:2004rg, Khodjamirian:2006st, DeFazio:2007hw, Khodjamirian:2013iaa, Straub:2015ica}, and Lattice QCD simulations for large \qsq (small recoil)~\cite{Gupta:1997nd,Hoelbling:2014uea,Detmold:2012vy,Horgan:2013hoa,Bouchard:2013pna,Bailey:2014tva,Lattice:2015rga,Aoki:2016frl}. In both regimes, a further expansion of the theory in powers of $\Lambda/m_\bquark$ (where $\Lambda$ corresponds to the typical scale of QCD of order 1\gev) can be performed. The resulting effective field theories (respectively Soft-Collinear Effective Theory~\cite{Charles:1998dr,Beneke:2000wa,Beneke:2001at,Beneke:2002ph}
and Heavy-Quark Effective Theory~\cite{Grinstein:2004vb,Bobeth:2010wg,Bobeth:2011gi}) can be used to provide approximate relationships between the form factors and thus reduce the number of independent hadronic inputs needed.
It should also be emphasised that most of these theoretical approaches treat the final state as a stable particle under the strong interaction, even though vector particles are unstable and with a finite width (in particular $D^*$ and $K^*$), which may require a specific assessment of the corresponding systematics.

\subsection{Lepton Universality observables}
\label{sec:observables}

As discussed in Sec.~\ref{sec:SMstructure}, LU is embedded in the SM and the violation of this symmetry is a clear signal of the presence of NP. It is thus natural to compare the same observable for processes differing only in the lepton flavours involved: this can be, for instance, the ratio of branching fractions for two decays ($b\to c\tau^-\bar{\nu}_{\tau}$ \vs $b\to ce^-\bar{\nu}_{e}$ and/or $b\to c\mu^-\bar{\nu}_{\mu}$), or the difference of the same angular observable describing the kinematics of the two decays ($b\to s\mu^+\mu^-$ \vs $b\to se^+e^-$).
From a theoretical point of view, it is often beneficial to choose observables based on ratios, so that common hadronic form-factors cancel out in the numerator and the denominator: the hadronic uncertainties are decreased and rely on the theoretical evaluation of ratios of form factors, which are often better determined than absolute form factors.
From an experimental point of view, the use of ratios is also advantageous since it alleviates the dependence on the absolute knowledge of efficiencies and thus reduces the size of the systematic uncertainties.  

%%%%%%%%%%%%%%%%%%%%%%%%%%%%%%%%%%%%
% !TEX root = main.tex
%%%%%%%%%%%%%%%%%%%%%%%%%%%%%%%%%%%%

%\clearpage
\section{Experiments at the \B-factories and at the Large Hadron Collider}
\label{sec:frameworkExp}

Present and future tests of LU in the \bquark-quark sector have been and will be obtained in two extremely different experimental environments: the \babar and \belle experiments running at \epem colliders (\B-factories), and the \lhcb experiment running at the \pp Large Hadron Collider (\lhc).
Both setups produce very large numbers of \bquark-hadrons, but differ in many aspects such as the \bquark-quark fragmentation fractions, the number of \bquark-hadron species produced, and more generally the experimental conditions.

\subsection{\B-factories}

Two experiments installed at two different \epem accelerators were built and operated during the same period (see Tab.~\ref{tab:BFactories}): \babar 
at SLAC (US) approved in 1993 and \belle at KEK (Japan) approved in 1994~\cite{Bevan:2014iga}. 
Both accelerators are \epem circular colliders that mostly operate at a centre-of-mass (CM) energy of 10.58\gev corresponding to the \FourS ~resonance. This particle decays either into a \Bp\Bm or a \Bz\Bzb pair, and thus represents a copious source of charged and neutral \B mesons in a clean, low background, environment.
In order to enable decay-time dependent measurements, in particular studies of CP violation, the \FourS ~resonance is boosted and thus the two beams have different energies: 9\gev (8\gev) and 3.1\gev (3.5\gev) for the electron and positron beam, respectively, at \babar (\belle).

\begin{table}[h!]
\centering
\resizebox{\textwidth}{!}{
\renewcommand\arraystretch{1.2}
\begin{tabular}{c|c|c|c|c}
\multirow{2}{*}{\textbf{Experiment}}	& \textbf{Data taking}	& \textbf{Average distance}	& \textbf{Integrated luminosity}	& \boldmath{\B\Bbar} \textbf{pairs} \\
	& \textbf{period}	& \textbf{between the two} \boldmath{\B}	& \textbf{at the} \boldmath{\FourS}	& \textbf{produced} \\	 \hline
\babar	& 1999--2008	& $\sim 270\mum$	& $\sim 433\invfb$ 	& $\sim 471 \times 10^6$ \\
\belle	& 1999--2010	& $\sim  200\mum$	& $\sim 711\invfb$ 	& $\sim 772 \times10^6$ \\					
\end{tabular}
}
\caption{Main characteristics of the data accumulated by the \babar and \belle experiments.}
\label{tab:BFactories}
\end{table}

A particular interesting feature of \bquark-hadrons is the large number of weak decay modes these can access, which is a consequence of the large \bquark-quark mass.
Therefore, in order to study decay modes with small branching fractions, the accelerators need to run at instantaneous luminosities in excess of $10^{33}~\cm^{-2}\sec^{-1}$ and thus have large beam currents.
The two detectors roughly cover the full solid angle, however, because of the asymmetric beam energies, these are more instrumented in the direction of the high-energy beam and are offset relative to the average interaction point by a few tens of centimetres in the direction of the low-energy beam.
Since most of the particles produced in \B-meson decays have relatively low momentum, multiple scattering is the main source of inaccuracy in the momentum measurement.
For this reason, both detectors (including the beam pipe) are characterised by a low-material budget. 
The vertex resolution is excellent ($10-20\mum$), both along the beam direction and in the transverse plane.
The reconstruction efficiencies for charged particles and photons, down to momenta of a few tens of \mevc, is very high and the momentum resolution is very good over a wide spectrum to help separating signal from background.
The detectors also have excellent particle identification (PID) capabilities (including $K/\pi$ separation for a momentum range between 0.6\gevc and 4\gevc) and allow the detection of photons down to an energy of 20\mev, which permits an efficient bremsstrahlung recovery. 
Finally, the trigger systems are more than 99\% efficient for events with \B\Bbar pairs.

In order to distinguish signal from background events, the \B invariant mass reconstructed from the measured decay products is exploited.
The experimental setup of the \B-factories enables setting additional kinematic constraints that improve the knowledge of the \B-meson momentum and allow for better discriminating signal from background.
Since the \FourS ~resonance decays uniquely into a pair of \B mesons, the energy of the \B-decay products is equal to half of the CM energy.
This enables defining two weakly correlated discriminating variables, $\Delta E$ and  $m_{\mathrm{ES}}$: the former compares the reconstructed \B-meson energy to the beam energy in the CM frame; the latter, also known as the beam-energy substituted mass (\babar), corresponds to the \B-meson mass reconstructed from the measured momenta of the decay products and the beam energy (the equivalent variable at \belle is the beam-energy constraints mass, $M_{\textrm{bc}}$).
The $\Delta E$ resolution depends on the detector performance and varies a lot depending on the final state (\eg for the decay \BdTopipi this is about 29\mevcc~\cite{Bevan:2014iga}).
The $m_{\mathrm{ES}}$ resolution depends mainly on the accelerator energy spread and is of the order of 2 to 3\mevcc. 
The well-constrained kinematics of the \B\Bbar pair is also extremely helpful to impose extra constraints to reconstruct final states with neutrinos.  

\begin{figure}[t!]
\centering
\includegraphics[width=0.9\textwidth]{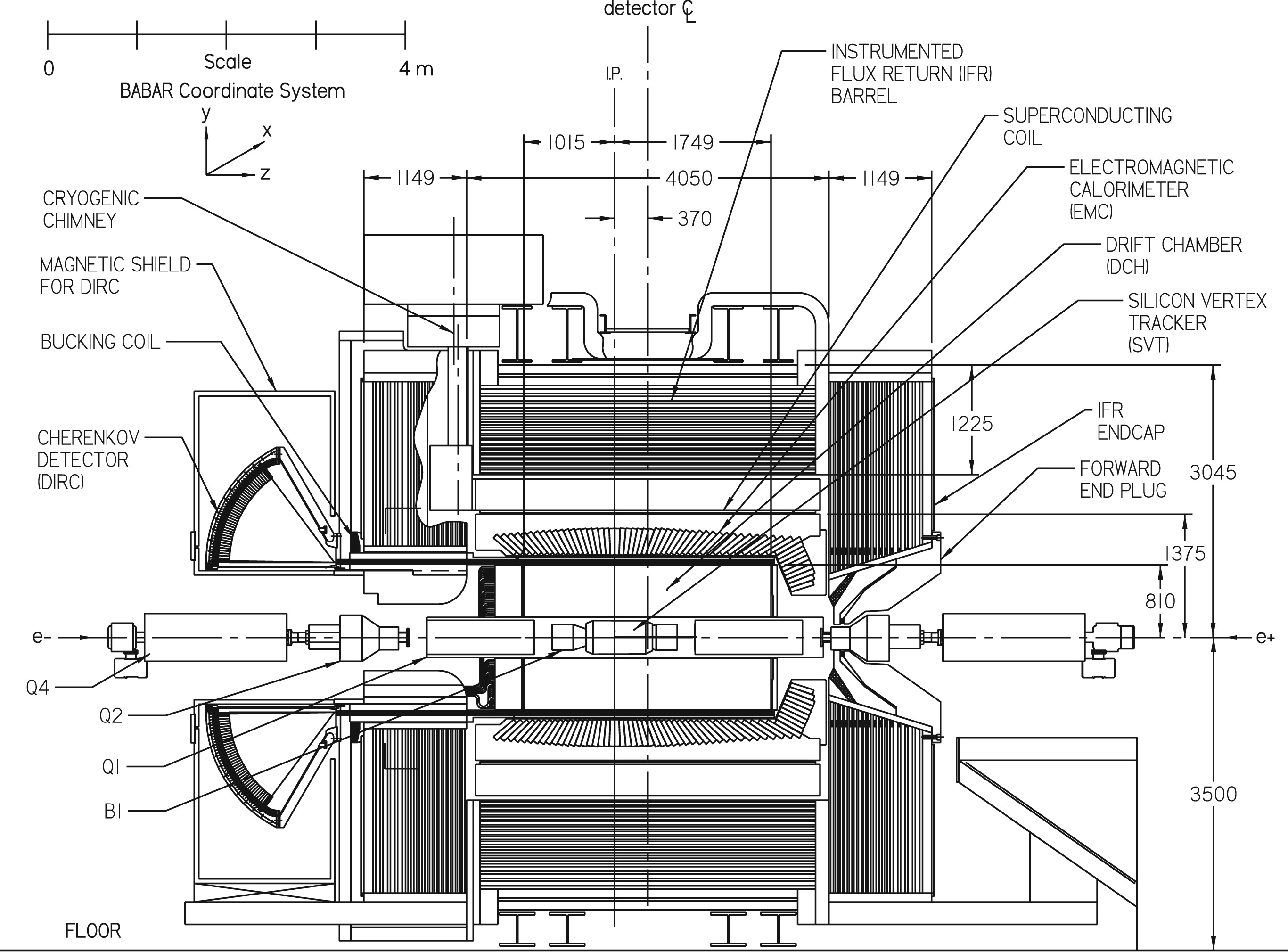}
\\ \vspace{1cm}
\includegraphics[width=0.9\textwidth]{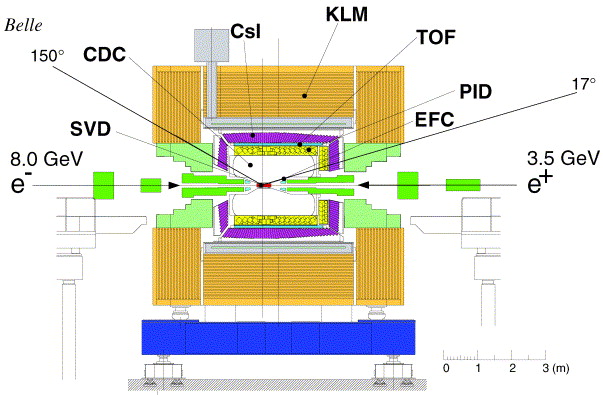}
\caption{Schematic view of the (top) \babar (taken from Ref.~\cite{AUBERT20021}) and (bottom) \belle (taken from Ref.~\cite{ABASHIAN2002117}) detectors. The $4\pi$ coverage and the asymmetric setup are clearly visible.}
\label{fig:BFact}
\end{figure}

In summary, despite different choices for the sub-detector technologies, the two \mbox{\B-factory} experiments, \babar (Fig.~\ref{fig:BFact}, top) and \belle (Fig.~\ref{fig:BFact}, bottom), are conceptually  similar.
The main differences are the amount of data recorded and the \FourS ~boost (Tab.~\ref{tab:BFactories}).
Some detector performances particularly relevant for LU tests are listed in Tab.~\ref{tab:BFactories-det}.

\begin{table}[h!]
\centering
\resizebox{\textwidth}{!}{
\renewcommand\arraystretch{1.2}
\begin{tabular}{c|c|c|c|c|c|c}
\textbf{Experiment}	& \boldmath{$e$} \textbf{efficiency}	& \boldmath{$e\to\pi$} \textbf{mis-ID}	& \boldmath{$\mu$} \textbf{efficiency}	& \boldmath{$\mu\to\pi$} \textbf{mis-ID}	& \boldmath{\kaon} 	& \boldmath{$\kaon\to\pi$} \textbf{mis-ID} \\ \hline
\textbf{\babar}	& $\sim 92\%$	& $\sim 0.1\%$	& $\sim 60\%$	& $\sim 1\%$	& $\sim 84\%$	& $\sim 1\%$ \\
\textbf{\belle}	& $\sim 90\%$	& $\sim 0.1\%$	&  $\sim 93\%$	& $\sim 3\%$	& $\sim 85\%$	& $\sim 5\%$ \\
\end{tabular}
}
\caption{Main characteristics of the \babar and \belle detectors relevant for LU tests~\cite{Bevan:2014iga}.}
\label{tab:BFactories-det}
\end{table}

Since 2018 a new experiment, \belleTwo, operating at the \FourS ~CM energy at \mbox{SuperKEKB} (Japan), is being commissioned (see for example Ref.~\cite{Riccardo:724}).
Data taking with a complete detector should start in 2019, with the goal to collect an integrated luminosity of about 50\invab by 2025. 
The detector performances are expected to be better than those of the \babar and \belle experiments, most notably for the impact parameter and secondary vertex resolutions, \KS reconstruction efficiency and PID.

\subsection{Large Hadron Collider}

%Since the only tests of LU in the \bquark-quark sector currently available from the \lhc have been performed by the \lhcb collaboration, its experimental environment is the topic of this section.  
%Proton-proton collisions at various CM energies have been collected by the \lhcb experiment at the \lhc starting from 2010 (see Tab.~\ref{tab:LHCdata}).
%The sample is divided into two datasets: Run1 which corresponds to \lhc energies of 7 and 8\tev and Run2 which covers data taken since 2015 at a CM energy of 13\tev. 

The only tests of LU in the \bquark-quark sector currently available from the \lhc have been performed by the \lhcb collaboration.
Proton-proton collisions at several CM energies have been collected by the \lhcb experiment starting from 2010 (see Tab.~\ref{tab:LHCdata}).
The recorded dataset is divided in two sub-samples: Run1, collected during 2011 and 2012 and corresponding to \lhc energies of 7 and 8\tev, and Run2, covering data taken since 2015 at a CM energy of 13\tev. 

\begin{table}[h]
\centering
\renewcommand\arraystretch{1.2}
\begin{tabular}{cc|c|c|c}
\multicolumn{2}{c|}{\textbf{Data taking}}	& \textbf{Centre-of-mass}	& \textbf{Integrated}	& \boldmath{\bbbar} \textbf{pairs} \\
\multicolumn{2}{c|}{\textbf{period}}		& \textbf{energy}		& \textbf{luminosity}	& \textbf{produced in \lhcb} \\ \hline
\multirow{2}{*}{\textbf{Run1}}	& 2011 		& 7\tev	& $\sim 1.1\invfb$	& $\sim 8 \times 10^{10}$  \\
						& 2012 		& 8\tev	& $\sim 2.1\invfb$	& $\sim 17 \times 10^{10}$ \\ \hline
\textbf{Run2}				& 2015--2017 	& 13\tev	& $\sim 3.7 \invfb$	& $\sim 49 \times 10^{10}$  \\
\end{tabular}
\caption{Summary of the data recorded by the \lhcb experiment until the end of 2017. The Run2 period will extend up to the end of 2018. }
\label{tab:LHCdata}
\end{table}

\begin{figure}[t!]
\centering
\includegraphics[width=0.8\textwidth]{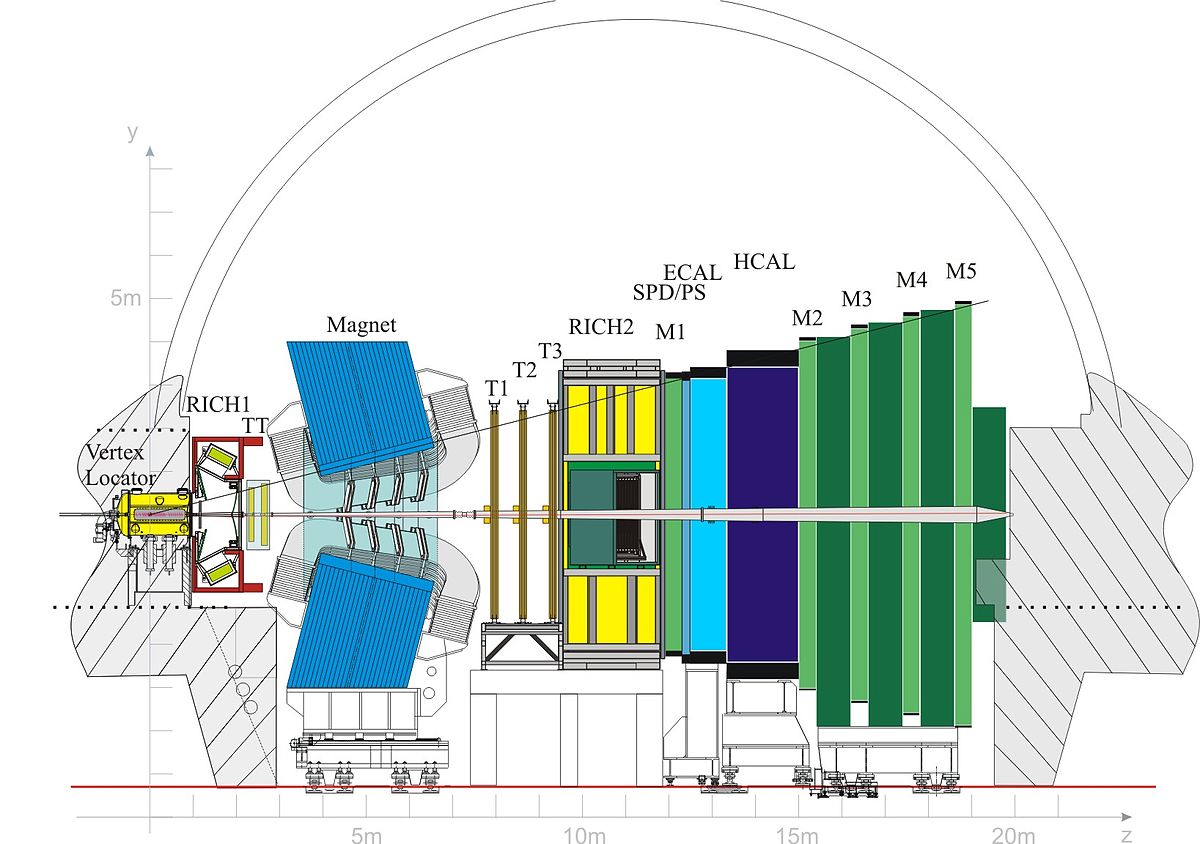}
\caption{Schematic view of the \lhcb detector.}
\label{fig:LHCb}
\end{figure}

The main mechanism for \bbbar production at the \lhc is gluon-gluon fusion, where the two \bquark-quarks are predominantly produced collinearly and close to the beam directions.
This characteristics has strongly influenced the design of the \lhcb detector~\cite{Alves:2008zz}, which is a single-arm forward spectrometer with a polar-angle coverage between 10 and 300\mrad in the horizontal plane and 250\mrad in the vertical one (Fig.~\ref{fig:LHCb}).
At the \lhc energies the \bbbar cross section is very large, $\sim 295~\mu\mathrm{b}$ at 7\tev and increasing roughly linearly with the CM energy.
This value is more than five orders of magnitude larger than the \FourS ~cross section ($\sim 1~\mathrm{nb}$), however, the ratio of the \bbbar cross section to the total inelastic one is of the order of $1/4$ at the \FourS~CM energy and about $1/300$ at the \lhc for a CM energy of 7\tev.
As a consequence of this small ratio, the collection of interesting data requires the usage of a complex trigger system consisting of a hardware stage, based on information from the calorimeter and muon systems, followed by a software stage, which applies a full event reconstruction.
Since the occupancy of the calorimeters is significantly higher than that of the muon stations, the constraints on the trigger rate require that higher thresholds should be imposed on the electron transverse energy than on the muon transverse momentum.
The most noticeable effect on LU tests is the difference in the efficiencies of the electron and muon hardware triggers for the kinematics of interest.

\begin{figure}[t!]
\centering
\includegraphics[width=0.47\textwidth]{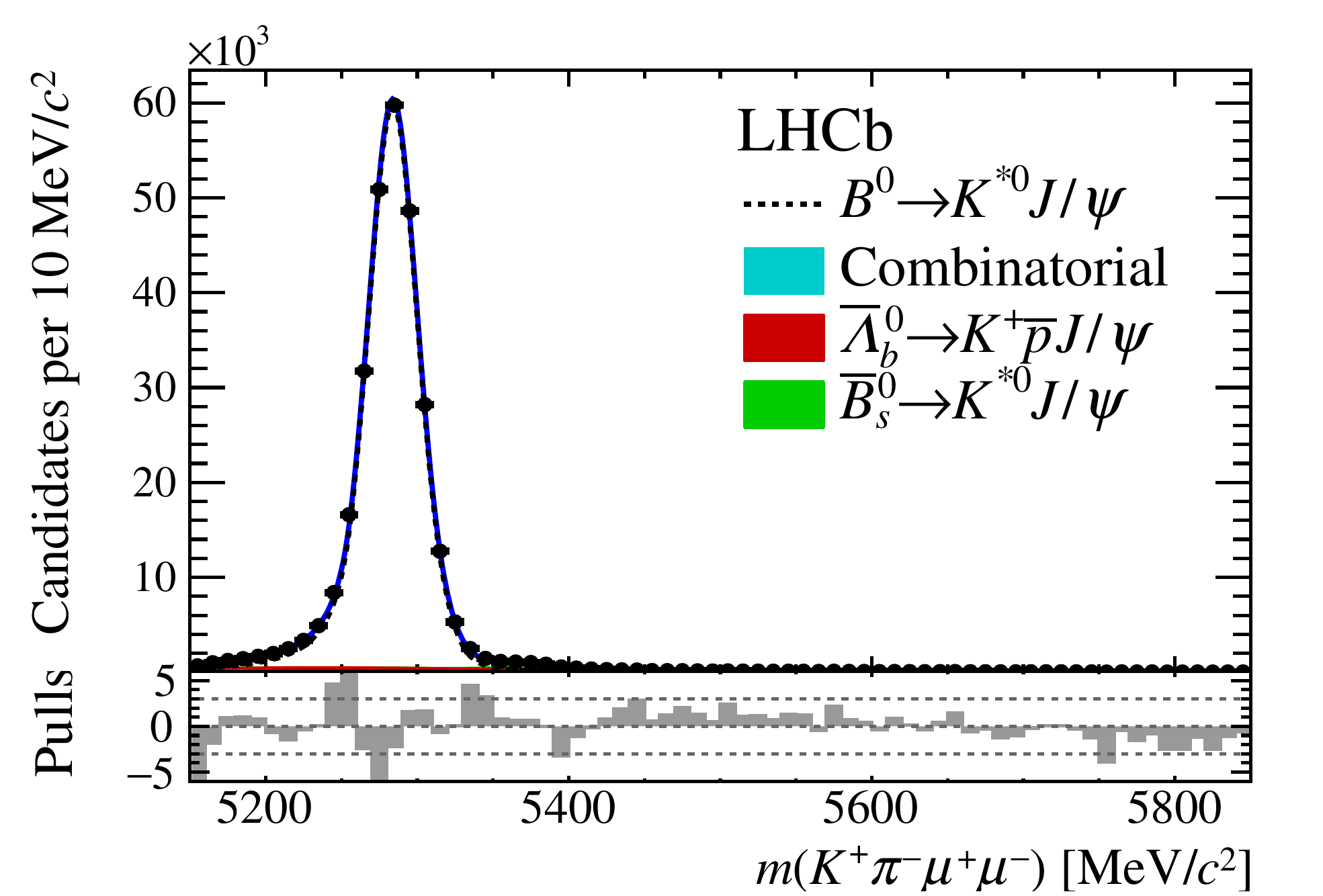}
\hspace{0.5cm}
\includegraphics[width=0.47\textwidth]{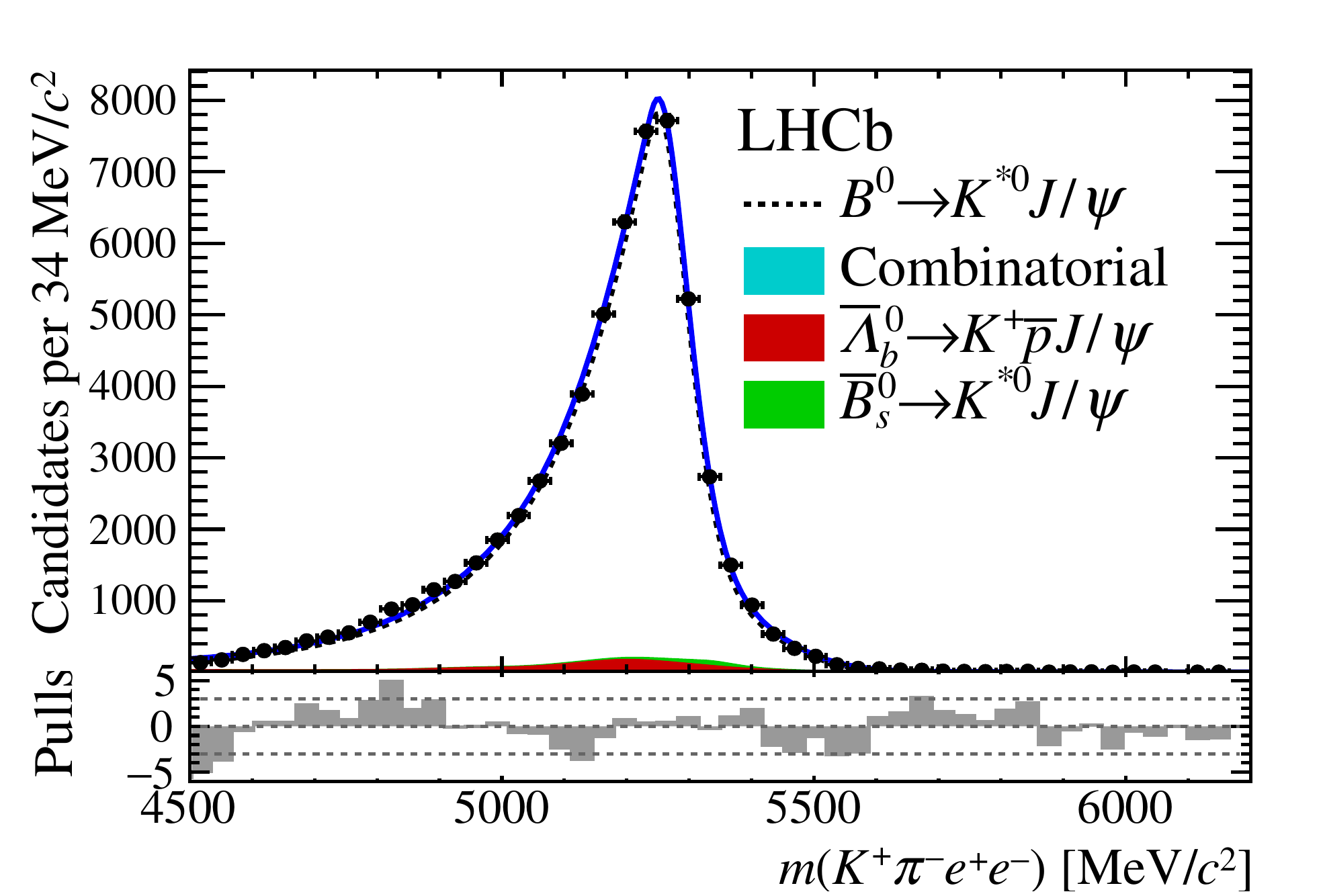}
\caption{Distribution of the four-body invariant mass in \BdToKstKpiJPsll decays for (left) $\ell=\mu$ and (right) $\ell=e$ reconstructed at \lhcb. Taken from Ref.~\cite{LHCb-PAPER-2017-013}.}
\label{fig:JPsiKst_LHCb}
%\end{figure}
\vspace{1cm}
%\begin{figure}[t!]
\centering
\includegraphics[width=0.47\textwidth]{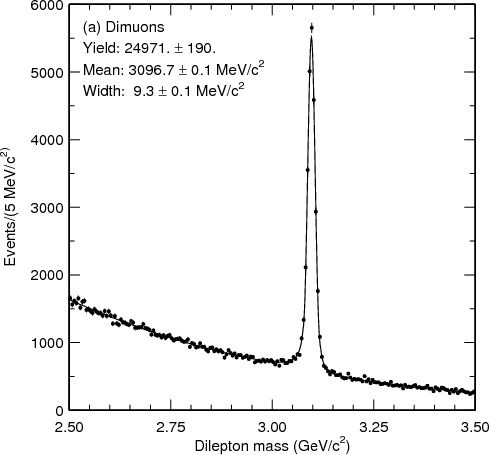}
\hspace{0.5cm}
\includegraphics[width=0.47\textwidth]{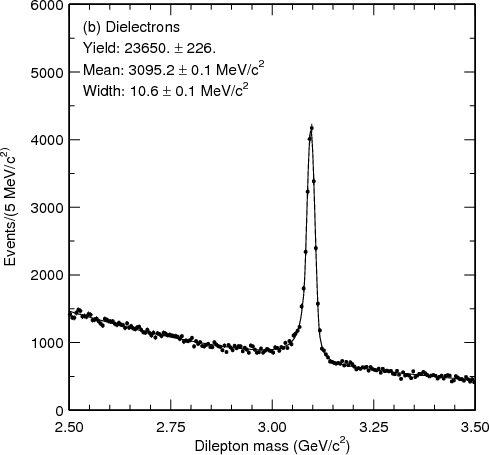}
\caption{Distribution of the dilepton invariant mass in the \jpsi region for (left) $\ell=\mu$ and (right) $\ell=e$ reconstructed at \belle. Taken from Ref.~\cite{Abe:2002rc}.}
\label{fig:JPsi_BELLE}
\end{figure}

Another significant difference compared to the \B-factories is the very large energy of the \B decay products.
Consequently, electrons emit a large amount of bremsstrahlung and a dedicated recovery procedure is in place in order to improve the momentum reconstruction.
Despite this method, the \B mass resolution is still degraded for final states involving electrons compared to final states including muons or charged hadrons, as illustrated in Fig.~\ref{fig:JPsiKst_LHCb}. 

Due to the large CM energy, \bquark-hadrons are produced with a significant boost and their decay length is of the order of a centimetre.
This results in significantly large values of the impact parameter (IP), which is the minimum distance of a track to a primary vertex.
The resolution on the IP is of about 25\mum for a particle with a momentum transverse to the beam of 3\gevc~\cite{LHCb-DP-2014-002}.
Since the average IP for tracks coming from \B decays is of the order of 800\mum, these can be well separated from tracks originating from the primary vertex. 
The resolution of the reconstructed \B invariant mass is similar to the resolution of $\Delta E$ at the \B-factories (for the decay \BdTopipi this amounts to 22\mevcc~\cite{LHCb-PAPER-2012-002}). 

The \lhcb experiment is also characterised by excellent PID performances for charged particles:
the muon efficiency is of the order of 97\%, for a pion mis-identification of approximately 1\%~\cite{LHCb-DP-2013-001};
the electron efficiency is of the order of 90\%, for a pion mis-identification of about 0.6\%~\cite{LHCb-DP-2014-002};
the PID efficiency is of the order of 90\% for charged kaons, while the $\pi$ mis-identification is around 3\%~\cite{LHCb-DP-2014-002}. 

Since the detector performances in the busy \lhc environment are challenging to model, simulated samples are corrected for using unbiased control samples selected from data, in particular for what concerns the \bquark-hadron kinematics, and the trigger and PID efficiencies. 
Finally, contrary to what happens when running at the \FourS ~resonance, all \bquark-hadron species (\eg ~\Bu, \Bd, \Bs, \Bc, and \Lb) are produced at the \lhc and LU tests can be performed exploiting all types of hadrons.

\subsection{Complementarity between the \B-factories and the \lhc}
\label{sec:complementarity}

As previously discussed, the experimental conditions at the \B-factories and at the \lhcb experiment are extremely different.
This section summarises the overall performances achieved for particular \bquark-hadron decays that illustrate the main similarities and differences between the two environments.  

The PID performances for pions and kaons at the three experiments are similar, as well as the \B mass resolution for fully reconstructed final states to charged hadrons.
The background contamination is also comparable for such decays, as can be concluded by examining, for example, the study of the \BmToDzK decay carried out by the three collaborations.
The \babar~\cite{delAmoSanchez:2010dz} and \belle~\cite{Belle:2011ac} experiments have similar performances, where the difference in yields is due to the different integrated luminosities analysed and a slightly more efficient selection for the \belle result.
The yields obtained by the \lhcb experiment on the Run1 dataset~\cite{LHCB-PAPER-2016-003} are more than five times larger than the combined yields at the \B-factories, with a lower level of combinatorial background and a similar \kaon/\pion cross-feed.
While the determination of the absolute scale of the experimental efficiencies is more challenging in the much busier environment of the \lhc compared to the \B-factories, measurements of ratios of observables enable their full control and exploitation of the entire statistical power of these samples.

The situation for electron and muon reconstruction is instead significantly different.
At the \B-factories, the \jpsi invariant mass peak reconstructed using different lepton flavours is similar, as shown in Fig.~\ref{fig:JPsi_BELLE}.
In particular, the event yields are similar indicating that the efficiency to reconstruct the two final states is comparable.
Due to the combined effect of reduced bremsstrahlung and of a performant recovery algorithm, the dilepton invariant mass distribution for the \epem final state only exhibits a slightly wider \jpsi peak and a small radiative tail compared to the \mumu final state. 
On the contrary, the \BdToKstJPs invariant mass reconstructed by \lhcb when the \jpsi is identified via its decay to two muons or two electrons manifests striking differences, as displayed in Fig.~\ref{fig:JPsiKst_LHCb}.
Despite the bremsstrahlung recovery procedure, the electron final state features a long radiative tail and its yield is about five times smaller than that of \BdToKstJPsmm.
The \BuToKJPsll decay mode can also provide interesting information to compare the \B-factories and the \lhcb experiment.
The yield reported in Ref.~\cite{Abe:2002rc} can be scaled to the full integrated luminosity of the \B-factories, amounting to about $81\times10^3$ events that are roughly evenly split between the electron and muon final states.
While this number is similar to the \BuToKJPsee yield ($88\times10^3$) reported in Ref.~\cite{LHCb-PAPER-2014-024} by \lhcb, the \BuToKJPsmm yield is more than seven times larger.   
These significant differences between final states with electron and muon led the \lhcb collaboration to exploit double ratios to measure the \RK and \RKst observables (see Sec.~\ref{sec:loop}). 

Decays of \bquark-hadrons to final states that include a \tauon lepton are more challenging to reconstruct due to the presence of neutrinos.
In order to handle the very large backgrounds resulting from the partial reconstruction of these final states, diverse tools that exploit the extremely different experimental environments of the \B-factories and of the \lhc have been developed.
In both cases, pure leptonic as well as hadronic decays of the \tauon are considered.
At the \B-factories, the missing information due to the undetected neutrinos is compensated by the full reconstruction of the other \B meson in the event, since the only tracks present in the event are originating from the decays of the two \B mesons produced by the \FourS~decay. 
At the \lhcb experiment instead, the additional information needed to constrain the final state is obtained from the precise reconstruction of the decay vertices involved, which, paired with the large boost due to the CM energy of the \lhc, allow the measurement of the flight distance of the particles and provide enough information to compensate for the missing neutrinos.
Despite the very different production rates and the stringent requirements that have to be applied in the more difficult \lhc environment, the results reported in Sec.~\ref{sec:tree} indicate that the statistical precision is similar between the \B-factories and the \lhcb thanks to the dedicated techniques that have been developed.
Moreover, the systematic uncertainties are typically of similar size as the statistical ones.

In summary, all the points discussed in this section highly motivate the necessity and importance of having (at least) two different, but complementary, experimental setups to perform precision tests of LU. 

%%%%%%%%%%%%%%%%%%%%%%%%%%%%%%%%%%%%
% !TEX root = main.tex
%%%%%%%%%%%%%%%%%%%%%%%%%%%%%%%%%%%%

%\clearpage
\section{Lepton Universality tests in \bTocln decays}
\label{sec:tree}

In the SM the electroweak couplings of the gauge bosons to the leptons are independent of the lepton generation involved in the decay (see Sec.~\ref{sec:introduction}).
Precision tests of LU can be performed by studying semileptonic \bTocln decays, where \ellm represents any of the three charged leptons (the electron, $e^-$, the muon, $\mu^-$, and the tau, $\tau^-$) and $\overline{\nu}_{\ell}$ is the corresponding antineutrino.

\subsection{Introduction}

Semileptonic \BToDxmunu and \BToDxenu decays are generally assumed to be free of NP contributions and they are used to perform measurements of the CKM matrix element $|\Vcb|$ and of the hadronic form factors involved in those decays.
Moreover, measurements of the branching fractions \BF{(\BzbToDpmunu)},
\BF{(\BzbToDpenu)}, \BF{(\BmToDzmunu)} and \BF{(\BmToDzenu)} are consistent with each other within experimental uncertainties and in agreement with LU (see Tab.~\ref{tab:tree_EandMuDecays}). 

\begin{table}[h!]
\centering
\renewcommand\arraystretch{1.2}
\begin{tabular}{c|c|c}
\textbf{Experiment (year)}  & \boldmath{\Hc} \textbf{type} & \textbf{Ref.} \\ \hline
CLEO (2002)	& \Dstarpm and \Dstarz	& \cite{Adam:2002uw} \\
\babar (2008)	& \Dstarpm			& \cite{Aubert:2007rs} \\
\babar (2009)	& \Dz and \Dstarz		& \cite{Aubert:2008yv} \\
\belle (2010)	& \Dstarpm			& \cite{Dungel:2010uk} \\
\belle (2016)	& \Dz and \Dp			&  \cite{Glattauer:2015teq} \\
\belle (2018) & \Dstarpm & \cite{Abdesselam:2018nnh} \\
\end{tabular}
\caption{Tests of LU in \HbToHclnu tree-level transitions using the first two generations of leptons, where \Hb and \Hc represent hadrons containing a \bquark and a \cquark quark, respectively.}
\label{tab:tree_EandMuDecays}
\end{table}

However, the large $\tau$ mass (about 17 and 3500 times heavier than the muon and electron mass, respectively) could make semileptonic \B decays to the third generation, referred to as semitauonic \B decays, more sensitive to the presence of NP effects.
Observables to probe NP contributions to such decays are ratios of branching fractions between the third and the first and second generation leptons:
\begin{eqnarray}
\RHc = \frac{\BF(\HbToHctaunu)}{\BF(\HbToHclpnu)} \, ,
\label{eq:RXc}
\end{eqnarray}
\noindent where \ellpr represents an electron or muon for the \B-factories, but only a muon at the \lhcb experiment due to experimental considerations.
This definition cancels a large part of the theoretical ($|\Vcb|$
and form factors) and experimental (branching fractions and reconstruction efficiencies) uncertainties. 

\begin{table}[h!]
\centering
\renewcommand\arraystretch{1.2}
\begin{tabular}{c|c}
\textbf{Decay}		& \boldmath{\BF} \textbf{[\%]} \\ \hline
\tauTomununu		& $17.39 \pm 0.04$ \\
\tauToenunu		& $17.82 \pm 0.04$ \\
\tauTopipiznu		& $25.49 \pm 0.09$ \\
\tauTopinu			& $10.82 \pm 0.05$ \\
\tauTopipipinu		& $9.02 \pm 0.05$ \\
\tauTopipipipiznu	& $4.49 \pm 0.05$ \\
\end{tabular}
\caption{Branching fractions of $\tau$ decays that have been used to perform measurements in semitauonic $H_b$ decays. The 3-prong hadronic modes do not include the $K^0$ contribution. Values taken from Ref.~\cite{PDG2017}.}
\label{tab:tau_decays}
\end{table}

Measurements of \RHc ratios can be performed using different $\tau$ decay
modes.
The final states with the largest branching fractions that have been used to perform measurements in semitauonic \Hb decays are listed in Tab.~\ref{tab:tau_decays}.
The different $\tau$ decay modes have specific advantages and disadvantages:
\begin{itemize}
\item Leptonic \tauTomununu and \tauToenunu decays include one light
charged lepton in the final state.
Experimentally, candidate events are selected by requiring the presence of a charmed hadron, \Hc, and a light charged lepton, $\mu^-$ or $e^-$, which results in a sample containing both signal, \HbToHctaunu, and normalisation, \HbToHcmunu or \HbToHcenu, modes.
This allows the extraction of \RHc from a fit to a single dataset that contains both contributions. 
Systematic uncertainties due to the detection of the light lepton will partially cancel due to the presence of the same lepton flavour both in the numerator and in the denominator of Eq.~\ref{eq:RXc}.
However, this method has to account for the presence of inclusive semileptonic
\HbToHcenuX and \HbToHcmunuX decays, where $X$ represents any possible undetected particle(s), whose branching fractions and form factors are not precisely known.

\item Hadronic $\tau$ final states contain only one neutrino compared to two of the leptonic case (besides the one produced by the \Hb decay) and therefore have a more constrained topology.
The \HbToHctaunu signal sample is composed by events with a charmed hadron, \Hc, and one or three charged pions, also referred to as 1-prong and 3-prong decays, respectively.
Hadronic 1-prong (\tauTopinu and \tauTopipiznu) decays are the most sensitive to the $\tau$ polarisation, while 3-prong (\tauTopipipinu and \tauTopipipipiznu) modes enable reconstructing the $\tau$ vertex, as defined by the three charged pions, in addition to the \Hb decay vertex.
This piece of information can be exploited to estimate the $\tau$ decay time, which improves the signal to background separation.
The main sources of background contamination are due to the presence of inclusive \Hb decays into a \Hc and three charged pions, and doubly-charmed decays \decay{\Hb}{\Hc \D(X)}, where the \D meson ($D_s^+$, $D^+$ or $D^0$) decays inclusively to three charged pions.
Contrary to the leptonic $\tau$ decays, these modes are not very sensitive to the knowledge of the \HbToHcenuX and \HbToHcmunuX backgrounds.  
The presence of different final states in the numerator and denominator of Eq.~(\ref{eq:RXc}) could be a source of systematic uncertainty, particularly in the \lhc environment.
However, this can be mitigated by redefining the \RHc ratio to use external inputs.
\end{itemize}

%Decays of \Bp and \Bz mesons can be studied both at the \B-factories and at a hadronic collider as the LHC, however, decays of \Bs, \Bcm and \Lb hadrons are only accessible at the latter.

\subsection{Results from the \B-factories}

The \babar and \belle collaborations have used different $\tau$ decay modes to
measure the \RD and \RDst ratios~\cite{Lees:2012xj,Lees:2013uzd,Huschle:2015rga}, where \D (\Dstar) indicates either \Dz or \Dp (\Dstarm or \Dstarz) mesons.
A summary of the measurements performed by the two experiments is reported in
Tab.~\ref{tab:RXc_Bfac}.

\begin{table}[h!]
\centering
\resizebox{\textwidth}{!}{
\renewcommand\arraystretch{1.2}
\begin{tabular}{c|c|c|c|c|c|c|c}
\textbf{Experiment (year)} & \boldmath{\B}\textbf{-tag} & \boldmath{$\tau$} \textbf{decay}	& \boldmath{\RD}	& \boldmath{\RDst}	& \textbf{Correlation} & \boldmath{$P_{\tau}(\Dstar)$} & \textbf{Ref.} \\ \hline
      \babar (2012)       & Had     & \tauTolpnunu                        & $0.440 \pm 0.058 \pm 0.042$ & $0.332 \pm 0.024\pm 0.018$      & $-0.27$     & --- & \cite{Lees:2012xj,Lees:2013uzd} \\
      \belle (2015)       & Had     & \tauTolpnunu                        & $0.375 \pm 0.064 \pm 0.026$ & $0.293 \pm 0.038\pm 0.015$      & $-0.49$     & --- & \cite{Huschle:2015rga} \\
      \belle (2016)       & SL      & \tauTolpnunu                        & ---                         & $0.302 \pm 0.030\pm 0.011$      & ---         & --- & \cite{Sato:2016svk} \\
      \belle (2017)       & Had     & $\tau^-\to\pim(\pi^0)\neu_{\tau}$ & ---                    & $0.270 \pm 0.035{~} ^{+0.028}_{-0.025}$ & --- & $-0.38 \pm 0.51{~}^{+0.21}_{-0.16}$ & \cite{Hirose:2016wfn} \\
    \end{tabular}
}
\caption{Measurements of \RD and \RDst performed by the \babar
and \belle experiments. Had (hadronic) and SL (semileptonic) indicate the tagging algorithm used in the analyses.}
\label{tab:RXc_Bfac}
\end{table}

Both collaborations have exploited the fact that the \FourS ~resonance decays
exclusively into a \B\Bb pair, either \Bp\Bm or \Bz\Bzb. One of the \B
mesons, referred as \B-tag, is reconstructed using hadronic (Had) or semileptonic (SL) decays.
The hadronic tagging algorithms search for an exclusive hadronic \B decay from a list of more than one thousand decays, while semileptonic tagging algorithms search for \BToDxlpnu decays, exploiting their large branching fractions.
The efficiencies of these algorithms are quite low, about $0.3\%$ for hadronic and $1\%$ for semileptonic \B-tagged decays.
Although the hadronic tagging has a lower efficiency, this algorithm is the most widely used.
In this case, in fact, the \B-tag decay is fully reconstructed and all remaining particles in the event are coming from the other (signal) \B decay.
This allows the precise measurement of the invariant mass squared of the undetected particles in the signal decay, $\mmsq=(p_{\ee}-p_{\textrm{tag}}-p_{D^{(*)}}-p_{\ellpr})^2$, where $p_{\ee}$, $p_{\textrm{tag}}$, $p_{D^{(*)}}$ and $p_{\ellpr}$ are the four-momenta of the colliding beam particles, the \B-tag candidate, the $\D^{(*)}$ meson and the light lepton.
The \mmsq distribution peaks around zero when only one neutrino is
present in the decay (this is the case for the normalisation \BToDxlpnu modes), while peaks at larger values when more neutrinos are present (signal decays), or any other particle is not detected.

\begin{figure}[t!]
\centering
\includegraphics[width=0.50\textwidth]{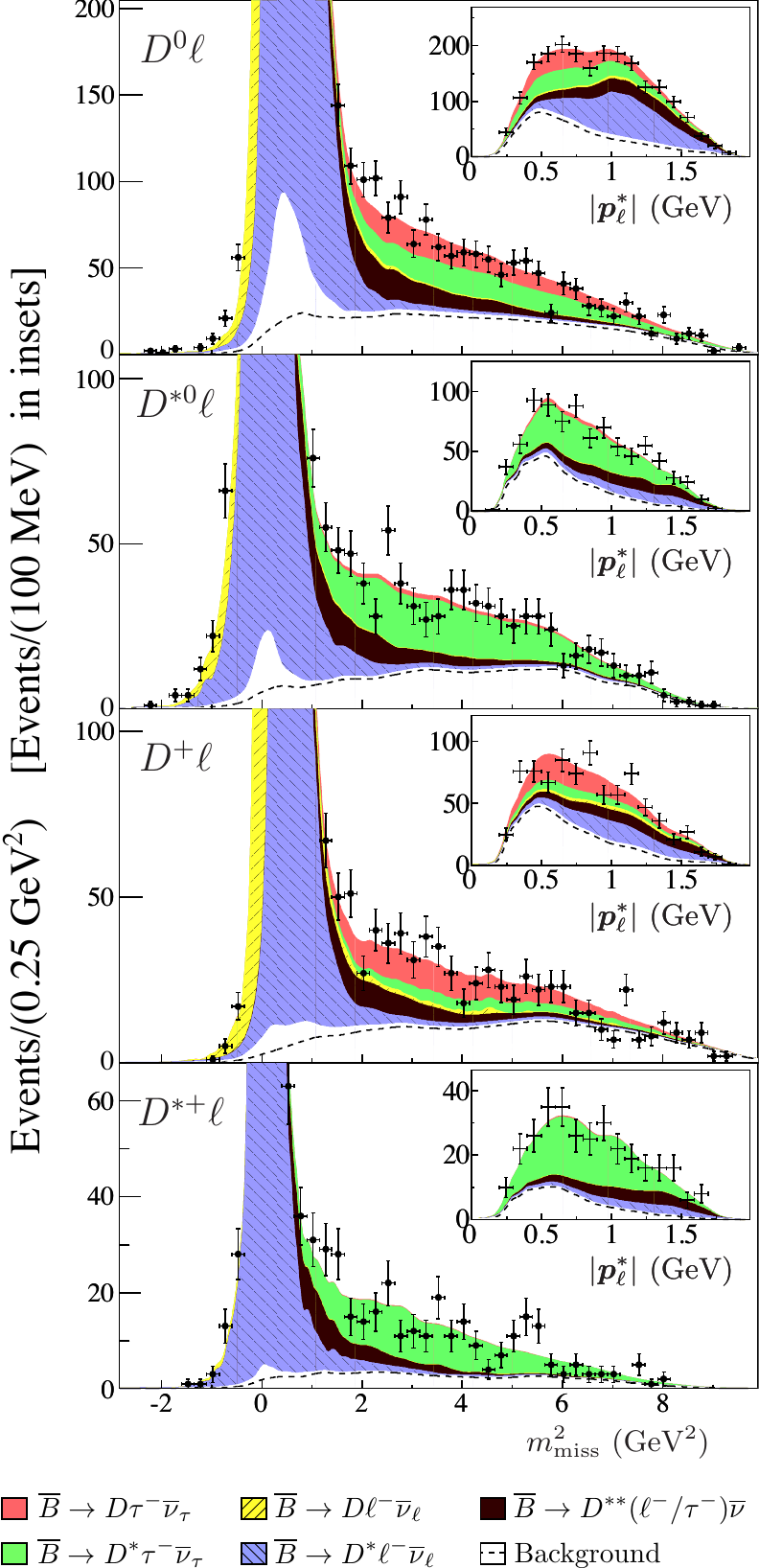}
\caption{Projections on \mmsq from a simultaneous fit to four $D^{(*)}\ellpr$ samples at \babar. Taken from Ref.~\cite{Lees:2012xj}.}
\label{fig:RD_BaBar}
\end{figure}

Using leptonic $\tau$ decays and the hadronic \B-tag, both experiments have exploited their full data sample to perform simultaneous analyses of \BzbToDstptaunu, \BzbToDptaunu, \BmToDstztaunu and \BmToDztaunu decays in the $\qsq>4\gevgevcccc$ region (\qsq is the invariant mass squared of the lepton-neutrino system), where the contribution from signal decays is enhanced. 
The \Dstarp candidate is reconstructed in both $\Dz\pip$ and $\Dp\piz$ decay modes, while for the \Dstarz the $\Dz\piz$ and $\Dz\gamma$ decay channels are used.
The \Dz and \Dp mesons are reconstructed using different two-, three- and four-body final states.
The analyses in both experiments assume isospin conservation ($\RDst=\RDstm=\RDstz$ and $\RD=\RDm=\RDz$).
The \babar collaboration has performed a two-dimensional fit to \mmsq and $E_{\ellpr}^*$, the energy of the light lepton in the \B rest frame~\cite{Lees:2012xj,Lees:2013uzd}.
The \belle experiment has split the data in two regions, below and above $\mmsq=0.85\gevgevcccc$, and performed a simultaneous fit to the two sub-samples~\cite{Huschle:2015rga}.
In the low region, only the \mmsq distribution is fitted, while in the high region, a two-dimensional fit is performed using the \mmsq and the output of a neural network, optimised to suppress contributions from background events.
The neural network includes \mmsq, $E_{\ellpr}^*$, kinematic variables and information from the calorimeter system.
The projection of the fit results on \mmsq for the \babar analysis is shown in Fig.~\ref{fig:RD_BaBar}.
An important background contribution arises from $\B\to D^{**}\ellprm\overline{\nu}_{\ellpr}$ decays, where $D^{**}$ represents excited charm-meson states with a mass greater than that of the \Dstarm and \Dstarz mesons.
These decays, whose branching fractions are not well known, pollute the \mmsq range between the peak at $\mmsq \sim 0$ and the signal region.

The \belle experiment has also performed two additional measurements of \RDst.
The first one using semileptonic \B-tag and leptonic $\tau$ decays~\cite{Sato:2016svk}, and the second with hadronic \B-tag and 1-prong hadronic $\tau$ decays~\cite{Hirose:2016wfn}.
The latter includes also a measurement of the $\tau$ polarisation, $P_{\tau}(\Dstar)=-0.38 \pm 0.51{~}^{+0.21}_{-0.16}$.
In order to tag semileptonic \B decays, a \Dstarp is combined with a light lepton.
The cosine of the angle between the \Bz momentum and the $\Dstarp\ellprm$ in the \FourS ~rest frame, $\cos\theta_{\B-\Dstar\ellpr}$, is then calculated under the hypothesis that only one massless particle is not reconstructed.
The final sample is composed by a 5\% (68\%) of signal (normalisation) events, estimated from simulation.
A neural network, $\mathcal{O}_{\textrm{NB}}$, is trained using simulated samples in order to separate the signal from the normalisation events.
This neural network uses as input $\cos\theta_{\B-\Dstar\ellpr}$, \mmsq and the visible energy $E_{vis}=\sum_iE_i$, where $E_i$ is the energy of the $i\textrm{th}$ particle in the
\FourS ~rest frame.
The ratio \RDst is obtained from a two-dimensional extended maximum-likelihood fit to $\cos\theta_{\B-D^*\ellpr}$ and $\mathcal{O}_{\textrm{NB}}$.
Three parameters are floating in the fit: the signal, the normalisation and the \decay{\B}{\D^{**}\ellprm\overline{\nu_{\ellpr}}} components.
The dominant systematic uncertainties are due to the limited statistics of the simulated samples and the knowledge of the \decay{\B}{\D^{**}\ellprm\overline{\nu_{\ellpr}}} decays. 
The hadronic \B-tag and 1-prong hadronic $\tau$ decay analysis includes semitauonic \B decays with a \Dstarm or a \Dstarz in the final state.
The separation of signal events from background sources is obtained through a fit to a single variable $E_{\textrm{ECL}}$, the energy of the electromagnetic calorimeter clusters not associated to the reconstruction of the \B-tag and the signal candidate.
The signal is normalised with respect to the average of \decay{\B}{\Dstar\ellprm\overline{\nu}_{\ellpr}} decays.
The most important backgrounds are due to events with incorrectly reconstructed \Dstar candidates and contributions from \decay{\B}{\D^{**}\ellprm\overline{\nu}_{\ellpr}} decays which are poorly known.

\subsection{Results from the Large Hadron Collider}

At a \pp collider such as the \lhc, \bbbar pairs are produced with a broad energy spectrum that does not allow using the missing-mass technique developed at the \B-factories to estimate the momentum of the \Hb hadron.
The \Hb direction of flight can be exploited instead.
The long flight distance of the \Hb hadron combined with the excellent resolution of the \lhcb vertex detector provides the measurement of its direction of flight, defined by the vector pointing from the PV to the \Hb decay vertex.
This information is employed not only to estimate the \Hb four-momentum, but also to suppress the background due to the additional particles produced at the PV. 

\lhcb has performed measurements of the ratios \RDst and \RJPs using leptonic \tauTomununu decays, and \RDst using 3-prong hadronic \tauTopipipixnu decays.
The results are summarised in Tab.~\ref{tab:RXc_LHCb}. 
The three analyses are based on the \lhcb Run1 data collected in \pp collision at energies of 7 and 8\tev and corresponding to an integrated luminosity of 3\invfb.

\begin{table}[h!]
\centering
\renewcommand\arraystretch{1.2}
\begin{tabular}{c|c|c|c}
\textbf{Observable}	& \boldmath{$\tau$} \textbf{decay}	& \textbf{Value}	& \textbf{Ref.} \\ \hline
\RDst	& \tauTopipipixnu	& $0.291 \pm 0.019 \pm 0.029$	& \cite{LHCb-PAPER-2017-027,LHCb-PAPER-2017-017} \\
\RDst	& \tauTomununu	& $0.336 \pm 0.027 \pm 0.030$	& \cite{LHCb-PAPER-2015-025} \\
\RJPs	& \tauTomununu	& $0.71 \pm 0.17 \pm 0.18$	& \cite{LHCb-PAPER-2017-035} \\
\end{tabular}
\caption{Measurements of \RDst and \RJPs performed by the \lhcb experiment.}
\label{tab:RXc_LHCb}
\end{table}

\begin{figure}[t!]
\centering
\includegraphics[height=0.21\textwidth]{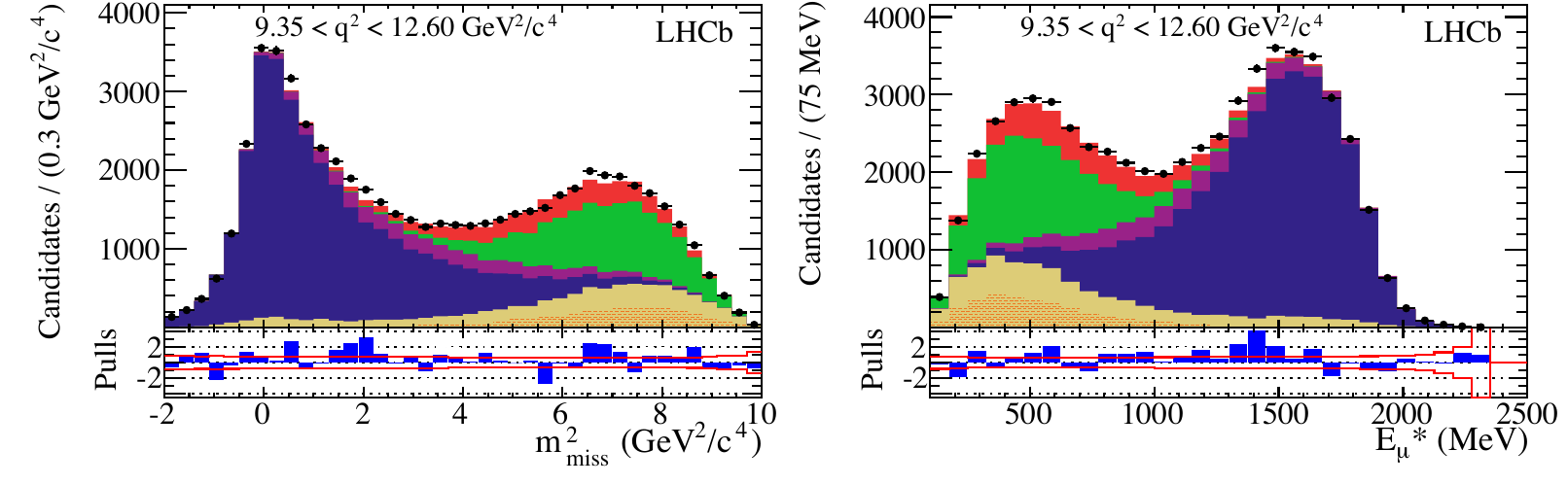}
\hspace{1mm}
\includegraphics[height=0.21\textwidth]{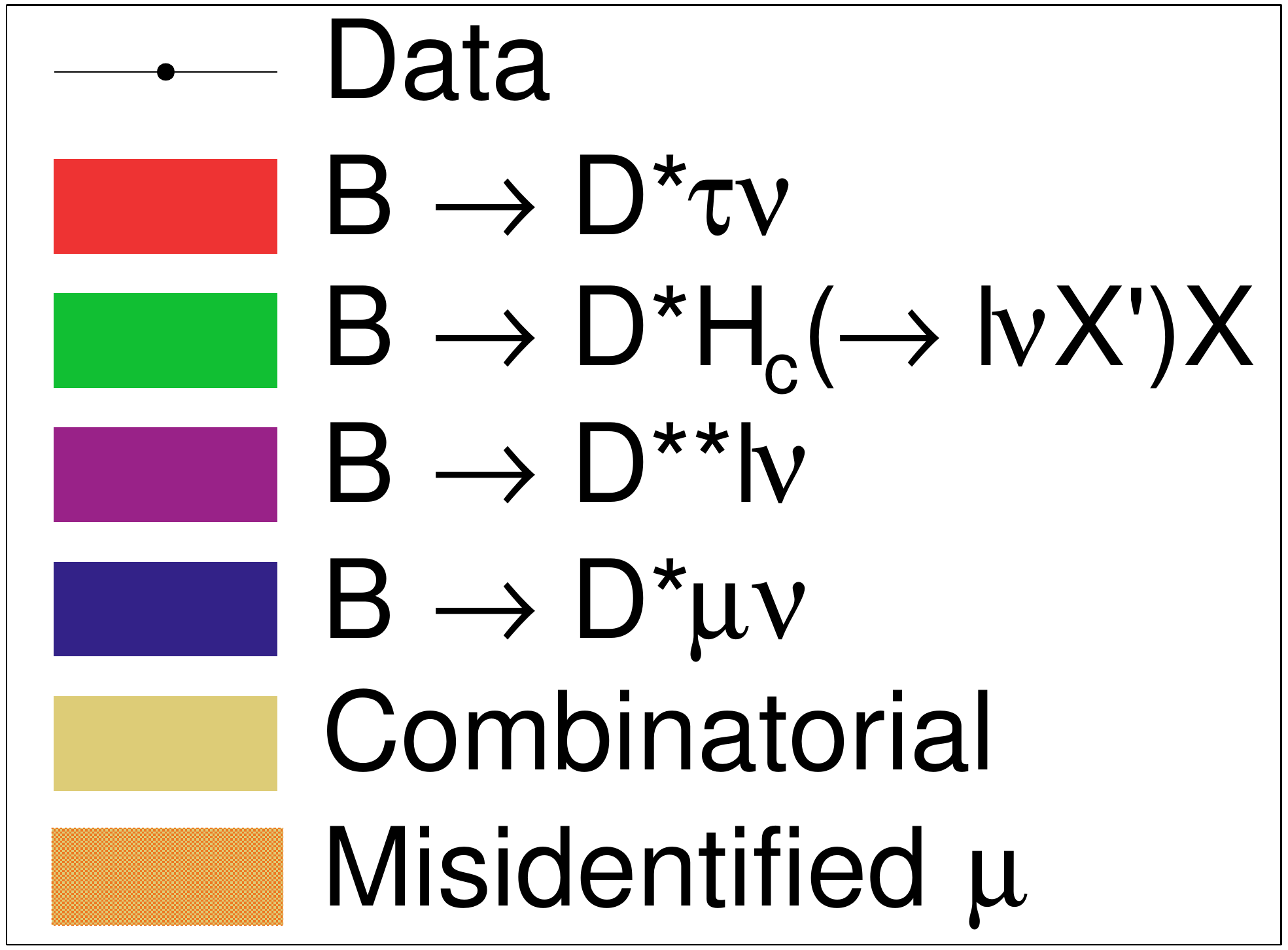}
 \caption{Fit projections on (left) \mmsq and (middle) $E_{\mu}^*$ in the region $9.35 < \qsq < 12.60\gevgevcccc$ for the measurement of \RDst using leptonic \tauTomununu decays at \lhcb. Taken from Ref.~\cite{LHCb-PAPER-2015-025}.}
\label{fig:RDst_muonic_LHCb}
\end{figure}

For the \RDst measurement using leptonic $\tau$ decays~\cite{LHCb-PAPER-2015-025}, the \Bz-hadron momentum, $p_\B$, is approximated by scaling the $\Dstarp\mun$ momentum by the ratio of the \Bz mass to the \mbox{\Dstarp\mun} mass.
The \RDst ratio is obtained from a three-dimensional fit to data, using templates obtained from simulation and validated using data control samples, including \qsq, $E_{\mu}^*$ (the energy of the muon in the \B CM system), and the missing mass squared, calculated as $\mmsq=(p_\B-p_{\Dstarm} - p_{\mu})^2$, where $p_\B$, $p_{\Dstarm}$ and $p_{\mu}$ are the four-momentum of the \B, the \Dstarm and the muon.
The fit projections are shown in Fig.~\ref{fig:RDst_muonic_LHCb}.
The main systematic uncertainties arise from the limited statistics of the simulated samples and the knowledge of the contribution from doubly-charmed \decay{\B}{\Dstarm\D(X)} decays.

\begin{figure}[t!]
\centering
\includegraphics[width=0.47\textwidth]{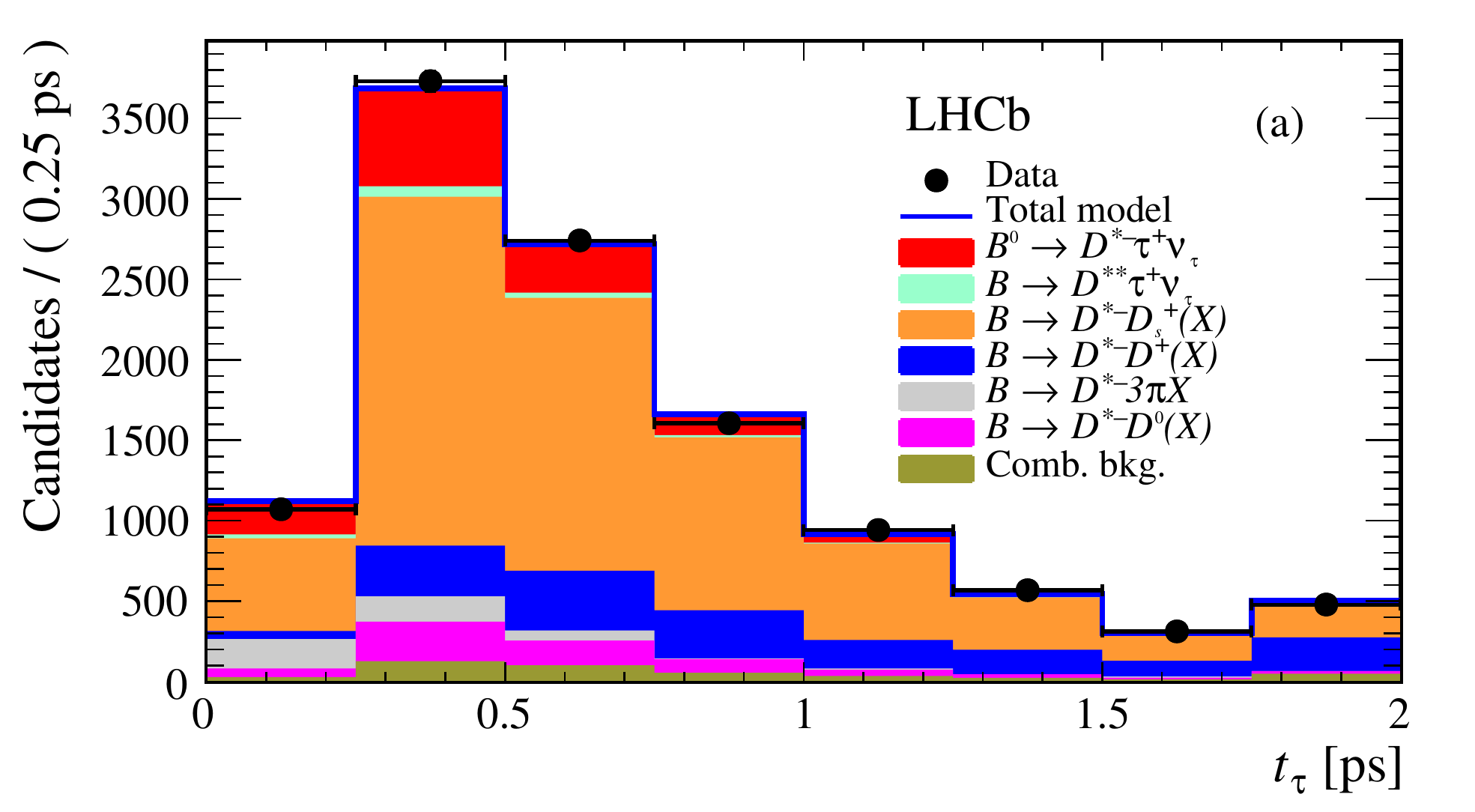}
\hspace{0.5cm}
\includegraphics[width=0.47\textwidth]{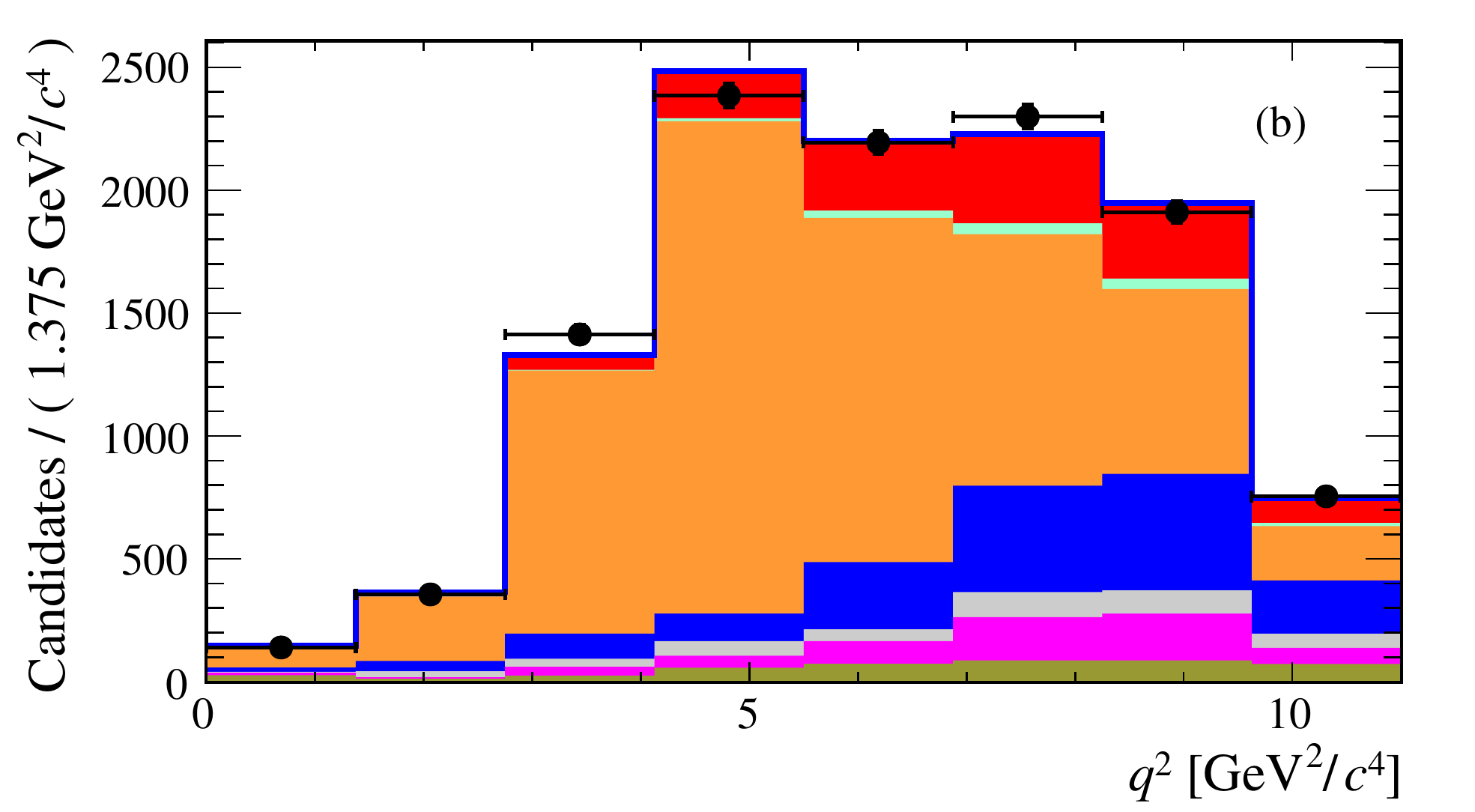}
\caption{Fit projections on (left) the $\tau$ decay time and (right) \qsq for the measurement of \RDst using 3-prong hadronic $\tau$ decays at \lhcb. Taken from Refs.~\cite{LHCb-PAPER-2017-027,LHCb-PAPER-2017-017}.}
\label{fig:RDst_3prong_LHCb}
\end{figure}

Using 3-prong hadronic $\tau$ decays the \lhcb collaboration has also measured the ratio of branching fractions \mbox{$\BF(\BzbToDstptaunu)/\BF(\Bz\to\Dstarm\pip\pim\pip)$}~\cite{LHCb-PAPER-2017-027,LHCb-PAPER-2017-017}. 
From this measurement, \RDst can be obtained using the \decay{\Bz}{\Dstarm\pip\pim\pip} and \BzbToDstpmunu branching fractions as external inputs~\cite{PDG2017}.
Thanks to the use of the \tauTopipipixnu decay, it is possible to exploit the $\tau$ decay vertex.
The \Bz and $\tau$ momenta are estimated imposing constraints on the \Bz and $\tau$ masses, in order to reconstruct the $\tau$ decay time and $\qsq=(p_\B-p_\Dstarm)^2$.
The dominant background due to inclusive \decay{\B}{\Dstarm\pip\pim\pip(X)} decays is highly suppressed by requiring the $\tau$ vertex to be significantly displaced from the \B vertex.
The remaining background is dominated by doubly-charmed \decay{\B}{\Dstarm\D(X)} decays, where \D represents a $D_s^+$, $D^0$ or $D^+$ meson.
A boosted decision tree (BDT) is trained to suppress this background.
The BDT includes information from the calorimeter system to suppress events where a photon or a \piz are present, the invariant masses of oppositely charged pion pairs, the invariant mass of the $\Dstarm\pip\pim\pip$ system and additional kinematic variables.
After applying a requirement on the BDT, a three-dimensional fit to the BDT, the $\tau$ decay time and the \qsq variable is performed. 
The fit projections are shown in Fig.~\ref{fig:RDst_3prong_LHCb}. The main systematic uncertainties are due to the limited size of the simulated samples and the knowledge of the doubly-charmed decays \decay{\B}{\Dstarm\D(X)}.

\begin{figure}[t!]
\centering
\includegraphics[width=0.47\textwidth]{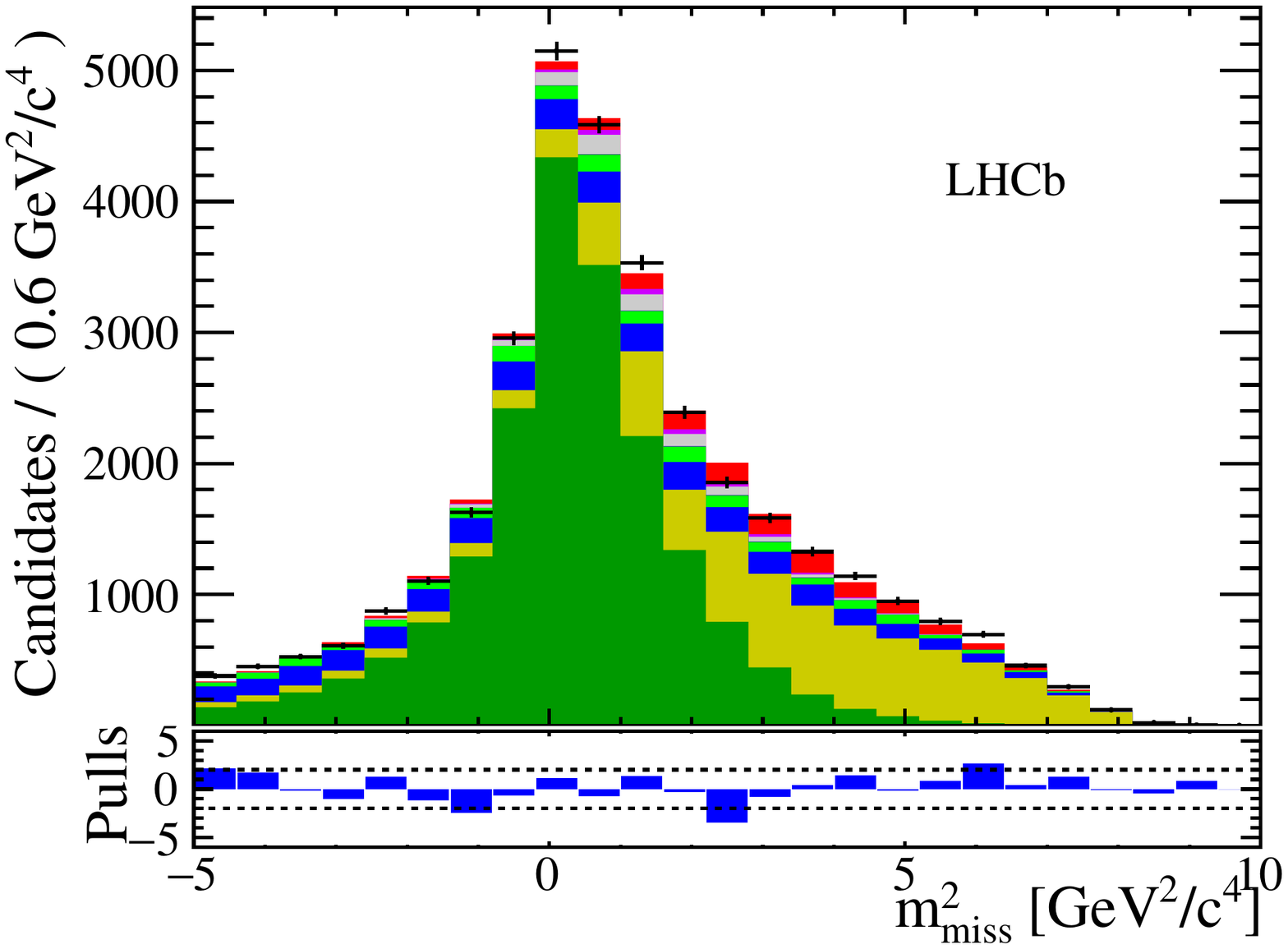}
\hspace{0.5cm}
\includegraphics[width=0.47\textwidth]{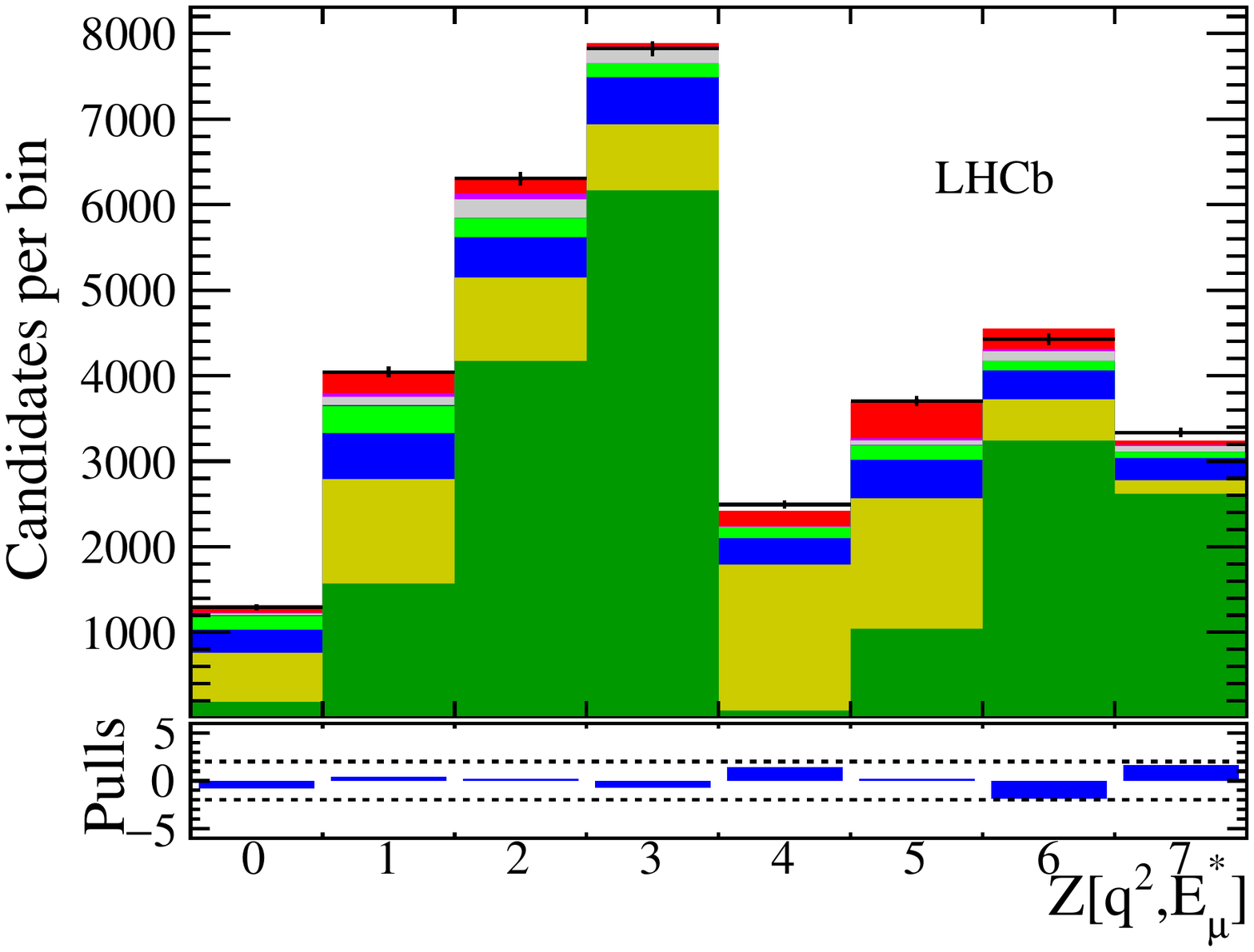}
\\ \vspace{0.5cm}
\includegraphics[width=0.47\textwidth]{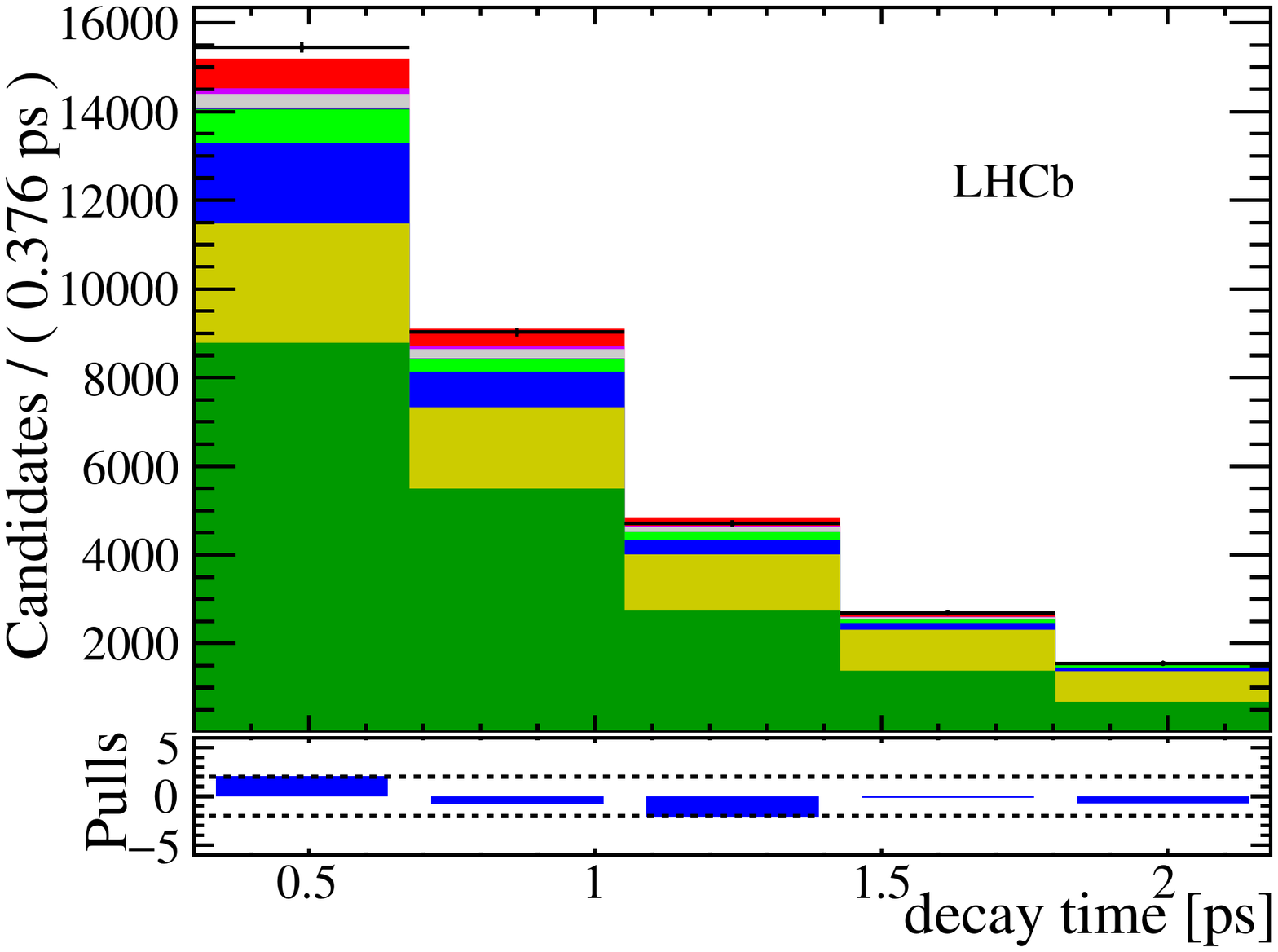}
\hspace{0.5cm}
\includegraphics[width=0.47\textwidth]{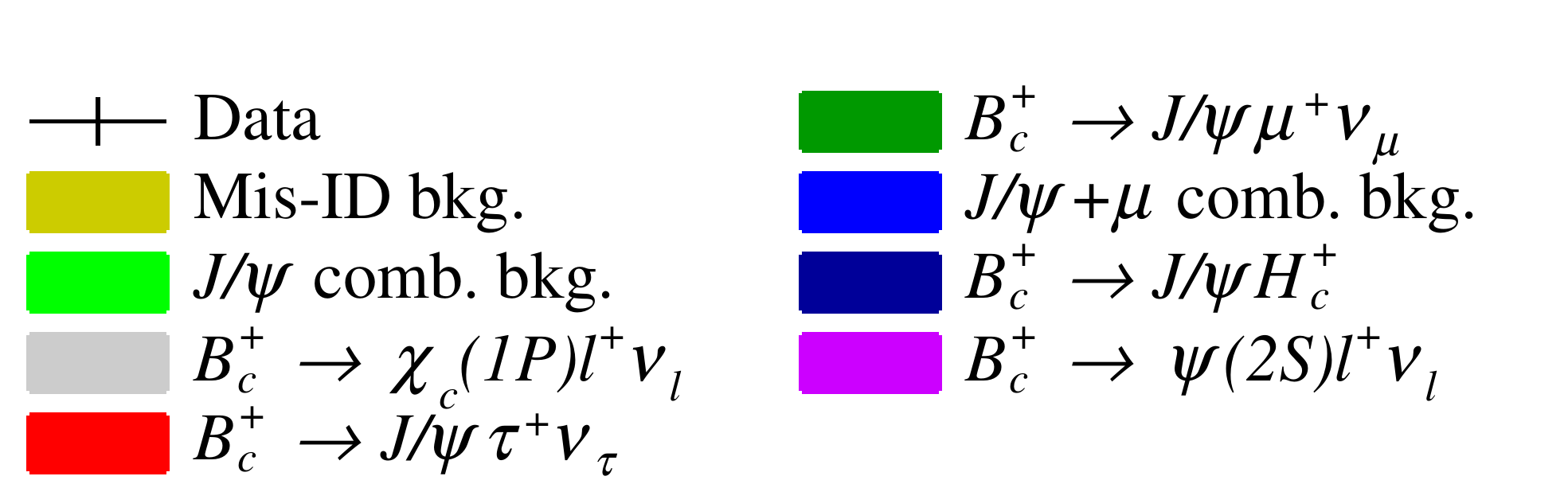}
\caption{Fit projections on (top left) \mmsq, (top right) $Z(\qsq,E_{\mu}^*)$ and (bottom left) \Bcp decay time for the measurement of \RJPs using leptonic \tauTomununu decays at \lhcb. Taken from Ref.~\cite{LHCb-PAPER-2017-035}.}
\label{fig:RJpsi}
\end{figure}

In addition to the \RDst measurements the \lhcb collaboration has taken advantage of the \Bc production at the \lhc to measure the ratio \RJPs using leptonic $\tau$ decays~\cite{LHCb-PAPER-2017-035}.
The \jpsi meson is reconstructed using the \mpmm final state, which offers a clear trigger signature that helps compensating for the small \Bc production. 
The \Bc momentum is approximated in the same way as in the \RDst measurement. 
The ratio \RJPs is obtained from a three-dimensional binned fit to \mmsq, the \Bc decay time and $Z[\qsq,E_{\mu}^*]$, a discrete variable representing eight bins
in $(E_{\mu}^*,\qsq)$.
The \Bc decay time helps to separate the signal from backgrounds due to \Bz and \Bp decays.
The signal and background components are described by templates obtained from simulated events and validated with data.
The fit projections are shown in Fig.~\ref{fig:RJpsi}.
As for the other analyses, a significant part of the systematic uncertainty is due to the size of the simulation samples, but contrary to the \RDst measurements, the systematic uncertainty due to the poor knowledge of the form factors involved in \Bc semileptonic decays and its consequence on the shapes extracted from the simulation and used in the fit is also important. 

\begin{figure}[t!]
\centering
\includegraphics[width=0.80\textwidth]{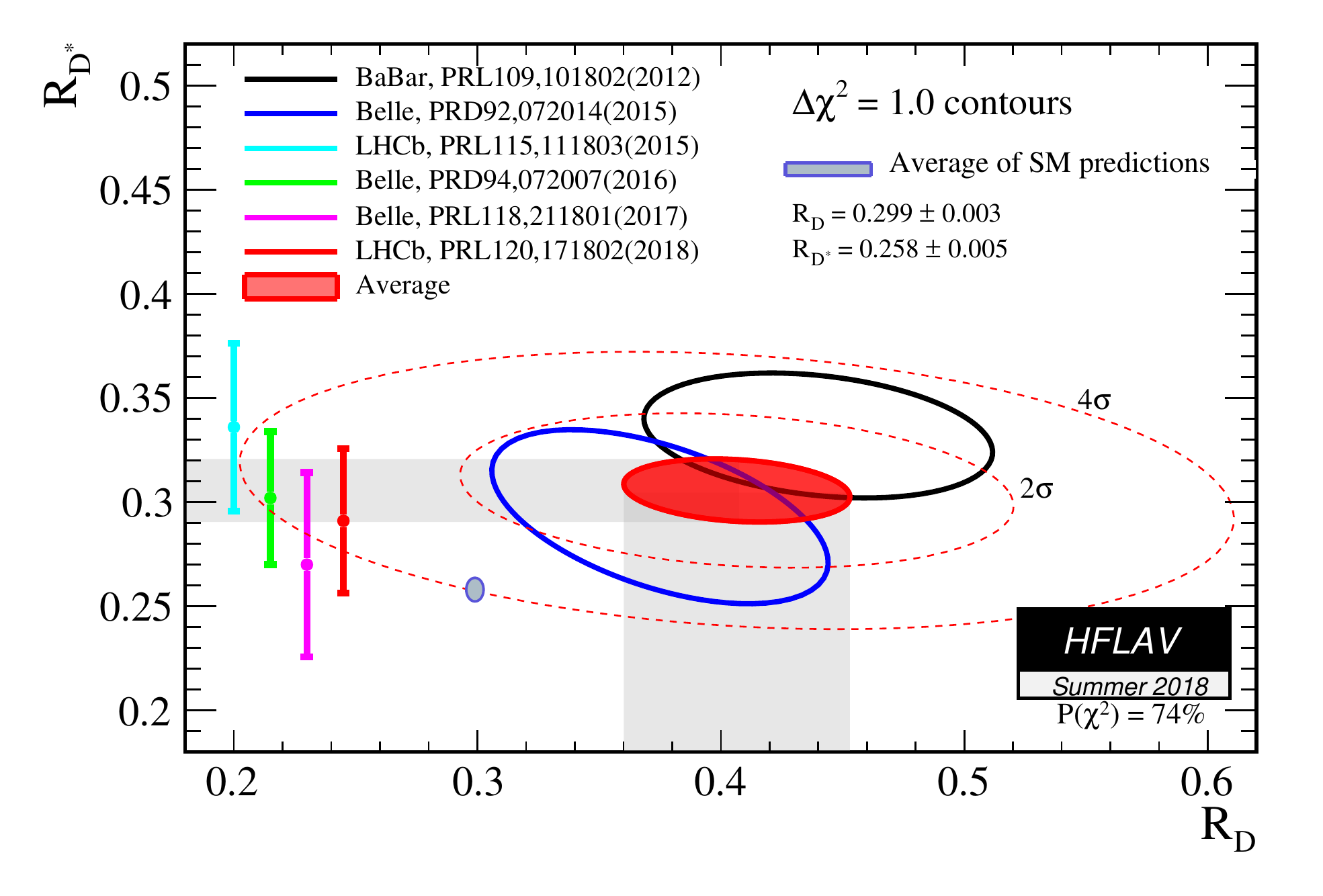}
\caption{Experimental results on \RD and \RDst and comparison with the SM prediction. Taken from Ref.~\cite{Amhis:2016xyh} (online update).}
\label{fig:RDRDst_HFLAV}
\end{figure}

A summary of the measurements of \RD and \RDst performed at the \B-factories and by \lhcb is shown in Fig.~\ref{fig:RDRDst_HFLAV}.

\subsection{Standard Model predictions}

As discussed in Sec.~\ref{sec:frameworkTh}, FCCC \bTocln transitions are described through an effective Hamiltonian that allows separating the amplitude into a leptonic tensor and a hadronic tensor of the form $\langle H_c|\bar{c}\gamma_\mu(1-\gamma_5)|H_b\rangle$.
The latter can be can be decomposed on a basis of possible Lorentz structures, which depend on the spin of the two mesons.
In the case of a final \D-meson (spin 0), the SM prediction involves 2 form factors, $f_+$ and $f_0$, whereas 4 form factors, $V$ and $A_{0,1,2}$, are involved in case
of the \Dstar meson (spin 1) (a similar decomposition occurs in the case of the decay \decay{\Bcm}{\jpsi\ellm\neulb}).
The same form factors are involved even in the presence of NP, unless tensor operators are present (requiring additional form factors).

The expressions of the differential decay rates (including the angular dependence) have been known for some time~\cite{Korner:1989qb}.
In the case of \decay{\B}{\Dstar}, it is possible to include the subsequent decay \decay{\Dstar}{\D\pi}, which adds further kinematics variables (in particular the angle between the \D and \Dstar mesons -- the corresponding expressions can be found in Refs.~\cite{Becirevic:2016hea,Alonso:2016gym,Ligeti:2016npd}).

The decay rates for the heavy $\tau$ lepton and the light $e$ and $\mu$ leptons differ by terms proportional to $m_\tau$, meaning that the ratios testing LU with these modes will involve specific ratios of form factors (\eg $f_0/f_+$ for the \D meson, $A_0/V$ for the \Dstar meson).
This implies that the SM predictions for the ratios \RD, \RDst and \RJPs will not be equal to 1, and that they will rely on information concerning ratios of form factors.
\begin{itemize}
\item For \decay{\B}{\D\ellm\neulb}, the form factors were evaluated by two different lattice collaborations, MILC and HPQCD~\cite{Lattice:2015rga,Na:2015kha}. In Ref.~\cite{Bigi:2016mdz}, the results were combined together with experimental information from \B factories on $f_+$ (assuming no NP in decays involving light leptons) that leads to very similar results for \RD (but not for other quantities like $|\Vcb|$). 
\item For \decay{\B}{\Dstar\ellm\neulb}, the strong decay of the \Dstar meson makes the theoretical evaluations of the form factors more complicated. In Ref.~\cite{Fajfer:2012vx}, these form factors were expressed using the Heavy-Quark Expansion (HQE) supplemented with estimates of higher-order corrections and combined with experimental results on \decay{\B}{\Dstar\en\neueb} and \decay{\B}{\Dstar\mun\neumb}, assuming that no NP is present in decays involving light leptons. Concerns have been raised recently about HQE-based parameterisations of the \decay{\B}{\D^{(*)}\ellm\neulb} form factors, potentially affecting the extraction of $|\Vcb|$~\cite{Bernlochner:2017xyx}. However, fits using different HQE-inspired parameterisations and combining experimental results on light-lepton decays, Light-Cone Sum Rules, and lattice inputs show very little dispersion in the predictions for \RD and \RDst~\cite{Bernlochner:2017jka}. These predictions do not take into account the width of the $D^*$-meson, whose effect is currently under investigation and could have a noticeable impact of a few percent on the measurement of \RDst~\cite{Chavez-Saab:2018lcn,Le-Yaouanc:2018ojj}.
\item Estimates of the electromagnetic radiative corrections~\cite{deBoer:2018ipi} suggest that soft-photon corrections could enhance the SM prediction for \RD and \RDst (comparing $\tau$ and $\mu$ modes) by 5\% and 3\% respectively, for a soft photon energy cut between 20 and 40\mev. 
\item The situation is less satisfactory in the case of \decay{\Bcm}{\jpsi\ellm\neulb} form factors, as the same methods cannot be used easily due to the \cquark-quark mass which is of the same order as the typical QCD scale $\Lambda$ and thus neither light nor heavy. Model computations exist~\cite{Ivanov:2005fd,Hernandez:2006gt,Wen-Fei:2013uea,Dutta:2017xmj,Watanabe:2017mip}, but there are (up to now) no lattice or Light-Cone Sum Rules estimates for these form factors, and the systematics attached to these predictions remain difficult to assess (see Refs ~\cite{Lytle:2016ixw,Colquhoun:2016osw,Huang:2018nnq} for preliminary results in this direction). 
\end{itemize}

\noindent This leads to the predictions listed in Tab.~\ref{tab:pred-SM-tree} and summarised in Fig.~\ref{fig:RDRDst_HFLAV}. \RD and \RDst exceed the SM predictions by around 2.3$\,\sigma$ and 3.0$\,\sigma$, respectively, leading to a combined pull around 3.6--3.8$\,\sigma$ depending on the particular choice for the SM prediction of \RDst.

\begin{table}[t!]
\centering
\renewcommand\arraystretch{1.2}
\begin{tabular}{c|c|c}
\textbf{Observable} & \textbf{SM prediction} & \textbf{Ref.} \\ \hline
\RD		& $0.299\pm 0.011$			& \cite{Lattice:2015rga} \\
\RD		& $0.300\pm 0.008$			& \cite{Na:2015kha} \\
\RD		& $0.299\pm 0.003$			& \cite{Bigi:2016mdz} \\
\RD     & $0.299 \pm 0.004$         &
\cite{Jaiswal:2017rve}\\
\RDst	& $0.252 \pm 0.003$			& \cite{Fajfer:2012vx} \\
\RDst   & $0.260 \pm 0.008$         &
\cite{Bigi:2017jbd}\\
\RDst   & $0.257 \pm 0.005$         &
\cite{Jaiswal:2017rve}\\
\RJPs	& $0.268^{+0.018}_{-0.001}$	& \cite{Hernandez:2006gt} \\
\RJPs	& $0.283\pm 0.048$			& \cite{Watanabe:2017mip} \\
\end{tabular}
\caption{SM predictions for the LU ratios.}
\label{tab:pred-SM-tree}
\end{table}

A last comment is in order here. The integrated branching fractions contain only a part of the information that can be extracted from these decays. 
The \qsq dependence of the differential decay rate for \decay{\B}{\D\ellm\neulb} provides also interesting information as various operators of the effective Hamiltonian can affect this shape differently.
The overall shape is currently compatible with the SM within errors~\cite{Freytsis:2015qca,Bernlochner:2017jka}.
There are also predictions for other observables, in particular the angular observables~\cite{Becirevic:2016hea,Alonso:2016gym} describing the kinematics of the decay products.

%%%%%%%%%%%%%%%%%%%%%%%%%%%%%%%%%%%%
% !TEX root = main.tex
%%%%%%%%%%%%%%%%%%%%%%%%%%%%%%%%%%%%

%\clearpage
\section{Lepton Universality tests in \bTosll decays}
\label{sec:loop}

Decays of \B hadrons involving a loop-level transition of the type \bTosll provide an ideal laboratory to test LU.
This class of decays presents different challenges compared to the more frequent, leading-order, \bTocln decays (see Sec.~\ref{sec:tree}), both experimentally and theoretically.

\subsection{Introduction}

Since there are no tree-level contributions in the SM, FCNC decays are highly suppressed and hence provide an increased sensitivity to the possible existence of NP.
The leading Feynman diagrams of a \bTosll transition proceed through the exchange of either a $\Z/\gamma$ penguin, or a $\Wp\Wm$ box, as illustrated in Fig.~\ref{fig:feynBdecay-btosll}.

Sensitive probes of LU that exploit \bTosll decays are ratios of the type~\cite{Hiller:2003js}:
\begin{equation}
\RHs = \frac{\bigintssss_{\qsqmin}^{\qsqmax} \frac{ d\Gamma(\Hb \to \Hs\mumu) }{d\qsq} \, d\qsq}{\bigintssss_{\qsqmin}^{\qsqmax} \frac{ d\Gamma(\Hb \to \Hs\epem) }{d\qsq} \, d\qsq} \, ,
\end{equation}
where \Hb represents a hadron containing a \bquark-quark, such as a \Bp or a \Bz meson, \Hs represents a hadron containing an \squark-quark, such as a \kaon or a \Kstar meson, and \qsq is the invariant mass squared of the dilepton system integrated between \qsqmin and \qsqmax.
As outlined in Sec.~\ref{sec:observables}, ratios of branching fractions allow very precise tests of LU, as hadronic uncertainties in the SM predictions are largely reduced, QED corrections are controlled at the level of $\sim1\%$~\cite{Bordone:2016gaq}, and experimental systematic uncertainties cancel to a large extent.
In the SM such ratios, expected to be close to unity above the dimuon threshold~\cite{Hiller:2003js,Bobeth:2007dw,Bouchard:2013mia}, are sensitive to contributions from new particles that can induce a sizeable increase or decrease in the rate of particular decays. These contributions can also modify the angular distribution of the final-state particles, which can provide further tests of LU.

As discussed in Sec.~\ref{sec:complementarity}, there are significant differences between electron and muon reconstruction performances for experiments operating at the \B-factories compared to the \lhc.
Nevertheless, some features are common: for example, the regions around the charmonium resonances are vetoed, as being dominated by SM hadronic physics (see Sec.~\ref{sec:ccbar}), and the \Kstar is reconstructed in a mass region dominated by the $\Kstar(892)$ state, where the pollution from partial waves other than the P-wave can be neglected.
The last assumption is supported by the fact that the fraction of the \kaon\pion S-wave component is relatively small~\cite{LHCB-PAPER-2016-012}, and similar between electrons and muons~\cite{LHCb-PAPER-2017-013} within the current statistical precision.
However, let us add that NP can potentially modify the relative fraction of \Kstar partial waves.

\subsection{Results from the \B-factories} \label{sec:loop-Bfactories}

The \babar and \belle experiments have exploited the clean environment of \epem colliders and the knowledge of the \B\Bbar initial-state provided by running at the $\PUpsilon(4S)$ resonance to make the first measurements of the \RKX ratios~\cite{Lees:2012tva,Wei:2009zv}.

\begin{figure}[t!]
\centering
\begin{overpic}[width=0.94\textwidth]{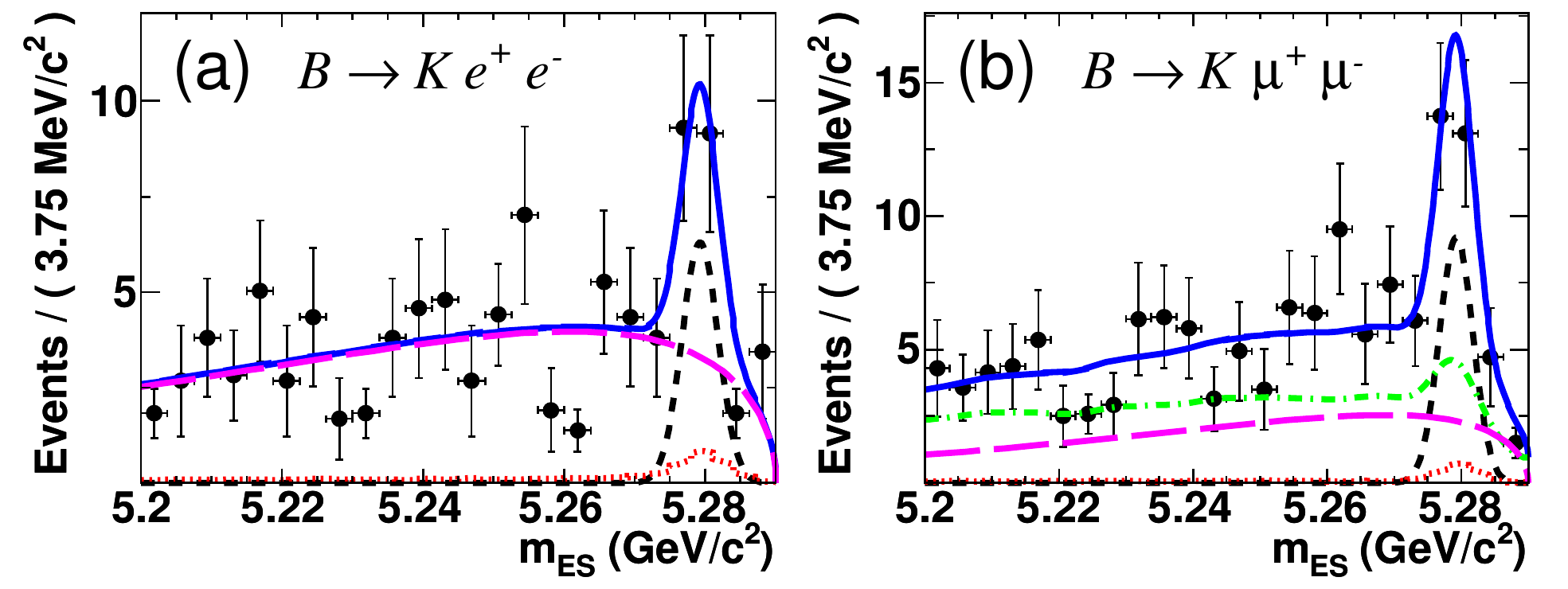}
\put(65,34){\color{white}\circle*{10}}
\put(62,34){\color{white}\circle*{10}}
\put(65,32){\color{white}\circle*{10}}
\put(62,32){\color{white}\circle*{10}}
\put(15,34){\color{white}\circle*{10}}
\put(12,34){\color{white}\circle*{10}}
\put(15,32){\color{white}\circle*{10}}
\put(12,32){\color{white}\circle*{10}}
\end{overpic}
\caption{Fits to (left) \BToKmm and (right) \BToKee candidates at low recoil ($\qsq > 10.11 \gevgevcccc$), but excluding  the \psitwos mass region, at \babar. Taken from Ref.~\cite{Lees:2012tva}.}
\label{fig:loopBFactories}
\end{figure}

Data samples corresponding to about 433 and 605\invfb of \epem collisions have been examined by the two collaborations, respectively.
Both charged and neutral \BToKXll decays are reconstructed, where the hadronic system can be a \KS, \Kpm, \Kstarz or \Kstarpm meson.
The \KS is reconstructed via its decay to a \pip\pim pair, and the \Kstar via its decay to \kaon\pion.
The knowledge of the initial-state energy allows the exploitation of beam-energy constraints to identify candidates consistent with \BToKXll decays and control the background contamination.
Peaking backgrounds are vetoed by means of specific requirements, while the combinatorial and continuum backgrounds are reduced by the use of multivariate analysis techniques that exploit several event properties to allow separation with respect to the signal.
When reconstructing electrons, bremsstrahlung photons are recovered by including neutral clusters consistent with the electron direction in the calorimeter.
Furthermore, background coming from photon conversions is typically rejected at low dielectron masses.
The yields are extracted by \babar in two regions of \qsq (Fig.~\ref{fig:loopBFactories} for the high-\qsq region), below the \jpsi and above the \psitwos resonances, while \belle determines the yields in six bins of \qsq, where the regions around the charm resonances are excluded. 

The \RKX measurement is presented by \babar in two regions of \qsq, while only in one by \belle. 
The results have a relative precision of 20 to 50\%, which is dominated by the statistical uncertainty.
Systematic uncertainties due to the hadronic system largely cancel, the only remaining ones stemming from the lepton identification and signal extraction.
A summary of the \RKX measurements is presented in Table~\ref{tab:RKX} and in Fig.~\ref{fig:loopSummary}.

\begin{figure}[t!]
\centering
\includegraphics[width=0.47\textwidth]{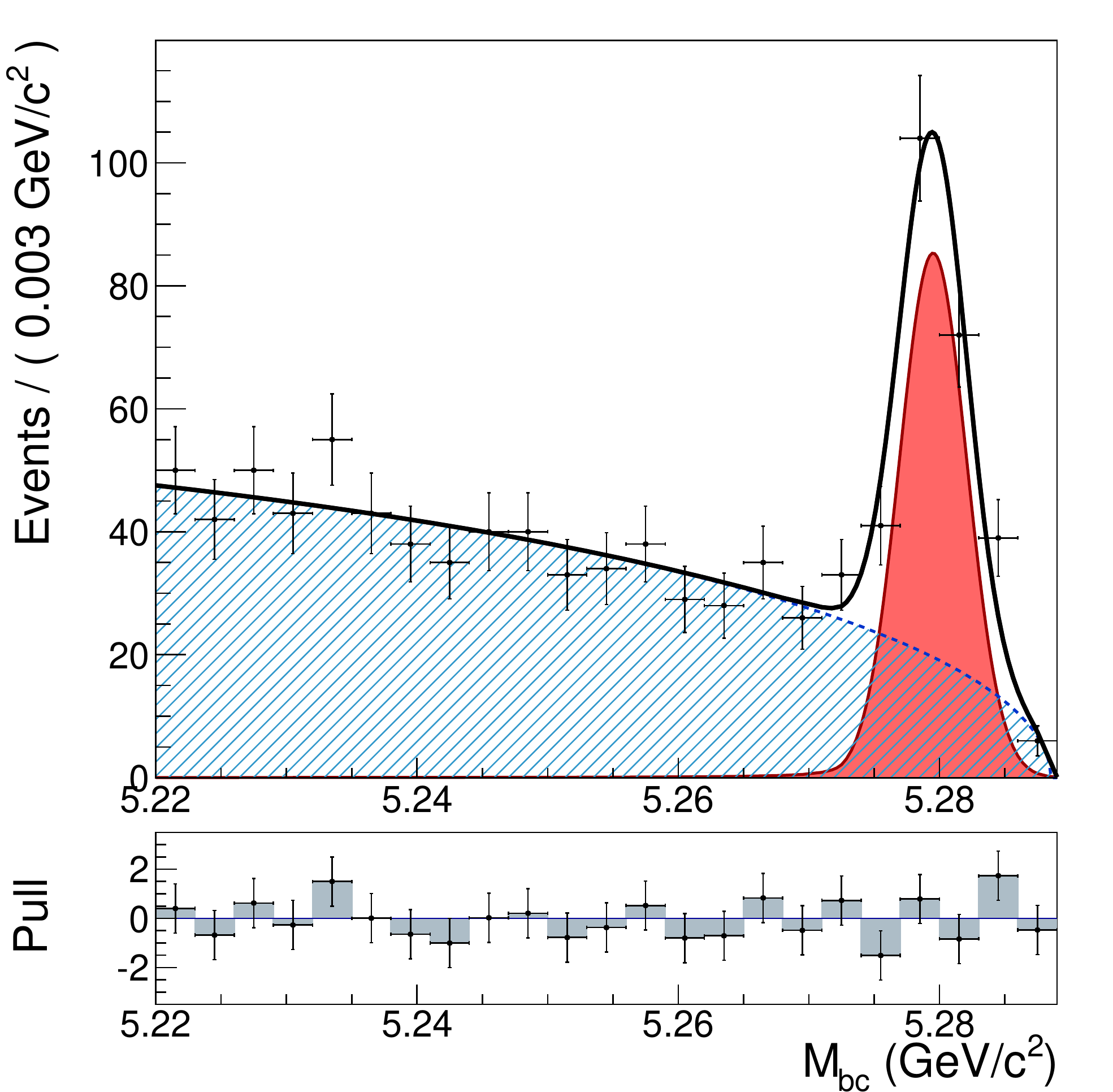}
\hspace{0.5cm}
\includegraphics[width=0.47\textwidth]{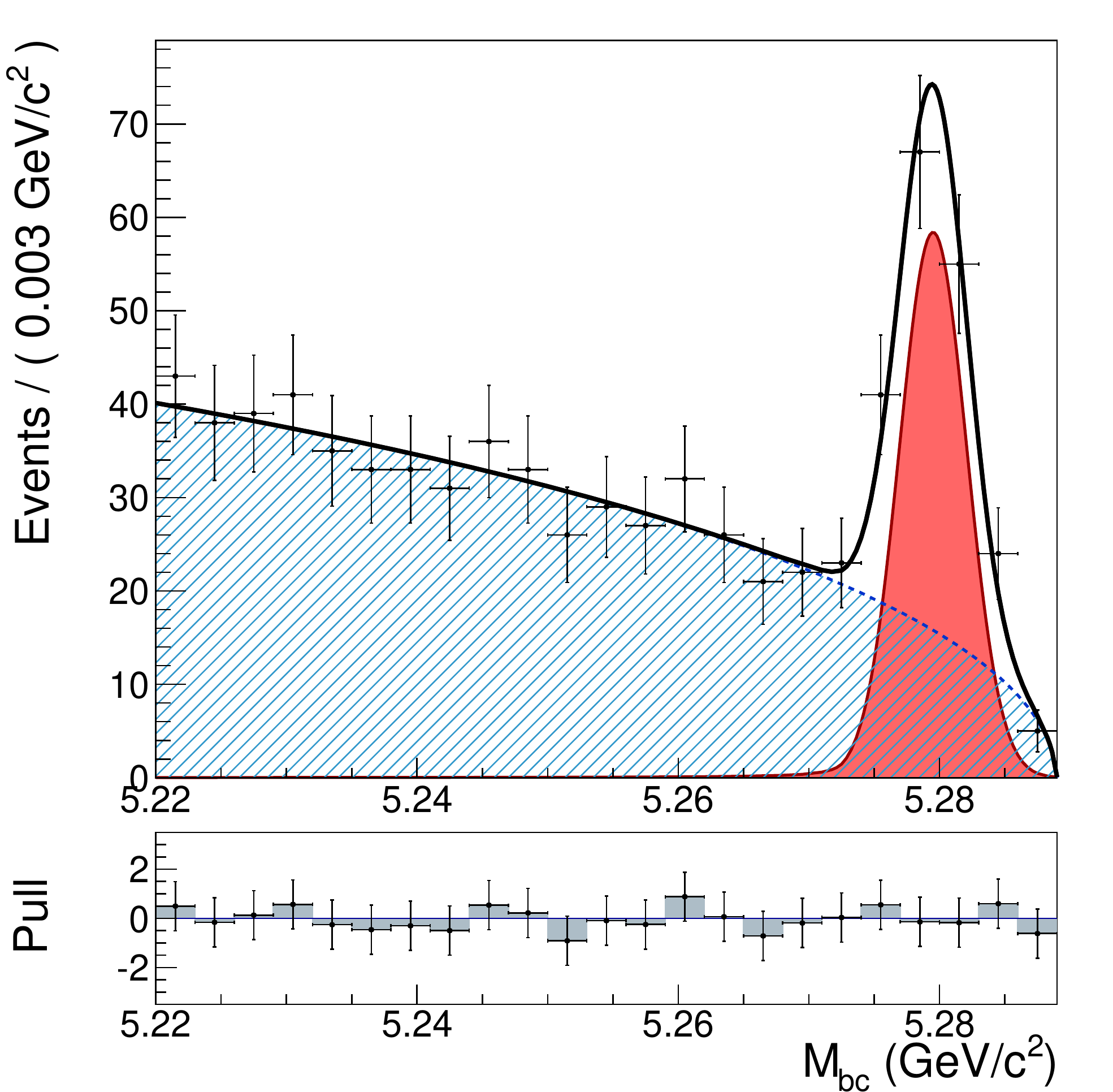}
\caption{Fits to (left) \BToKstmm and (right) \BToKstee candidates in the full \qsq range, but excluding  the \jpsi and the \psitwos mass regions, at \belle. Taken from Ref.~\cite{Wehle:2016yoi}.}
\label{fig:loopBelleA}
%\end{figure}
\vspace{1cm}
%\begin{figure}[t!]
\centering
\includegraphics[width=0.47\textwidth]{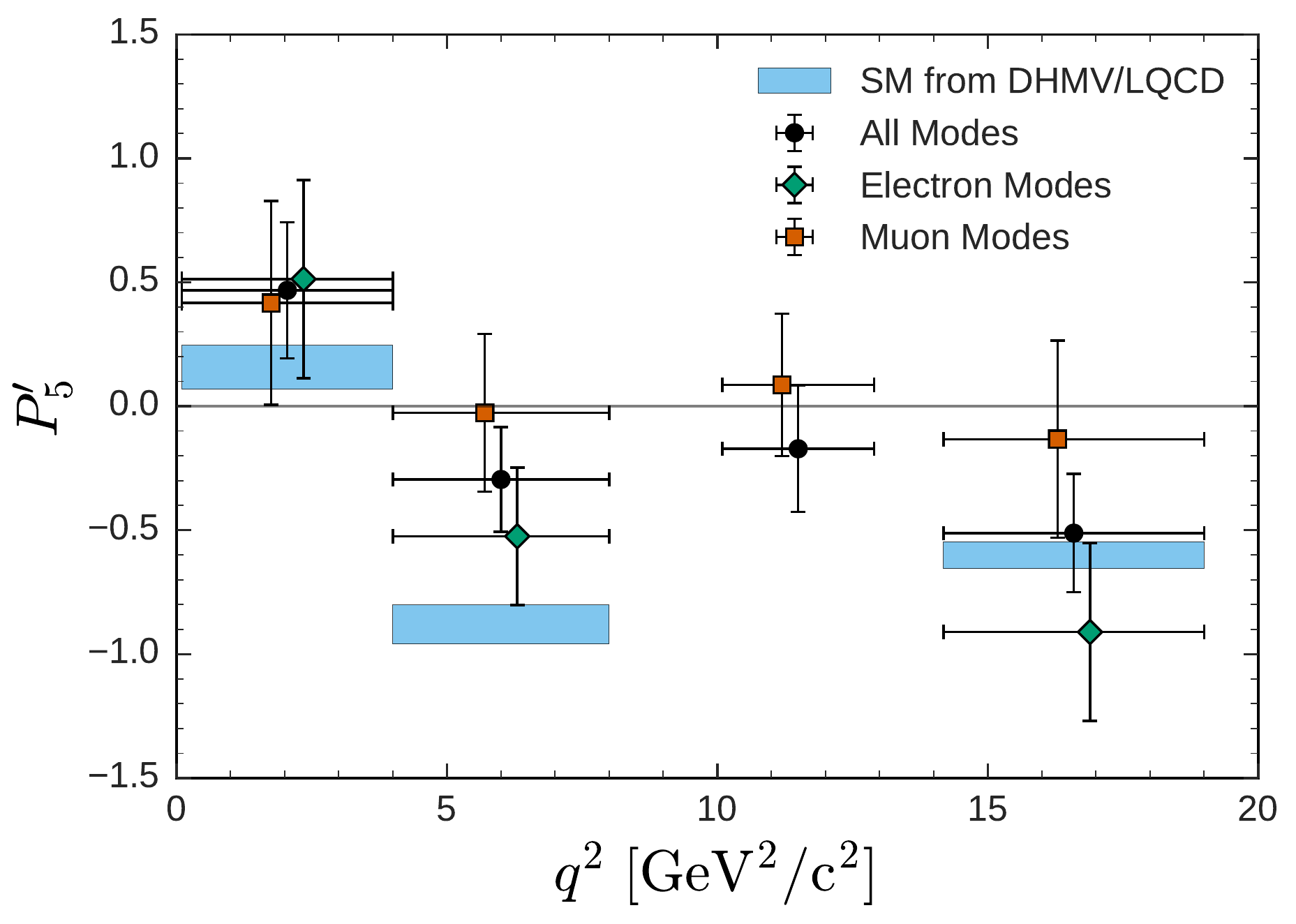}
\hspace{0.5cm}
\includegraphics[width=0.47\textwidth]{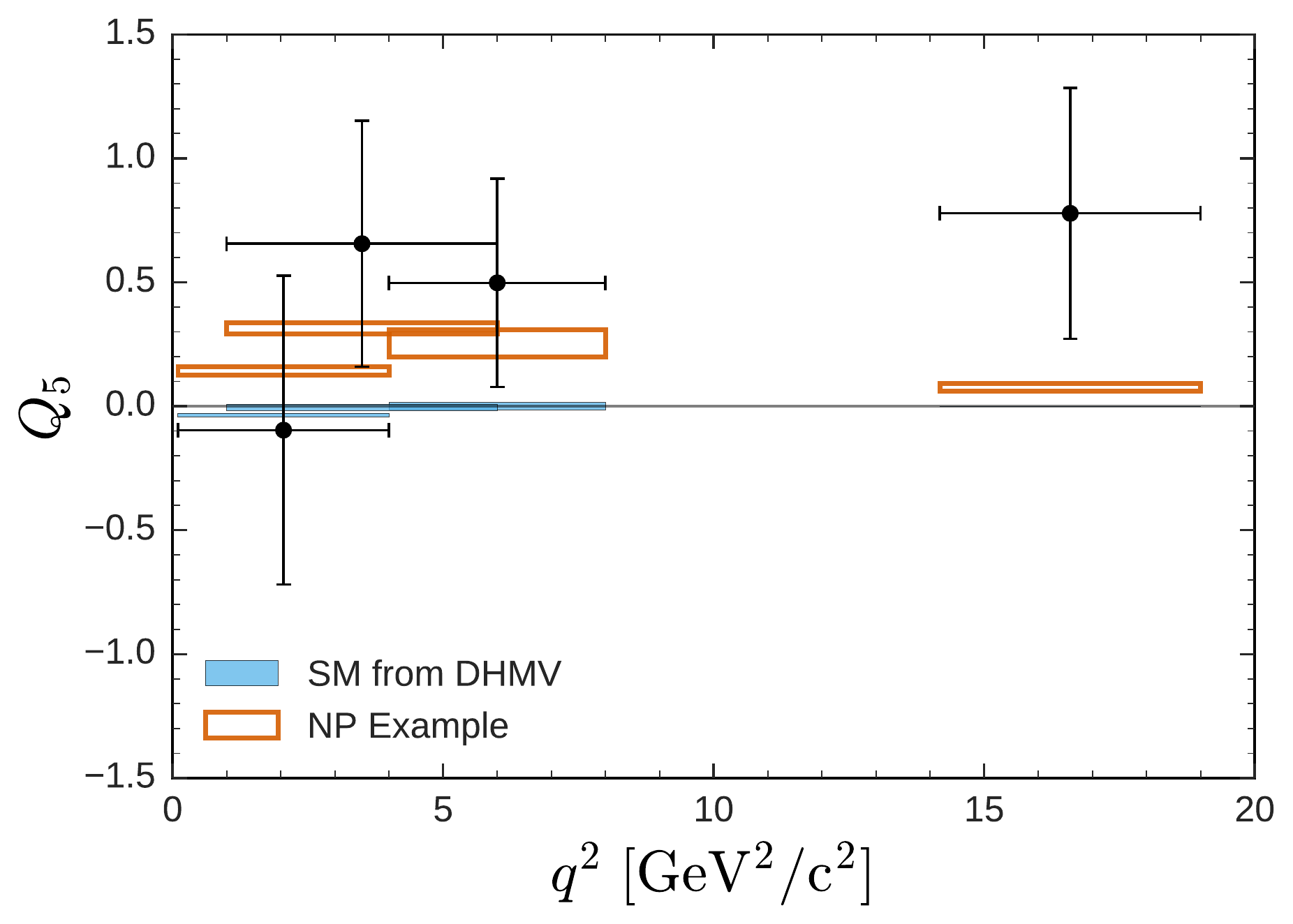}
\caption{Distribution of the (left) $P_{5}^{'}$ and (right) $Q_{5}$ observables at \belle. Taken from Ref.~\cite{Wehle:2016yoi}.}
\label{fig:loopBellePQ}
\end{figure}

The \belle experiment has analysed about 711\invfb of \epem collisions and made the first test of LU using angular observables~\cite{Wehle:2016yoi}.
A combined analysis of \BToKstmm and \BToKstee decays, with both charged and neutral \B meson decays, has been performed.
Signal yields are extracted by fitting the beam-energy constrained mass, $M_{bc}$ (Fig.~\ref{fig:loopBelleA}).
The differential decay rate of these transitions can be described by three angles and \qsq.
The coefficients that appear in the differential decay-rate expression can be combined to define new angular observables, $P_{i}^{'}$~\cite{Matias:2012xw,DescotesGenon:2012zf,Descotes-Genon:2013vna}, that are largely free of form-factor uncertainties.
To take full advantage of the limited statistics, a folding technique that exploits the symmetries of the decay rate can be adopted, allowing the measurement of $P_{4}^{'}$ and $P_{5}^{'}$.
In order to test LU, the difference between such observables for the two leptonic modes, $Q_{i}$~\cite{Capdevila:2016ivx}, is determined.
The analysis is performed in four independent bins of \qsq, as well as in an extended region that covers the range 1--6\gevgevcccc.
Maximum likelihood fits to the angular distributions are performed to measure each observable in each bin \qsq.
The data combines all charged and neutral decay channels, but is split between the muon and electron modes to test LU.
%.
The precision of the results is limited by the size of the data sample.
The largest systematic uncertainty is due to the modelling of the efficiency.
The distribution of the angular observables as a function of \qsq is shown in Fig.~\ref{fig:loopBellePQ}.

\subsection{Results from the Large Hadron Collider}

The \lhcb experiment at the \lhc has made the first measurements of the \RKX ratios at a hadron collider using the Run1 dataset corresponding to about 3\invfb of \pp collisions at centre-of-mass energies of 7 and 8\tev~\cite{LHCb-PAPER-2014-024,LHCb-PAPER-2017-013}.

\begin{figure}[t!]
\centering
\includegraphics[width=0.47\textwidth]{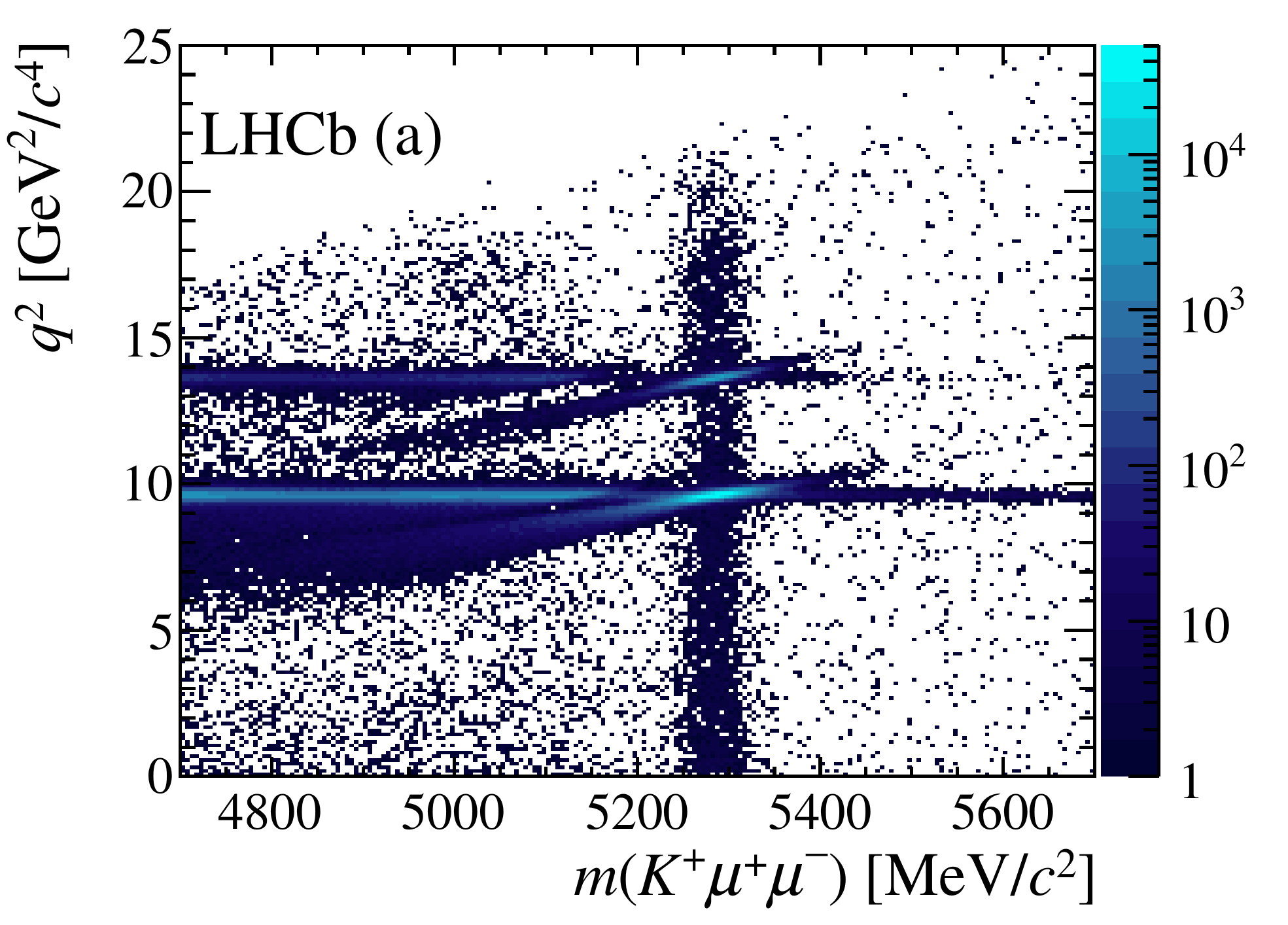}
\hspace{0.5cm}
\includegraphics[width=0.47\textwidth]{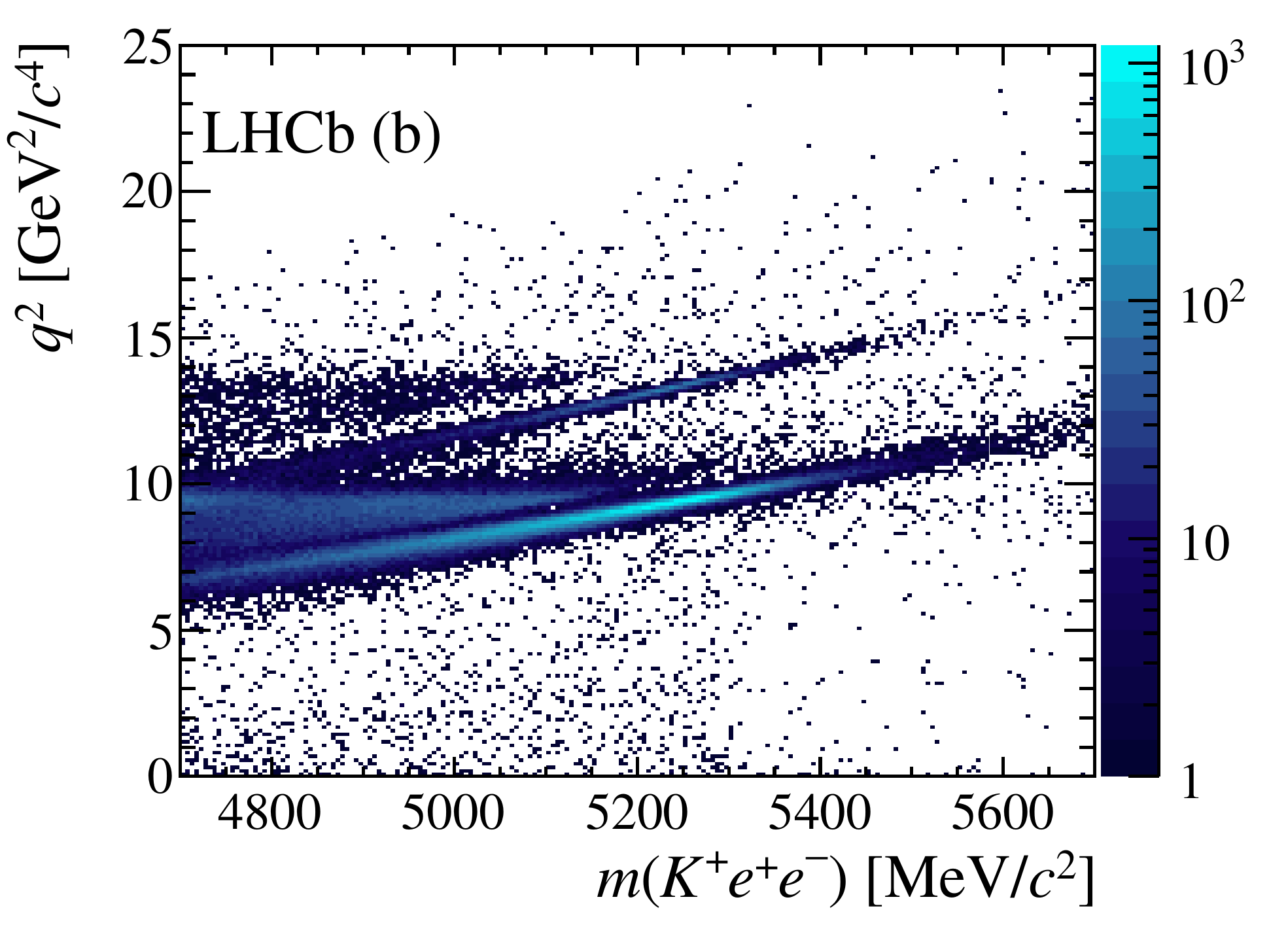}
\caption{Two-dimensional distribution of \qsq as a function of the reconstructed (left) \Kp\mumu and (right) \Kp\ee mass at \lhcb. Taken from Ref.~\cite{LHCb-PAPER-2014-024}.}
\label{fig:loopBrem}
\end{figure}

Compared to \epem machines, a hadronic collider introduces significant challenges due to the unknown initial state and the much busier environment that causes differences in the treatment of decays involving muons or electrons in the final state.
%However, the much larger \bbbar production cross section at a \pp machine compared to an \ee collider results in the world's largest sample of \bquark-hadron decays that boosts measurements of their properties to an unprecedented precision.
The high bremsstrahlung rate for electrons due to the higher energies, and the low efficiency of the hardware trigger due to the high occupancy of the detector complicate the analysis procedure, in particular making it more difficult to separate signal from backgrounds.
A dedicated bremsstrahlung recovery procedure is used to improve the electron momentum reconstruction, and signal candidates are classified depending on how many calorimeter clusters consistent with bremsstrahlung photons were recovered.
However, the limited performances still degrade the resolution of the reconstructed invariant masses of both the dielectron pair and the \B candidate (as shown in Fig.~\ref{fig:loopBrem}).
In order to mitigate the signal loss due to the low efficiency of the calorimeter hardware trigger, decays with electrons in the final state are also selected through the hadron hardware trigger, using clusters associated with the decay products of the hadronic system, or by any hardware trigger from particles in the event that are not associated with the signal candidate.
Several control samples identified on data are used to study the reconstruction performances (\eg bremsstrahlung recovery, trigger and PID), in particular \BToKXJPsee decays, and ultimately the precision of the measurements is dominated by the statistical uncertainty of the decays involving electrons.

\begin{figure}[t!]
\centering
\includegraphics[width=0.47\textwidth]{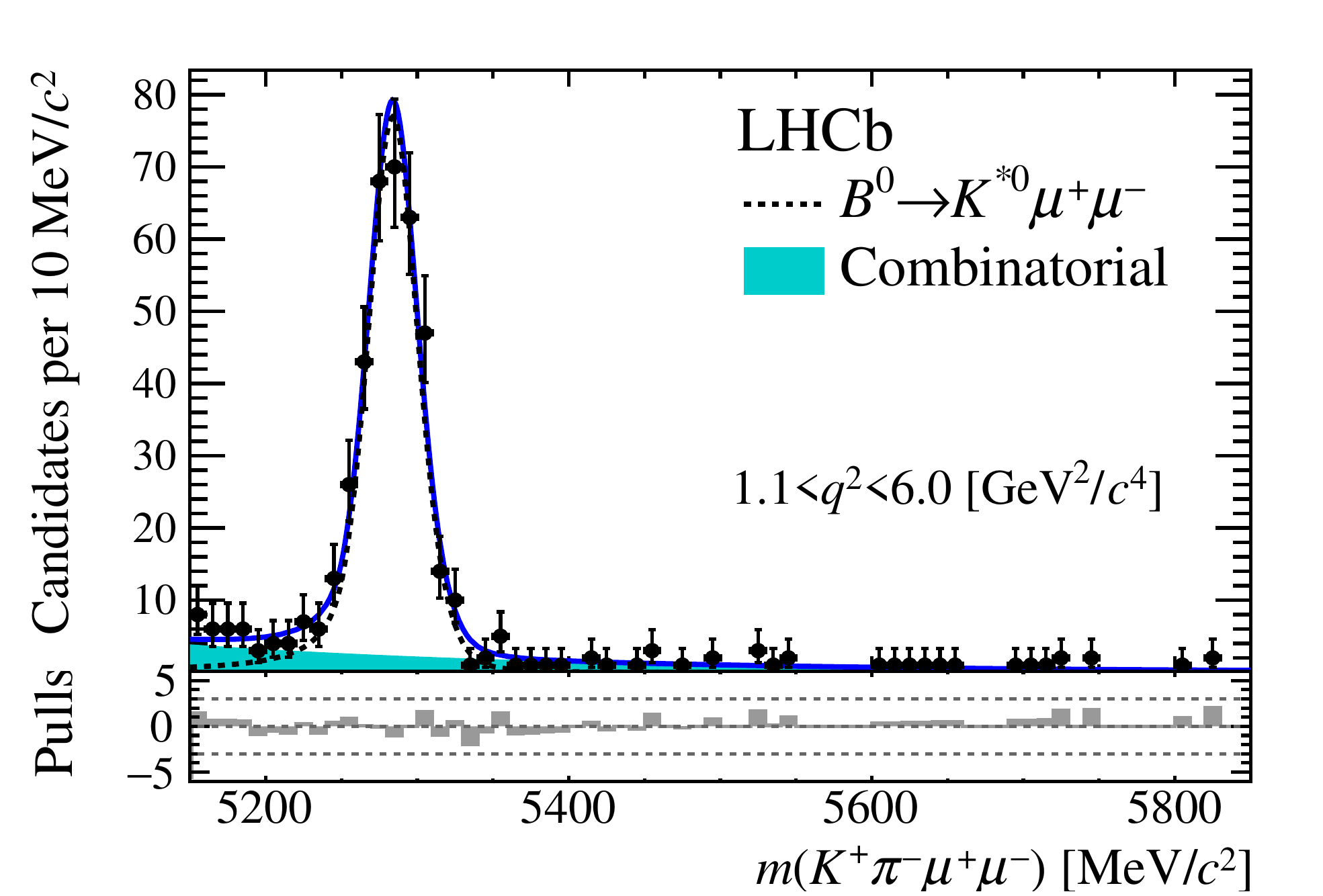}
\hspace{0.5cm}
\includegraphics[width=0.47\textwidth]{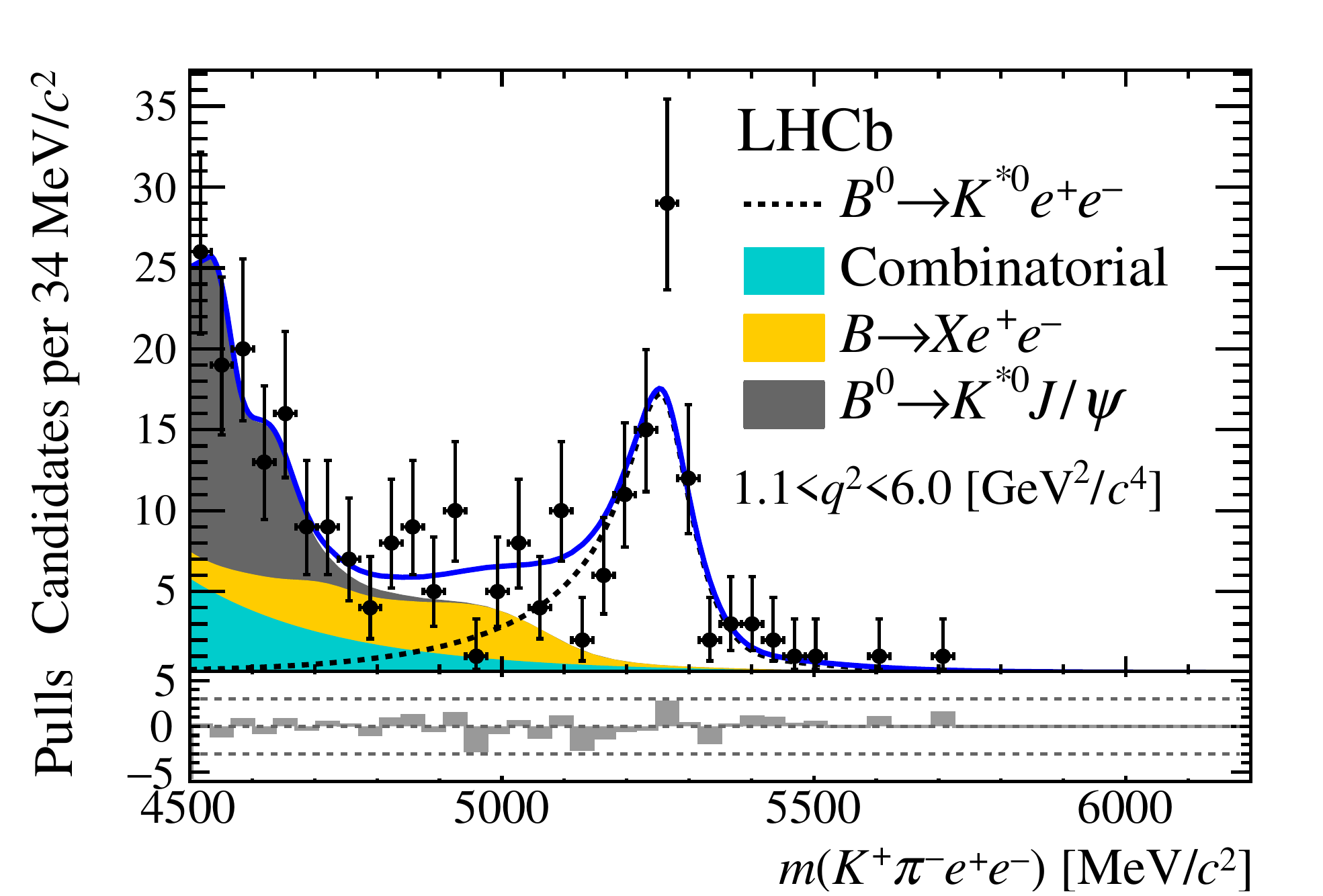}
\caption{Fit to (left) \BdToKstmm and (right) \BdToKstee candidates in the region \mbox{$1.1<\qsq<6.0\gevgevcccc$} at \lhcb. Taken from Ref.~\cite{LHCb-PAPER-2017-013}.}
\label{fig:loopLHCb}
\end{figure}

Both charged and neutral \B decays are studied, but the hadronic system is only reconstructed using charged \kaon and \pion mesons to exploit the very good performances of the tracking and vertexing systems.
Specific vetoes for peaking backgrounds are applied and the combinatorics is controlled using BDT or neural-network classifiers, while residual background due to decays where one or more of the products of the \B decay are not reconstructed is modelled in the fit.
The yields are extracted using one-dimensional fits to the reconstructed \B mass, separately in trigger and bremsstrahlung categories (Fig.~\ref{fig:loopLHCb}).

\begin{figure}[t!]
\centering
\includegraphics[width=0.80\textwidth]{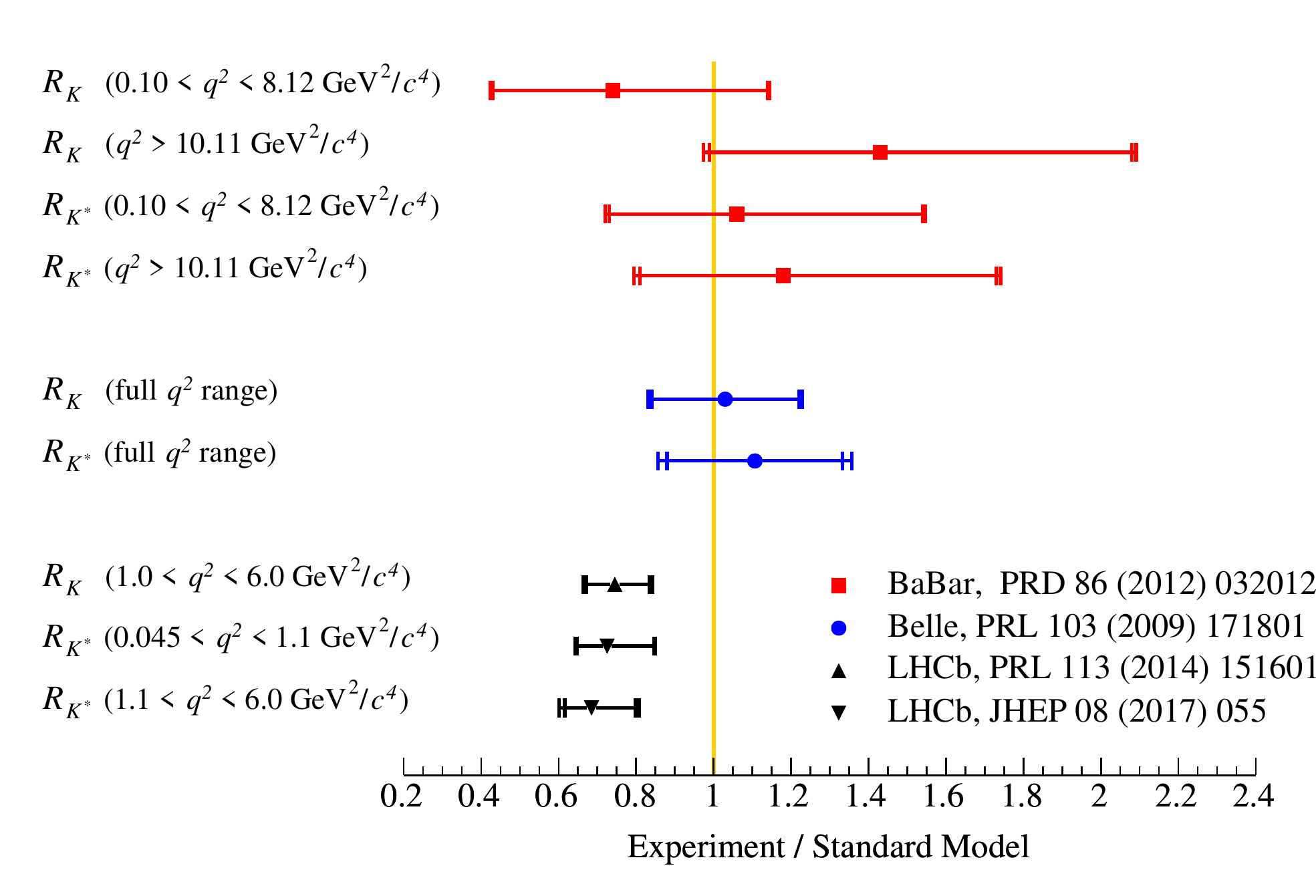}
\caption{Summary of the \RKX measurements performed at the \B-factories and by the \lhcb experiment. Results are presented using different coloured markers. The (yellow) vertical line corresponds to the SM prediction.}
\label{fig:loopSummary}
\end{figure}

The \RK ratio has been measured in the \qsq region between 1.0 and 6.0\gevgevcccc, while \RKst in two regions of \qsq (between 0.045 and 1.1\gevgevcccc, and between 1.1 and 6.0\gevgevcccc).
Measurements have been performed as double ratios to the resonant \BToKXJPsll modes to help mitigate the effect of significant differences in reconstruction performance between decays with muons or electrons in the final state.
Due to the similarity between the reconstruction efficiencies of \BToKXll and resonant decay modes, many sources of systematic uncertainty are substantially reduced.
The main residual experimental systematics are due to the modelling of the trigger response and description of the bremsstrahlung emission.
The \RK measurement has a relative precision of about 12\%, while for \RKst this amounts to about 17\% in either \qsq bins.
A summary of the \RKX measurements is presented in Tab.~\ref{tab:RKX} and in Fig.~\ref{fig:loopSummary}.

\begin{table}[h!]
\centering
\renewcommand\arraystretch{1.2}
\begin{tabular}{c|c|c|c|c}
\textbf{Experiment (year)}	& \boldmath{\Hs} \textbf{type}	& \boldmath{\qsq} \textbf{range [\gevgevcccc]}	& \textbf{Value}	& \textbf{Ref.} \\ \hline
\belle (2009)	& \PK 	& 0.0 $\--$ kin. endpoint	& $1.03 \pm 0.19 \pm 0.06$			& \cite{Wei:2009zv}\\
\belle (2009)	& \Kstar 	& 0.0 $\--$ kin. endpoint	& $0.83 \pm 0.17 \pm 0.08$			& \cite{Wei:2009zv}\\
\babar (2012)	& \PK	& 0.10 $\--$ 8.12		&  $0.74 ^{+0.40}_{-0.31} \pm 0.06$		& \cite{Lees:2012tva} \\
\babar (2012)	& \PK	& $ > 10.11$			&  $1.43 ^{+0.65}_{-0.44} \pm 0.12$		& \cite{Lees:2012tva} \\
\babar (2012)	& \Kstar	& 0.10 $\--$ 8.12		&  $1.06 ^{+0.48}_{-0.33} \pm 0.08$		& \cite{Lees:2012tva} \\
\babar (2012)	& \Kstar	& $ > 10.11$			&  $1.18 ^{+0.55}_{-0.37} \pm 0.11$		& \cite{Lees:2012tva} \\
\lhcb (2014)	& \Kp 	& 1.0 $\--$ 6.0			& $ 0.745  ^{+0.090}_{-0.074} \pm 0.036$	& \cite{LHCb-PAPER-2014-024}  \\
\lhcb (2017) 	& \Kstarz	& 0.045 $\--$ 1.1		& $ 0.66  ^{+0.11}_{-0.03} \pm 0.05$		& \cite{LHCb-PAPER-2017-013} \\
\lhcb (2017) 	& \Kstarz	& 1.1 $\--$ 6.0			& $ 0.69  ^{+0.11}_{-0.07} \pm 0.05$		& \cite{LHCb-PAPER-2017-013}  \\
\end{tabular}
\caption{Summary of the \RKX measurements performed at the \B-factories and by the \lhcb experiment. The first uncertainty is statistical and the second is systematic.}
\label{tab:RKX}
\end{table}

\subsection{Standard Model predictions}

As discussed in Sec.~\ref{sec:frameworkTh} interesting LU observables can be designed by comparing the branching fractions or the angular observables between final states with electrons and muons. 
In contrast with the tree-level case ($\tau$ lepton versus light leptons), the two leptons have light masses that can be neglected over most of the phase space, apart from the region of very small \qsq.

Theoretical predictions in the SM for \RK and \RKst are extremely accurate~\cite{Hiller:2014yaa,Hiller:2014ula} as hadronic uncertainties cancel to a large extent.
\RK needs only $f_+/f_0$ for the $m_\ell^2/\qsq$ suppressed part of the branching fraction.
\RKst involves several helicity amplitudes with a large number of form factors, however, some of them are suppressed due to the $V$--$A$ structure of the SM and the smallness of the lepton masses, leading to a good theoretical control of these quantities.
Moreover, electromagnetic corrections have been estimated to be small and under control~\cite{Bordone:2016gaq}.
Some SM predictions are listed in Tab.~\ref{tab:pred-SM-loop}. Finite-width effects and the underlying $S$-wave component of the $K\pi$ system were also discussed, with no indications of significant contributions~\cite{Matias:2014jua,Hofer:2015kka,Aaij:2016flj}.

\begin{table}[h!]
\resizebox{\textwidth}{!}{
\centering
\renewcommand\arraystretch{1.2}
\begin{tabular}{c|c|c|c|c}
\textbf{Observable} & \textbf{Ref.~\cite{Capdevila:2017bsm}} & \textbf{Ref.~\cite{Geng:2017svp}} & \textbf{Ref.~\cite{Altmannshofer:2017fio}} &
\textbf{Ref.~\cite{Bordone:2016gaq}} \\ \hline
\RK ($1.0<\qsq<6.0\gevgevcccc$)		& $1.00\pm 0.01$	& $1.0004^{+0.0008}_{-0.0007}$	& ---					& $1.000 \pm 0.010$ \\
\RKst ($0.045<\qsq<1.1\gevgevcccc$)	& $0.92\pm 0.02$	& $0.920^{+0.007}_{-0.006}$		& $0.9259\pm 0.0041$	& $0.906 \pm 0.028$ \\
\RKst ($1.1<\qsq<6.0\gevgevcccc$)		& $1.00 \pm 0.01$	& $0.996^{+0.002}_{-0.002}$		& $0.9965\pm 0.0006$	& $1.000 \pm 0.010$ \\
\end{tabular}
}
\caption{SM predictions for the \RHs LU ratios.}
\label{tab:pred-SM-loop}
\end{table}

The \RK and \RKst measurements performed by the \lhcb collaboration are found to be 2.6$\,\sigma$ and 2.1--2.5$\,\sigma$ lower than the SM expectation (Fig.~\ref{fig:loopSummary}), respectively~\cite{LHCb-PAPER-2014-024,LHCb-PAPER-2017-013}.
The \babar and \belle measurements, which have a more limited precision, are found to be in agreement with the SM expectation.
Significant deviations with respect to the SM predictions would indicate a violation of LU that cannot be due to hadronic effects, as these would impact in the same way \bTosmm and \bTosee transitions.
Such effects, however, are particularly relevant in regard to the deviations observed in \bTosmm alone, where there is an ongoing discussion concerning the size of hadronic effects (see Sec.~\ref{sec:loop-possible-interpretations}).
It is sometimes stated that \RHs observables have always limited hadronic uncertainties, but this statement must be modulated.
More specifically, the hadronic uncertainties remain small as long as there are no significant LU-violating effects.
If these are present, interferences between LU-violating and conserving contributions may spoil the cancellation of hadronic uncertainties.
These effects might come from NP or from lepton-mass effects in the SM.
The latter are only important at low \qsq, wherever $m_\ell^2/\qsq$ is not small compared to 1 (\eg below $\qsq \sim 1\gevgevcccc$), and affect in particular the first measured bin in \RKst.
In this bin one thus expects larger theoretical uncertainties than in the region above 1\gevgevcccc, as well as at any value of \qsq in the presence of LU-violating NP~\cite{Capdevila:2017ert,Capdevila:2016ivx}.

As discussed in Sec.~\ref{sec:loop-Bfactories}, LU violation can be probed using angular observables.
This was in particular proposed through the difference of angular coefficients or their optimised versions~\cite{Capdevila:2016ivx,Altmannshofer:2015mqa,Serra:2016ivr}.
The latter correspond to the difference of muon and electron optimised observables $P_i$ that exhibit a limited sensitivity to hadronic effects~\cite{Matias:2012xw,DescotesGenon:2012zf,Descotes-Genon:2013vna}.
These LU observables are predicted to vanish in the SM (apart from the very low \qsq region where phase-space effects lead to slight differences between muon and electron modes), and exhibit a limited enhancement of hadronic uncertainties in the presence of NP contributions.
The current \belle measurements are compatible with the SM within large uncertainties (Fig.~\ref{fig:loopBellePQ}), with the largest deviation from the SM observed for muons in the \qsq region 4--8\gevgevcccc at the 2.6$\,\sigma$ level (only 1.3$\,\sigma$ for electrons in the same region)~\cite{Wehle:2016yoi}.

%%%%%%%%%%%%%%%%%%%%%%%%%%%%%%%%%%%%
% !TEX root = main.tex
%%%%%%%%%%%%%%%%%%%%%%%%%%%%%%%%%%%%

%\clearpage
\section{Interpretation of Lepton Universality tests in \bquark-quark decays}
\label{sec:interpretation}

As discussed in Sec.~\ref{sec:tree} and Sec.~\ref{sec:loop} there are hints of LU violation in both \bTocln and \bTosll transitions:
\RD and \RDst exceeds the SM predictions by around 2.3$\,\sigma$ and 3.0$\,\sigma$, respectively;
\RK lies about 2.6$\,\sigma$ below the SM expectation ($1.0<\qsq<6.0\gevgevcccc$);
and \RKst lies around 2.1--2.5$\,\sigma$ below the SM prediction ($0.045<\qsq<1.1\gevgevcccc$ and $1.1<\qsq<6.0\gevgevcccc$).
These deviations from LU observed in semileptonic \bTocln and \bTosll transitions may not be explained within the SM and suggest a NP origin.
This can be analysed either in a general Effective Field Theory approach or through specific NP models.
The challenge does not consist only in explaining the deviations from LU observed in both decays, but also to reproduce other related measurements.
More specifically, these NP explanations should reproduce the pattern of deviations observed in \bTosmm observables, such as the branching fractions in \BToKmm, \BToKstmm and \BsToPhimm~\cite{LHCb-PAPER-2014-006,LHCb-PAPER-2015-023,Aaij:2016flj} as well as the \BToKstmm angular observables~\cite{LHCb-PAPER-2013-037,LHCb-PAPER-2015-051,Abdesselam:2016llu,Wehle:2016yoi,ATLAS:2017dlm,CMS:2017ivg}. On the other hand, these NP explanations should not contradict constraints on LU in other sectors discussed in Sec.~\ref{sec:history}, they should not generate lepton-flavour violating processes above the current experimental bounds and they should not disagree with the experimental bounds on direct production of heavy NP particles at the LHC.

\subsection{Effective Field Theory approach}

In order to explain the deviations of measurements from the SM expectations, one can perform a model-independent analysis by considering the relevant effective Hamiltonian, determining the values of the short-distance Wilson coefficients from the data, and comparing them with respect to the SM computation.
This can be achieved by either only assuming the SM basis or including further operators.
The described method is usually referred to as the Effective Field Theory (EFT) approach.

Additional NP contributions are often assumed to be real, as there have been no signs of CP violation in these processes up to now. Even though there is no theoretical reason for the alignment of SM and NP sources of CP violation,
the absence of additional sources of CP violation due to NP is empirically supported by the current data where CP-asymmetries have been measured as compatible with zero for \BuToKmm and \BdToKstmm~\cite{LHCb-PAPER-2014-032}, as well as by the good consistency of the CKM global fit with SM expectations~\cite{CKMfitter2015,UTfit-UT,Koppenburg:2017mad}. Under these assumptions, a specific scenario of NP is defined by adding NP contributions to some of the Wilson coefficients ${\cal C}_i={\cal C}_i^{SM}+{\cal C}_i^{NP}$, and a global fit to all relevant observables is performed to constrain the value of ${\cal C}_i^{NP}$.
For simplicity and due to the limited set of data available, these global fits are often performed assuming NP contributions to only one or two Wilson coefficients following motivated NP scenarios, even though more ambitious studies are also available.

\subsubsection{\bToctaun}

\begin{figure}[t!]
\centering
\includegraphics[width=0.94\textwidth]{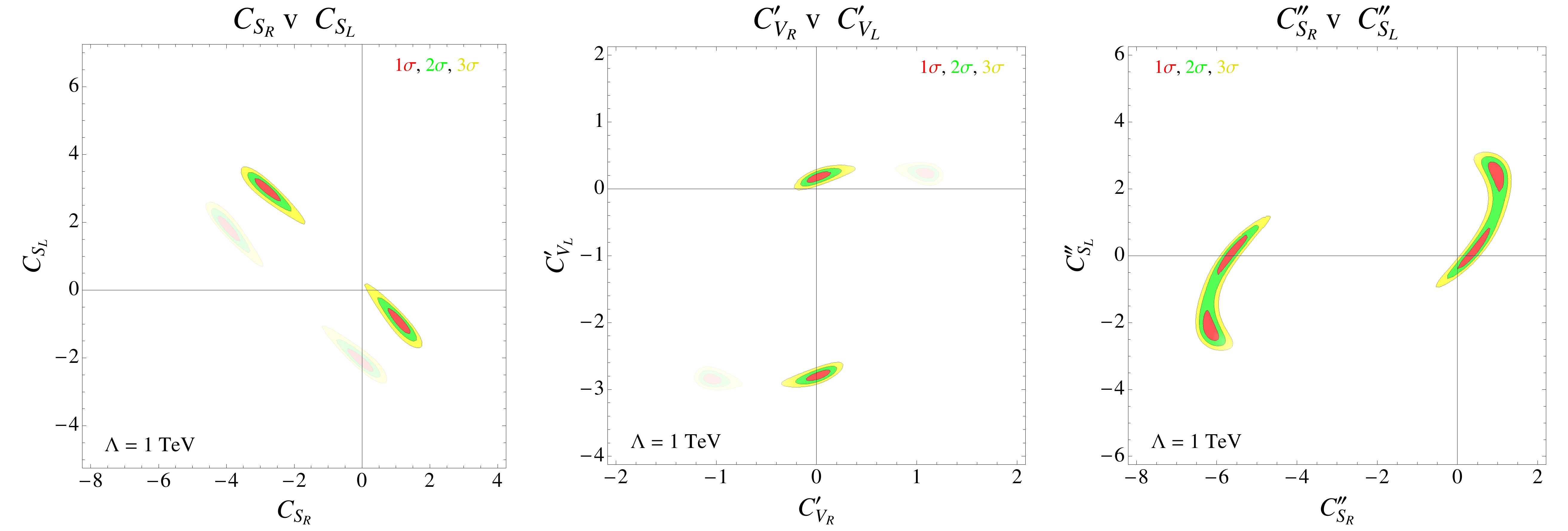}
\caption{Constraints on various Wilson coefficients for \bToctaun (from left to right: scalar, vector and scalar). Taken from Ref.~\cite{Freytsis:2015qca}.}
\label{fig:treeEFT}
\end{figure}

Most of the EFT analyses assume that NP is present mainly in \bToctaun transitions, as there are no signs of violation of LU (see Sec.~\ref{sec:tree}), and more generally no signs of NP contributions~\cite{Jung:2018lfu} for electron and muon \bTocln decays.
Several types of contributions have been considered:
\begin{itemize}
\item Left-handed contributions to vector and axial operators are favoured.
In fact, the relative deviations of the measured LU ratios compared to the SM are of similar sizes for \D, \Dstar and \jpsi, and the \qsq-dependence of $\BR(B\to D\taum\neutb)$ is in agreement with the SM.
A particularly simple explanation would consist in having an additional NP contribution to the Wilson coefficient of the SM operator $O_{V\tau}=(\bar{c}\gamma_\mu P_L b)(\bar\tau\gamma^\mu P_L \nu)$, changing only the overall normalisation  (but not the \qsq shape) for \bToctaun transitions.
This solution is recovered in the global fits~\cite{Freytsis:2015qca,Alok:2017qsi}.
\item Right-handed currents for \bToctaun are in principle an alternative to generate \RD and \RDst through contributions $O_{V'\tau}=(\bar{c}\gamma_\mu P_R b)(\bar\tau\gamma^\mu P_L \nu)$.
However, the constraints from the total lifetime of the \Bcm meson~\cite{Li:2016vvp,Alonso:2016oyd,Akeroyd:2017mhr} (which puts a bound on \mbox{\decay{\Bcm}{\taum\neutb}}) and from differential distributions in $B\to D^{(*)}\tau^- \bar\nu_\tau$~\cite{Freytsis:2015qca,Ivanov:2017mrj} suppress right-handed contributions very significantly in the global fits~\cite{Alok:2017qsi}, so that they cannot be considered as viable alternatives to explain LU violation.
\item Scalar and pseudoscalar couplings could in principle also mimic similar changes in \RD and \RDst.
However, the constraints described above (total lifetime of the \Bcm meson, differential distributions in \decay{\B}{\D^{(*)}\taum\bar\nu_\tau}) are in tension with a simultaneous explanation of \RD and \RDst~\cite{Celis:2016azn,Bardhan:2016uhr} (these constraints could be avoided with right-handed couplings~\cite{Li:2016vvp}).
Moreover, some ultraviolet complete models with an additional neutral spin-0 particle are ruled out by LHC direct searches for resonances in the \tt final state~\cite{Faroughy:2016osc}.
\item A tensor contribution has a distinct impact on \RJPs and on the $\tau$ polarisation in \decay{\B}{\Dstar\taum\neutb}, compared to other contributions, and provides a decent description of the data~\cite{Freytsis:2015qca,Alok:2017qsi}.
However, such operators are not easily generated by NP theories at the electroweak scale: in some cases, they can appear due to the evolution from the electroweak scale down to the \bquark-quark scale, with strong correlations with scalar operators~\cite{Jung:2018lfu}.
\end{itemize}
A typical outcome of global fits~\cite{Freytsis:2015qca,Alok:2017qsi} in terms of constraints on Wilson coefficients is shown in Fig.~\ref{fig:treeEFT}.

More elaborate scenarios may include several of the above contributions~\cite{Azatov:2018knx}.
As indicated before, these analyses depend significantly on the hadronic uncertainties coming from the form factors, with a rather satisfying situation for \decay{\B}{\D\ell\nu} transitions, but not so much for \decay{\B}{\Dstar\ell\nu} (and also for \decay{\Bcm}{\jpsi\ell\nu}, even though the theoretical predictions are less precise in the latter case).
However, the detailed analysis of the assumptions used in order to constrain the form factors (lattice inputs, HQET relations, estimate of $1/m_\bquark$ suppressed corrections) indicate that there is a good consistency between the form factor inputs, and that underestimated theoretical systematics do not appear sufficient to explain the large deviations of \RD and \RDst from the SM~\cite{Bernlochner:2017jka}.

\subsubsection{\bTosll}
\label{sec:loop-possible-interpretations}

Compared to \bTocln, \bTosll transitions are currently richer due to the large number of observables already considered: some of them are able to test directly LU, whereas others provide separate information for muon and electron modes.
Hints for LU violation are found in both \RK and \RKst, in conjunction with deviations in \bTosmm observables such as branching fractions in \BToKmm, \BToKstmm and \BsToPhimm~\cite{LHCb-PAPER-2014-006,LHCb-PAPER-2015-023,Aaij:2016flj} as well as \BToKstmm angular observables~\cite{LHCb-PAPER-2013-037,LHCb-PAPER-2015-051,Abdesselam:2016llu,Wehle:2016yoi,ATLAS:2017dlm,CMS:2017ivg}.

\begin{figure}[t!]
\centering
\includegraphics[height=0.43\textwidth]{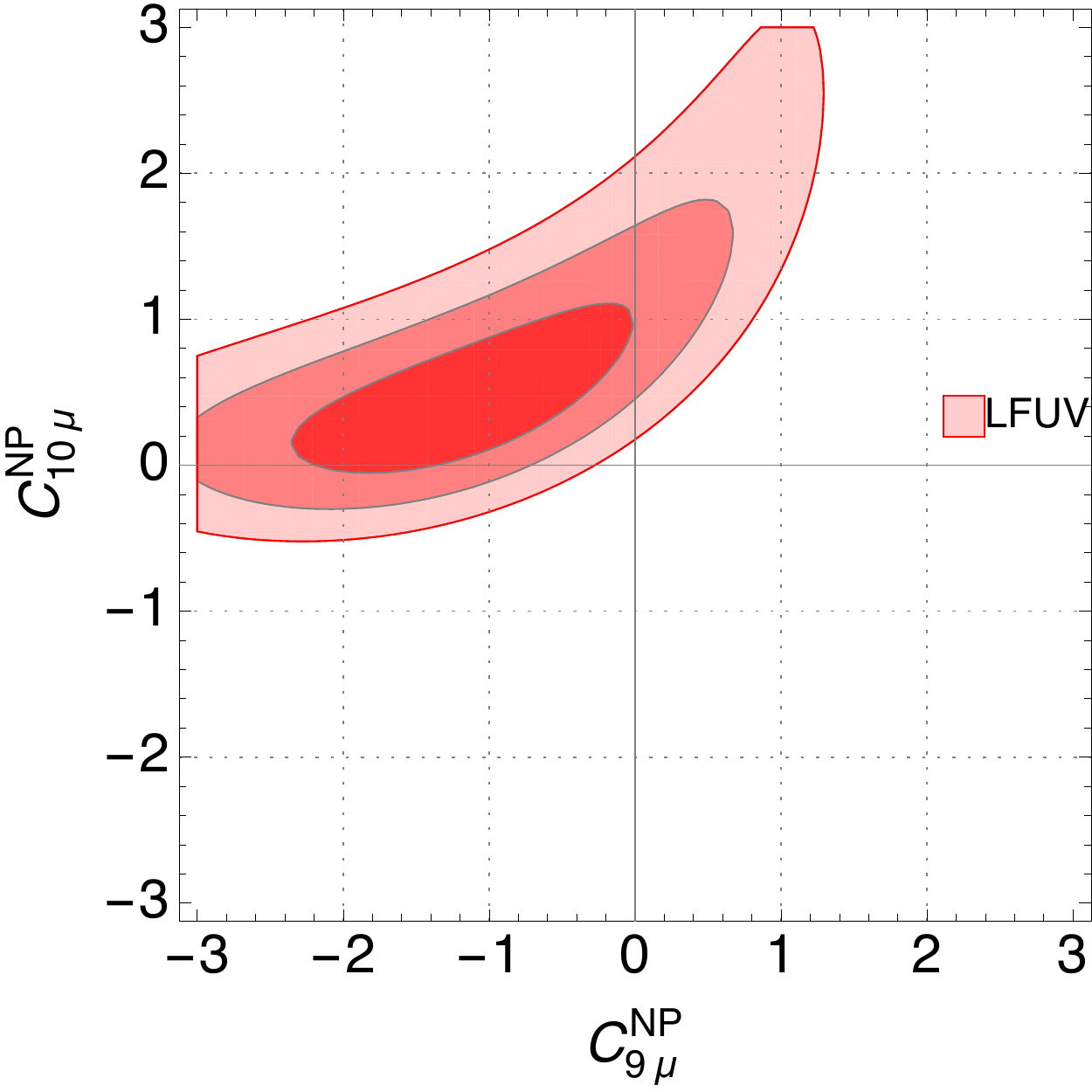}
\hspace{0.5cm}
\includegraphics[height=0.43\textwidth]{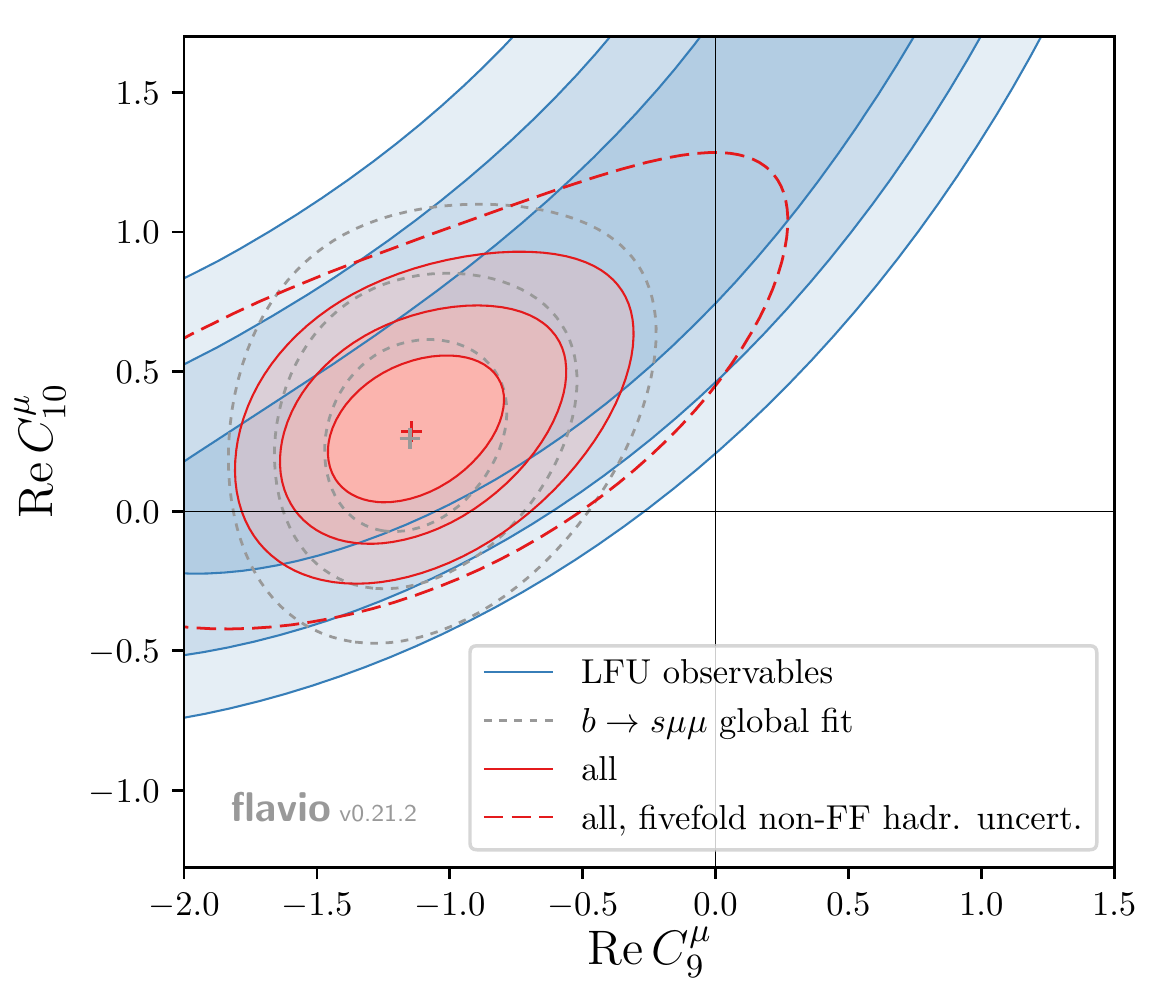}
\caption{Constraints on the NP contribution to the Wilson coefficients ${\cal C}_{9\mu}$ and ${\cal C}_{10\mu}$ for \bTosll (left) using only LU observables and (right) using all \bTosll data. Taken from Refs.~\cite{Capdevila:2017bsm, Altmannshofer:2017yso}}
\label{fig:loopEFT}
\end{figure}

LU observables comparing \bTosmm and \bTosee provide very powerful tests of the SM, with ratios of branching fractions (\RK, \RKst) very close to unity~\cite{Hiller:2014yaa,Hiller:2014ula} and difference of angular observables vanishingly small~\cite{Capdevila:2016ivx,Serra:2016ivr}: if significant deviations are observed, these must be due to NP since no SM effects (and in particular no long-distance hadronic effects) can result in a violation of LU.
However, the interpretation of these deviations in terms of NP contributions, and in particular the global fits of the Wilson coefficients, require accurate understanding of hadronic contributions in order to disentangle SM and NP effects in the measured observables. 

As already discussed in Sec.~\ref{sec:frameworkTh}, the EFT approach leads to two kinds of contributions for these decays: local (involving form factors for the hadronic part) and non-local (describing the propagation of \ccbar loops, and in particular the \jpsi and excited charmonium resonances). 
For the local part, form factors have been estimated in different kinematic regimes using several kinds of LCSR computations and lattice QCD simulations, with a good agreement and control over uncertainties~\cite{Ball:2004rg,Khodjamirian:2010vf,Horgan:2013hoa,Bouchard:2013pna,Straub:2015ica}, leading to a limited impact on the extraction of the Wilson coefficients~\cite{Descotes-Genon:2015uva,Altmannshofer:2014rta,Capdevila:2017ert}. 
For the non-local part, the contribution of charm loops can be taken into account at low-meson recoil relying on quark-hadron duality~\cite{Beylich:2011aq}. At large-meson recoil,
one can use the available computations from LCSR~\cite{Khodjamirian:2010vf,Khodjamirian:2012rm}, exploit these computations as an order-of-magnitude estimate, or take simpler estimates based on dimensional analysis.
Another approach to assess these contributions consists in adding a general parametrisation with parameters fitted to the data. This has been implemented,
for example, by adding to the decay amplitudes either polynomials in \qsq~\cite{Ciuchini:2017mik} fitted to branching fractions and angular observables,
or polynomials in the conformal variable $z(\qsq)$ fitting not only \bTosll observables but also theoretical and experimental information on charmonia~\cite{Bobeth:2017vxj}.
The charm-loop contribution obtained yield results for various observables in good agreement with the predictions obtained using the estimates in Refs.~\cite{Khodjamirian:2010vf,Khodjamirian:2012rm}.

Global fits of NP contributions performed for LU observables only lead to equivalently good fits if \bTosmm Wilson coefficients are modified in one direction or the corresponding \bTosee Wilson coefficients in the opposite direction.
The significance for the favoured NP scenarios considering only the LU observables \RK and \RKst  (and excluding other \bTosmm processes) is at the 3--4$\,\sigma$ level~\cite{Capdevila:2017bsm,Altmannshofer:2017yso,DAmico:2017mtc,Geng:2017svp,Ciuchini:2017mik,Hiller:2017bzc}.
The violation of LU between muons and electrons is indeed significant, around $25\%$ at the level of the ${\cal C}_{9\ell}$ coefficients.
More specifically, the fit favours NP in the Wilson coefficients corresponding to left-handed quark currents: negative ${\cal C}_{9\mu}^{NP}$ and positive ${\cal C}_{10\mu}^{NP}$ decrease both \BToKmm and \BToKstmm whereas positive ${\cal C}_{9e}^{NP}$ and negative ${\cal C}_{10e}^{NP}$  increase both \BToKee and \BToKstee, and thus improving the agreement with the data.
Moreover, good fits are obtained along the directions ${\cal C}_{9\ell}^{NP}=-{\cal C}_{10\ell}^{NP}$, which arises naturally in models obeying $SU(2)_L$ invariance. A typical outcome of these global fits in terms of constraints on Wilson coefficients is shown in Fig.~\ref{fig:loopEFT}.

A comment is in order concerning the lowest bin of \RKst, which is dominated by the photon pole at $\qsq=0$.
In the EFT language, this photon contribution is described by the LU coefficients ${\cal C}_{7,7'}$ related to $b\to s\gamma$ transitions, combined with the coupling of the photon to the lepton pair (which is LU in the SM and indeed measured as such to a very high accuracy, see for instance  Sec.~\ref{sec:Jpsi}).
LU-violating contributions could only come from other operators (like ${\cal C}_{9\ell}$), which are kinematically suppressed compared to the photon pole contribution.
Therefore, NP contributions can improve the agreement with experiment slightly, but they would not be able to reproduce the central value currently measured for the \RKst in this bin without contradicting measurements at higher \qsq (this is however not a problem within the current experimental uncertainties).

The degeneracy between NP contributions in \bTosmm and \bTosee for LU observables is lifted once separate \bTosmm and \bTosee observables are included (in particular angular observables).
Recent combined analyses~\cite{Capdevila:2017bsm,Altmannshofer:2017yso,Geng:2017svp,Arbey:2018ics} single out some NP scenarios preferred over the SM with a significance at the 5$\,\sigma$ level, consistently with earlier analyses that were mainly restricted to \bTosmm processes~\cite{Descotes-Genon:2013wba,Descotes-Genon:2015uva,Altmannshofer:2015sma,Hurth:2016fbr}.
These scenarios agree with the ones presented above for fits restricted to LU observables.
This is particularly the case where NP contributions enter only ${\cal C}_{9\mu}$ and in both ${\cal C}_{9\mu}$ and ${\cal C}_{10\mu}$, for which constraints can be extracted much more precisely than from the restricted set of LU observables. 
With this extended set of observables, there is still the possibility of adding NP also in electron Wilson coefficients, which however is not very much constrained, and suggests a much smaller NP coupling to electrons compared to muons.
Scalar, pseudoscalar and tensor contributions are not favoured by the fits to the data~\cite{Capdevila:2017bsm,Altmannshofer:2017yso,Geng:2017svp,Arbey:2018ics}.

\subsubsection{Common explanations}

The EFT approach provides a framework to correlate the LU deviations in the tree- and loop-level \bquark-quark transitions.
This can be achieved within the Standard Model Effective Field Theory (SMEFT), which is the effective theory obtained at the electroweak scale assuming that NP occurs at a sufficiently high scale, and yields only additional higher-dimensional operators to the SM~\cite{Buchmuller:1985jz,Grzadkowski:2010es,Alonso:2014csa,Celis:2017doq}.
Since both LU deviations are well explained by a NP vector contribution to left-handed fermions, it is quite natural to investigate the fairly general structure~\cite{Buttazzo:2017ixm,Kumar:2018kmr}:
\begin{equation}
{\mathcal L}_{NP}\propto\lambda^q_{il}\lambda^\ell_{\alpha\beta}
     [C_T(\bar{Q}^i_L\gamma_\mu \sigma^a Q^j_L)(\bar{L}^\alpha_L\gamma^\mu \sigma^a L^\beta_L)
     +C_S(\bar{Q}^i_L\gamma_\mu  Q^j_L)(\bar{L}^\alpha_L\gamma^\mu  L^\beta_L)]
\end{equation}
with the fermion doublets expressed as $Q^i_L\sim (V^*_{ji} u^j_L,d_L)$ and $L^\alpha_L= (\nu^\alpha_L,\ell^\alpha_L)$, $C_S$ and $C_T$ are singlet and triplet couplings (with respect to $SU_L(2)$) in this EFT extension, and $\lambda^q$ and $\lambda^\ell$ flavour matrix couplings. 

In this framework, the same four-fermion operator $(\bar{Q}^i_L\gamma_\mu  Q^j_L)(\bar{L}^\alpha_L\gamma^\mu  L^\beta_L)$ yields both FCNC and FCCC processes:
the effect in \RDDst is correlated to \bTosll and/or to $b\to s\nu\overline\nu$, following the pattern of NP contributions $C_{9\mu}^{NP}=-C_{10\mu}^{NP}$.
Since \bTocln processes are mediated already at tree level in the SM, a rather large NP contribution is required and in principle large contributions to \decay{\bquark}{\squark\nu\bar\nu} processes also appear. 
These bounds from \decay{\B}{K^{(*)}\nu\bar\nu} can be avoided if the coupling structure is mainly aligned to the third generation, but this enters in conflict with direct LHC searches~\cite{Faroughy:2016osc} and electroweak precision observables~\cite{Feruglio:2016gvd}.
Important constraints come also from lepton-flavour violating decays~\cite{Kumar:2018kmr} which can constrain some of the couplings significantly (for instance \decay{\B}{K\mu\tau}~\cite{Lees:2012zz}). Another solution to this problem consists in assuming $C_T\simeq C_S$~\cite{Grzadkowski:2010es,Celis:2017doq}. 
Due to $SU(2)_L$ invariance, a large enhancement of \bTostt processes can then be expected~\cite{Capdevila:2017iqn}.

Flavour symmetries can be used to ensure the proper hierarchy of couplings to the other families and to avoid too large lepton-flavour violating processes.
In many analyses, only one generation (in the interaction eigenbasis) is assumed to have non-vanishing NP couplings so that LU is violated through the misalignment between the interaction and mass bases~\cite{Bhattacharya:2014wla,Alonso:2015sja,Calibbi:2015kma}.
This approach ensures in particular that there is no significant lepton-flavour violation in FCNC such as \decay{\bquark}{\squark\ell_1^+\ell_2^-} with $\ell_1 \neq \ell_2$
(as could be expected in cases where no symmetries are invoked~\cite{Glashow:2014iga}).
More extensive analyses aim at providing models for the whole matrix of fermion couplings on the basis of flavour symmetries.
For instance, assuming an $U(2)_q\times U(2)_\ell$ symmetry among the first two generations which is broken in order to follow the hierarchy of Yukawa couplings in the SM, it is possible to provide a hierarchical structure for the flavour coupling matrices $\lambda^q$ and $\lambda^\ell$ without conflicting with other low-energy constraints or high-$p_\textrm{T}$ constraints~\cite{Buttazzo:2017ixm}.

\subsection{Specific models}
\label{sec:specificNPmodels}

The EFT approach provides constraints on Wilson coefficients encoding the short-distance contributions to \bTocln and \bTosll processes.
Discussing all possible models of NP able to reproduce these patterns of short-distance physics is beyond the scope of this review, but it is already interesting to discuss simple models, where only one or two heavy intermediate particles are introduced.
The quantum numbers involved in \bTocln and \bTosll processes correspond to exchanges of 
two heavy colourless vector bosons (\Wprime, \Zprime), heavy colourless scalars (charged or neutral heavy Higgs-like bosons), or heavy coloured vector or scalar bosons (leptoquarks)~\cite{Buttazzo:2017ixm}. 
In these simple models, both LU and lepton-violating effects are often generated together (for instance in leptoquark models), unless specific symmetries are imposed to prevent lepton-flavour violating couplings (for instance in \Zprime models). In both cases, information on LU as well as lepton-flavour violation (such as \decay{\B}{K^{(*)}\mu\tau}~\cite{Lees:2012zz}) provides interesting guides for model building, in conjunction with constraints from direct production at the LHC.

\subsubsection{\bTocln}

As the FCCC \bTocln transition occurs at tree level in the SM, NP explanations also occur typically at tree level.
Various models with single mediators have been proposed, with heavy charged vector bosons (\Wprime)~\cite{Greljo:2015mma,Boucenna:2016wpr,Boucenna:2016qad,Megias:2017ove}, heavy charged scalars (\Hprime)~\cite{Crivellin:2012ye,Tanaka:2012nw,Celis:2012dk,Crivellin:2013wna,Crivellin:2015hha,Chen:2017eby,Iguro:2017ysu,Chen:2018hqy}, or coloured vector or scalar bosons (leptoquarks)~\cite{Fajfer:2012jt,Deshpande:2012rr,Sakaki:2013bfa,Alonso:2015sja,Calibbi:2015kma,Bauer:2015knc,Fajfer:2015ycq,Barbieri:2015yvd,Deshpand:2016cpw,Li:2016vvp,Sahoo:2016pet,Becirevic:2016yqi,Dumont:2016xpj,Das:2016vkr,Barbieri:2016las,Chen:2017hir,Altmannshofer:2017poe}.
A last possibility allowed by the quantum numbers of the fermions involved consists in vector-like quarks that mix with chiral SM quarks and change the couplings to the \W boson (including right-handed couplings), but unless light right-handed neutrinos are introduced, they do not affect the couplings to the leptons and thus yield LU contributions which are not of interest here. 

Heavy vector and scalar bosons are disfavoured by the global fits (scalar) and
by the  direct LHC searches (vector)~\cite{Greljo:2015mma,Faroughy:2016osc}.
Leptoquark models provide a larger set of possibilities, with six different type of representations (lumping together representations differing only through hypercharge)~\cite{Dorsner:2016wpm,Buttazzo:2017ixm,Jung:2018lfu}: three are scalar leptoquarks $(S_1,R_2,S_3)$ and three are vector leptoquarks $(U_1,V_2,U_3)$. All of them are triplets under $SU_C(3)$, but they have different representations under $SU_L(2)$: singlet $(S_1, U_1)$, doublet $(R_2, V_2)$ or triplet $(S_3,U_3)$.
Two of them, $S_1$ and $R_2$, are able to generate tensor contributions (correlated with scalar contributions).
Leptoquark models are also bounded by high-$p_\textrm{T}$ constraints~\cite{Faroughy:2016osc}, but the bounds can be avoided by assuming large couplings to the second generation~\cite{Alonso:2015sja,Crivellin:2017zlb,Calibbi:2017qbu}.

The extension of this discussion to more general frameworks is possible and leaves room for other types of models. For instance, one may consider supersymmetric models allowing for couplings violating $R$-parity~\cite{Altmannshofer:2017poe}.

\subsubsection{\bTosll}

The discussion of \bTosll deviations has involved mainly \Zprime models, leptoquarks and composite models, that are generally tuned in order to reproduce a particular pattern of NP contributions to Wilson coefficients. Interestingly, due to their quantum numbers, leptoquarks contribute only at the loop level, typically of the same size as the SM contribution.

The general outcome of global fits is a significant contribution to ${\cal C}_{9\mu}$.
\Zprime models with couplings to leptons can easily yield such ${\cal C}_{9\mu}^{\rm NP}$-like solution avoiding gauge anomalies.
In this context, models gauging the $L_\mu-L_\tau$ number~\cite{Altmannshofer:2014cfa,Crivellin:2015mga,Crivellin:2015lwa,Crivellin:2016ejn} have been investigated in detail as not generating effects in electron channels.
Concerning leptoquarks, a ${\cal C}_{9\mu}^{\rm NP}$-like solution can only be generated by adding two scalar or two vector representations~\cite{Buttazzo:2017ixm}. 

Another popular scenario is to require that ${\cal C}_{9\mu}^{\rm NP}=-C_{10\mu}^{\rm NP}$, as  it obeys the $SU(2)_L\times U(1)_Y$ symmetry expected from SMEFT.
This pattern can be achieved in \Zprime models with loop-induced couplings~\cite{Belanger:2015nma} or in \Zprime models with heavy vector-like fermions~\cite{Boucenna:2016wpr,Boucenna:2016qad}, or in $Z'$ models with a gauged horizontal symmetry~\cite{Guadagnoli:2018ojc}.
Concerning leptoquarks, here a single representation (either $R_2$ or $U_1$) can generate a ${\cal C}_{9\mu}^{\rm NP}=-C_{10\mu}^{\rm NP}$ solution~\cite{Gripaios:2014tna,Fajfer:2015ycq,Varzielas:2015iva,Alonso:2015sja,Calibbi:2015kma,Barbieri:2015yvd,Sahoo:2016pet}.
This pattern can also be obtained in models with loop contributions from heavy new scalars and fermions~\cite{Gripaios:2015gra,Arnan:2016cpy,Mahmoudi:2014mja}.
Composite Higgs models are also able to achieve this pattern of deviations~\cite{Niehoff:2015bfa}.

Supersymmetric models like the MSSM experience difficulties to generate large NP contributions to ${\cal C}_{9\ell}$, as these are due to loop diagrams of supersymmetric particles (and thus significantly suppressed compared to the SM)~\cite{Altmannshofer:2014rta,Mahmoudi:2014mja}.
More elaborate supersymmetric models may be able to accomodate these data, for instance the ones violating $R$-parity which contain leptoquarks in their spectrum.

\subsubsection{Common explanations}

Since hints of LU deviations are only seen  in \bquark-quark transitions involving leptons, it is tempting to build models explaining both FCCC and FCNC deviations simultaneously. 

The LU ratios \RDDst could be explained with a NP contribution of $\order(10\%)$ of the tree-level SM contribution.
On the other hand, \bTosmm should be affected by a NP contribution of 
approximately 25\% of the loop- and CKM-suppressed SM contribution.
This suggests NP models able to generate $\order(1)$ couplings to the third generation, moderate couplings to the second generation and small or vanishing couplings to the first generation, reminiscent of the hierarchy present for the SM Yukawa couplings.
Moreover, these NP models must not generate FCNC processes violating lepton flavour, like \decay{\B}{K{(^*)}e\mu} or \decay{\B}{K{(^*)}\tau\mu} at a level already measured (it is also true for FCCC \bTocln processes, but not possible to test experimentally due to the impossibility to test the flavour of the neutrino emitted).

Simple models through the exchange of a single particle are very constrained by both sectors~\cite{Buttazzo:2017ixm} as well as by low-energy processes, in particular \Bs-\Bsb mixing, \decay{\B}{K{(^*)}\neu\neub}, $(g-2)_\mu$, \decay{\taum}{\mun\mup\mun}, and direct production at the LHC~\cite{DiLuzio:2017fdq,Buttazzo:2017ixm,Kumar:2018kmr}:
\begin{itemize}
\item Colourless vectors do not seem to be able to accommodate all the data (hitting either constraints from \Bs-\Bsb mixing or from direct production at the LHC).
It is possible to avoid these problems by introducing additional matter fields. 
For instance, one can promote the difference $L_\mu-L_\tau$ into a gauge symmetry that yields NP contributions both in the muon and tau sectors and must be with one generation of vector-like fermions (these models result in substantial modifications of the \decay{\taum}{\mun\mup\mun} and \decay{h}{\mumu} decay rates~\cite{Altmannshofer:2016oaq,Altmannshofer:2014cfa,Crivellin:2015mga,Crivellin:2015lwa,Crivellin:2016ejn}).
\item Scalar leptoquarks may in principle provide a good fit to the data~\cite{Crivellin:2017zlb}, but the radiative corrections to \ZTott and \decay{\Z}{\neu\neub} are significant and create a mild tension with \RDst.
This can be solved at the price of involving right-handed neutrinos~\cite{Becirevic:2016yqi}. Another possibility consists in combining the contributions from two different scalar leptoquarks. The combination of $R_2$ and $S_3$ provides a particularly good description of all available data at low and high energies generating both left- and right-handed currents.
It features a complex scalar contribution to \bToctaun and can be embedded in a $SU(5)$ Grand Unified Theory~\cite{Becirevic:2018afm}. Another option consists in combining $S_1$ and $S_3$, leading to solutions with only left-handed currents~\cite{Crivellin:2017zlb,Buttazzo:2017ixm,Marzocca:2018wcf}.
\item Finally, the vector leptoquark singlet $U_{1}$ %with quantum numbers $U_{1}\sim(\bar{3},1,-4/3)$
gives a particularly good fit to data~\cite{Barbieri:2015yvd,Alonso:2015sja,Calibbi:2015kma,Buttazzo:2017ixm,Kumar:2018kmr}. However, such a massive vector leptoquark requires additional fields to build an ultraviolet complete model, for which several possibilities have been investigated further. For instance, models based on $SU(4)$ Pati-Salam symmetry rely on partially unified gauge groups where the hypercharge and the colour charge are combined according to a single $SU(4)$ gauge group. Stringent bounds coming from light-quark processes require different couplings of the leptoquark to the different generations, leading to the introduction of a Pati-Salam gauge group for each of the three generations~\cite{DiLuzio:2017vat,Calibbi:2017qbu,Bordone:2017bld,Barbieri:2017tuq,Blanke:2018sro,Assad:2017iib}. Another possible ultraviolet completion corresponds to composite models where an extended global symmetry group is broken spontaneously to give rise to the Higgs boson (as a composite pseudo Nambu-Goldstone boson) as well as many other particles (Higgs doublets, leptoquarks, colourless and coloured vector bosons)~\cite{Buttazzo:2017ixm,Buttazzo:2016kid}. Related models have been proposed, based on partial compositeness and feature heavy vector-like quarks and leptoquarks~\cite{Marzocca:2018wcf,Sannino:2017utc,Carmona:2017fsn,Niehoff:2016zso,Barbieri:2016las,Niehoff:2015iaa}.
\end{itemize}

%%%%%%%%%%%%%%%%%%%%%%%%%%%%%%%%%%%%
% !TEX root = main.tex
%%%%%%%%%%%%%%%%%%%%%%%%%%%%%%%%%%%%

%\clearpage
\section{Future measurements}
\label{sec:future}

The present status in \bquark-quark semileptonic decays urges for new measurements to clarify the various deviations from the SM expectations that have been observed.
In the (optimistic) scenario where these tensions are confirmed, more precise measurements and new observables will permit to determine more accurately their pattern and source.
These sets of new measurements are at the core of the research programme of the \belleTwo and \lhcb experiments, which have both published documents presenting their expected performances for the coming years~\cite{Kou:2018nap,LHCb-PII-Physics}.
All figures and numbers in this section are based on these sources.
The main milestones in terms of years of data taking and integrated luminosities are summarised in Tab.~\ref{tab:future_timeline}.

\begin{table}[h!]
\centering
\renewcommand\arraystretch{1.2}
\begin{tabular}{c|c|c|c|c|c}
\textbf{Experiment} & \textbf{2018}	& \textbf{2021}	& \textbf{2024}	& \textbf{2025} & \textbf{2037} \\ \hline
\belleTwo	& ---		& 5\invab	&		& 50\invab	& \\
\lhcb		& 9\invfb	&		& 23\invfb	&			& 300\invfb \\
\end{tabular}
\caption{Summary of the considered years of data taking and corresponding integrated luminosities for the \belleTwo and \lhcb experiments. The \lhcb Upgrade-I has been approved to run between approximately 2021 and 2029. A further Upgrade-II has been proposed to run after 2031.}
\label{tab:future_timeline}
\end{table}

\subsection{Experimental prospects for \bTocln decays}

The current measurements already exhibit a significant tension with the SM prediction, and an important contribution to their uncertainties is of systematic nature.
It is thus particularly important to have accurate results from two diverse experimental environments that will be affected by different systematic effects, but the extrapolation of such uncertainties is a non-trivial task.  

The measurements of the \RDst ratio by the \lhcb experiment have systematic uncertainties that are comparable to the statistical ones, and therefore the extrapolation to future sensitivities should take into account their evolution as well.
The two main contributions to the systematic uncertainty are due to the modelling of the fit components and to the limited size of the simulated samples. 
The background modelling is estimated using dedicated control samples from data and larger datasets will improve its precision.
However, in order to simulate larger statistics and reduce the associated uncertainties, challenging software developments are mandatory to speed up production times without affecting their accuracy.

More information on \bTocln transitions will be provided by the \lhcb experiment that can measure \RHc ratios using \BsbToDsxlnu, \LbToLcxlnu and \BcmToJpsilnu decays.
These processes have different background structures and should be affected by different systematic uncertainties.
Decays of \Lb baryons are of particular interest since these will allow testing for new structures of NP operators.

\begin{figure}[t!]
\centering
\includegraphics[width=0.7\textwidth]{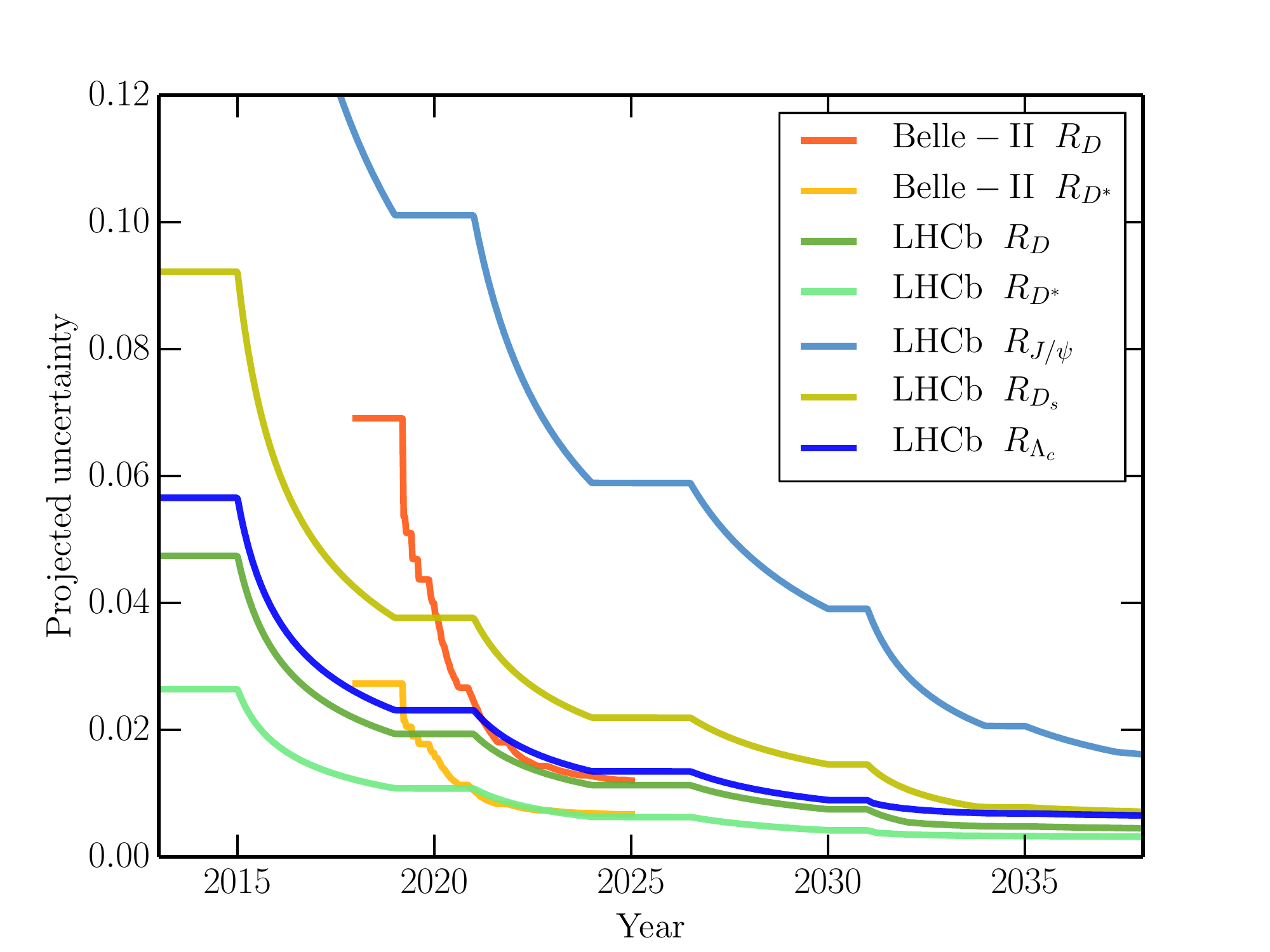}
\caption{Projected uncertainty for various \RHc ratios from the \belleTwo and \lhcb experiments (years are indicative). The \belleTwo uncertainties include estimates of the evolution of the systematic uncertainties. The systematic uncertainties at \lhcb are assumed to scale with the accumulated statistics until they reach limits at $0.003$, $0.004$ and $0.012$ for \RDst, \RD and \RJPs, and $0.006$ for both $R_{D_s}$ and $R_{\Lambda_c}$.}
\label{fig:Future_RHc}
%\end{figure}
%\begin{figure}[t!]
\centering
\includegraphics[width=0.7\textwidth]{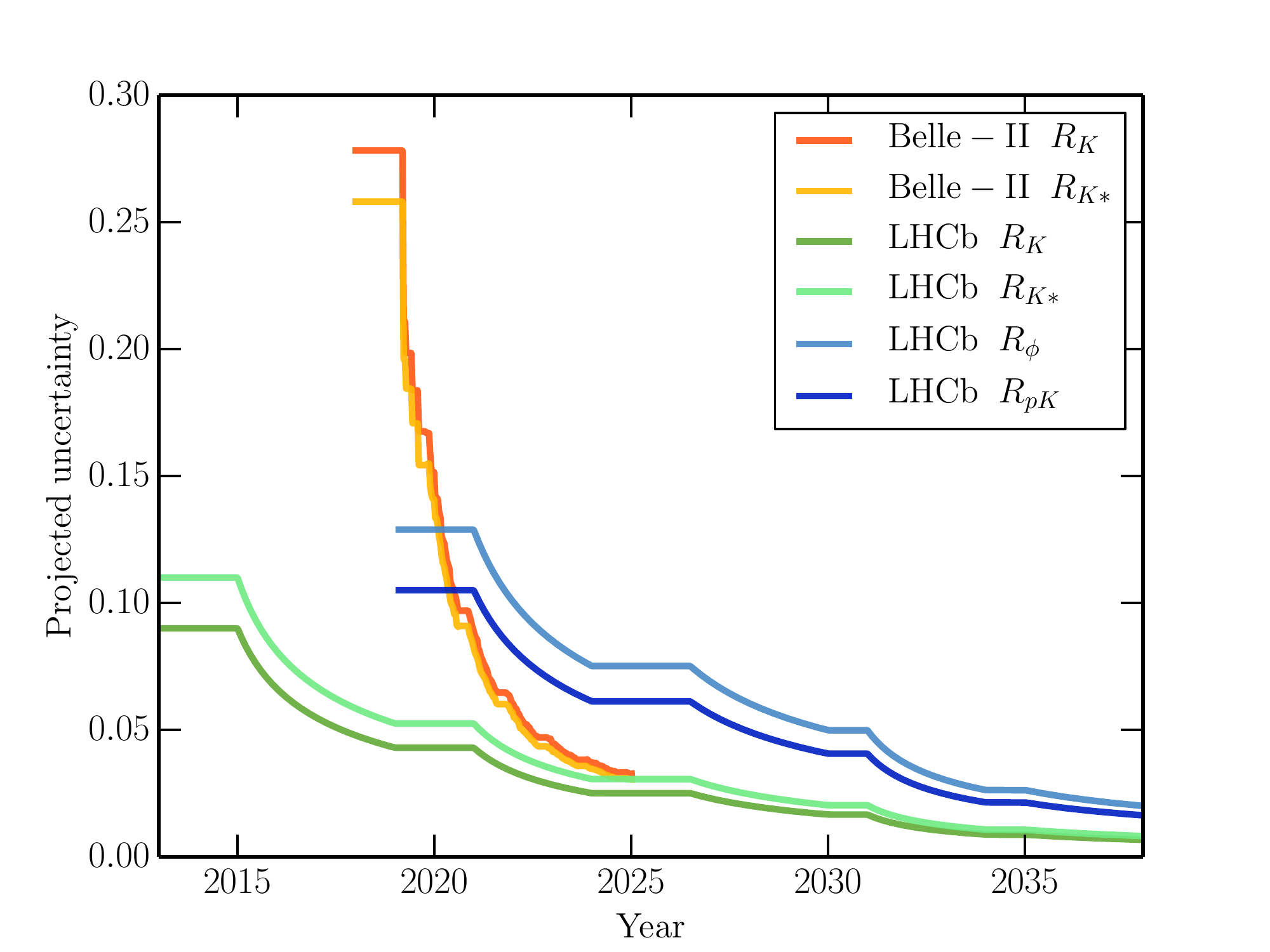}
\caption{Projected uncertainty for various \RHs ratios from the \belleTwo and \lhcb experiments (years are indicative) in the range \mbox{$\sim 1 < \qsq < 6\gevgevcccc$}. The \belleTwo values include estimates of the evolution of the systematic uncertainties (for \RKst, the charged and neutral channels have been combined). The \lhcb uncertainties are statistical only (the precision of all measurements will be dominated by the size of the available data samples except for \RK and \RKst at 300\invfb).}
\label{fig:Future_RHs}
\end{figure}

The \belleTwo expected precision on \RD, \RDst and on the $\tau$ polarisation are given in Tab.~\ref{tab:future_RDst}.
A significant part of the systematic uncertainty for all \RDDst ratios arises from the knowledge of the \decay{\B}{\D^{**}\ellprm\neub_{\ellpr}} decays, and their study is thus mandatory to reach the ultimate precision.
The projected precision for various \RHc ratios from the \belleTwo and \lhcb experiments is presented in Fig.~\ref{fig:Future_RHc}, where the precision on \RDDst is expected to be comparable between the two experiments in the future.

\begin{table}[h!]
\centering
\renewcommand\arraystretch{1.2}
\begin{tabular}{c|c|c}
\textbf{Observable}	& \textbf{5}\boldmath{\invab} & \textbf{50}\boldmath{\invab} \\ \hline
\RD				& $(\pm\, 6.0 \pm 3.9)\%$	& $(\pm\, 2.0 \pm 2.5)\%$ \\
\RDst			& $(\pm\, 3.0 \pm 2.5)\%$	& $(\pm\, 1.0 \pm 2.0)\%$ \\ 
$P_{\tau}(\Dstar)$	& $\pm\, 0.18 \pm 0.08$	& $\pm\, 0.06 \pm 0.04$ \\
\end{tabular}
\caption{Expected precision on \RD, \RDst (relative) and on the $\tau$ polarisation (absolute) from \belleTwo~\cite{Kou:2018nap}. The first uncertainty is statistical and the second is systematic.}
\label{tab:future_RDst}
\end{table}

\subsection{Experimental prospects for \bTosll decays}

The projected precision for various \RHs ratios from the \belleTwo and \lhcb experiments is shown in Fig.~\ref{fig:Future_RHs}, where the precision on \RKX is expected to be comparable between the two experiments at the end of the \belleTwo data taking (around 2025).
The \belleTwo measurements are expected to be completely dominated by the statistical uncertainty.
For \lhcb, a linear dependence of the \bbbar production cross section on the centre-of-mass energy and unchanged detector performances compared to Run1 are assumed.
This is a conservative approximation since from 2021 the bottleneck of the hardware trigger level will be removed and efficiencies for channels with electrons and hadrons in the final state will increase significantly.
However, the larger pile up in the Upgrade-I could increase the background contamination.

The \RK and \RKst ratios are the only tests of LU in \bTosll decays that have been performed so far.
The same underlying quark transition can also be investigated with other \bquark-hadrons at the \lhc.
In particular, the \RPhi (\BsToPhill) and \RpK (\LbTopKll) ratios can be measured to a good precision thanks to the large production cross sections of \Bs and \Lb hadrons.
The \RPhi ratio is particularly interesting since the background level is expected to be low due to the absence of higher hadronic resonances decaying into final states with a $\phi$ meson.
The \RK and \RHs ratios involving vector particles (\eg ~\Kstar and $\phi$) are differently affected by NP contributions, which will put important constraints on the structure of these contributions~\cite{Hiller:2014yaa,Hiller:2014ula}.

In the case that there will be no change in the central value currently observed for \RK and \RKst by the \lhcb experiment, around 2025 both \belleTwo and \lhcb will be able to confirm LU violation in \bTosll transitions with a precision significantly larger than 5$\,\sigma$.
For \lhcb, the current systematic uncertainties are of the order of 4--5\%, but a significant part of these is statistical in nature.
Starting from 2018, the \atlas and \cms experiments are recording data with a \B trigger in order to measure the \RK and \RKst ratios, however, the expected precisions are not yet known.

Concerning angular observables, the \lhcb collaboration has measured the $P_{4}^{'}$ and $P_{5}^{'}$ observables in the \BdToKstmm decay mode~\cite{LHCb-PAPER-2015-051} with a precision of the order of 11\%.
An angular analysis of the \BdToKstee decay has been performed in a different \qsq region with an observed yield of about 120 events~\cite{LHCb-PAPER-2014-066}.
An extrapolation of this work indicates that it should be possible to distinguish between different NP scenarios using these modes~\cite{LHCb-PII-Physics}.
In the case of \belleTwo, the precision on the $Q_4$ and $Q_5$ observabls is expected to reach about 5\% for the three \qsq bins (1--2.5, 2.5--4 and 4--6\gevgevcccc)~\cite{Kou:2018nap}. 

The size of the \lhcb data sample after the Upgrade-II (300\invfb) makes this a unique place to search for LU breaking in \bTodll due to the smallness of their branching fractions.
In particular, the expected statistical precision for \mbox{$\RPi = \BR(\Bpm\to \pipm \mumu)/\BR(\Bpm\to \pipm \epem)$} in the \qsq range between 1.1 and 6\gevgevcccc is around 4\%. 

%On a longer term, the FCC-ee collider, which may run at the \Z pole and collect of the order of $10^{12}$ collisions, could help clarifying the situation.
%In this environment a new study shows that a \BToKsttt branching fraction at the SM value could be measured with a precision of about 5\%~\cite{Kamenik:2017ghi}.
%Few thousands of \BToKsttt are expected to be reconstructed and could also be used to study the $\tau$ polarisation in \bTostt transition and thus discriminate among different NP scenarios.

\subsection{Additional observables}

The LU deviations observed in both FCCC and FCNC \bquark-quark decays can be analysed by invoking NP occurring through different kinds of operators (in the EFT approach) or different types of underlying dynamics (for UV complete models).
%Increasing the number of observables would thus be beneficial, by  measuring further observables for each lepton flavour separately, or by determining further LU quantities, \ie ratios or differences of the same observable for different lepton flavours, in order to confirm the LU deviations already observed.
Increasing the number of observables would thus be beneficial to shed light on the NP that could be at play.

More specifically, \bTocln transition to each lepton flavour could be investigated by considering not only ratios, but also angular observables~\cite{Becirevic:2016hea,Alonso:2016gym}, the polarisation of the $\tau$ lepton or that of the \Dstar meson (for \decay{\B}{\Dstar\ell\nu} transitions)~\cite{Fajfer:2012vx,Alok:2016qyh}, final states with mesons of higher spins like $\D^{**}$~\cite{Bernlochner:2017jxt} or baryon decays~\cite{Boer:2018vpx}.
As far as LU is concerned, one could look at ratios or differences of all these observables. Moreover,
ratios of branching fractions for electron and muon modes should be tested more precisely in order to determine if these also exhibit LU violation.

For \bTosll, a fairly comprehensive set of angular observables, $J_i$ or $P_i$, has already been measured for $\ell=\mu$.
The same quantities could be measured for $\ell=e$ in order to build LU observables such as $D_i=J_i^{\mu}-J_e^{\mu}$~\cite{Serra:2016ivr} or $Q_i=P_i^\mu-P_i^e$~\cite{Capdevila:2016ivx}, the latter offering smaller hadronic uncertainties and ensuring an easier interpretation in terms of NP.
Similar observables should also be studied in other modes, \eg \BsToPhill or \LbTopKll, to confirm the hints of LU violation.
Time-dependent measurements could also be considered~\cite{Descotes-Genon:2015hea}. Another possible test of LU violation is provided by $B_s\to\ell^+\ell^-\gamma$~\cite{Guadagnoli:2017quo}.
Furthermore, it would be very important to study \bTostt transitions, since general EFT considerations suggest that these could be significantly enhanced compared to electron and muon modes~\cite{Bobeth:2011st,Kamenik:2017ghi,Capdevila:2017iqn}. 
A sensitivity study~\cite{Kamenik:2017ghi} has been performed for the FCC-ee collider, which may run at the \Z pole and collect of the order of $10^{12}$ \Z bosons. 
In this environment a \BToKsttt branching fraction at the SM value could be measured with a precision of about 5\%.
Since a few thousands \BToKsttt decays are expected to be reconstructed, these could also be used to study the $\tau$ polarisation in \bTostt transition and thus discriminate among different NP scenarios.

As indicated in Sec.~\ref{sec:specificNPmodels}, the NP models proposing a common interpretation of the LU deviations at tree- and loop-level must in general indicate how the new, heavy, particles couple to the various generations of quarks and leptons.
Further LU tests in \bTodll and \bTouln transitions could thus be used as constraints for these models.
In a similar way, more stringent bounds on lepton-flavour violating decays \decay{\bquark}{\squark\ell_1^+\ell_2^-} (with $\ell_1 \neq \ell_2$) could also constrain NP models, as lepton-flavour violation occurs often in models violating LU~\cite{Glashow:2014iga}.
The \BToKKstemu and \BToKtaul (\ellpr = $e,\mu$) decays have been studied by the \babar collaboration and limits are of the order of few $10^{-7}$ and $10^{-5}$, respectively~\cite{Aubert:2006vb,Lees:2012zz}.
The expected limits by \lhcb with an integrated luminosity of about 9\invfb are of the order of $10^{-9}$ and $10^{-6}$ for the \BToKKstemu and \BToKKsttaumu decays, respectively.
These limits are expected to scale linearly with luminosity for \BToKKstemu and with its square root for the \BToKKsttaumu decays~\cite{LHCb-PII-Physics}.

%%%%%%%%%%%%%%%%%%%%%%%%%%%%%%%%%%%%
% !TEX root = main.tex
%%%%%%%%%%%%%%%%%%%%%%%%%%%%%%%%%%%%

%\clearpage
\section{Conclusions}
\label{sec:conclusions}

In recent years, several observations of tree- and loop-level \bquark-hadron decays hint at a possible violation of LU.
This article has discussed the relevant measurements as well as their potential implications.

LU plays a peculiar role in the SM of particle physics, as the same interactions and couplings characterise all three fermion generations.
As a consequence, a violation of LU would be an unambiguous sign of the existence of physics beyond the SM.
This property has been tested throughout the years by using a variety of different probes: the production and the decay of electroweak gauge bosons, the decay of quarkonia, the leptonic and semileptonic decays of mesons with light quarks (including the \cquark quark), however, no significant signs of deviations from the SM predictions have been observed.

A possible violation of LU is hinted at in two different classes of semileptonic \bquark-quark decays.
The measurements are obtained from experiments at the \B-factories (\babar, \belle and soon \belleTwo) as well as at the \lhc (\lhcb). 
For these processes the SM predictions can be computed using an effective Hamiltonian approach that separates short- and long-distance contributions, and require non-perturbative inputs (\eg form factors) obtained through diverse theoretical methods.

Tensions at the level of 4--5$\,\sigma$ are observed in \bTocln decays, which are mediated at tree level through a \Wpm boson in the SM, when the branching ratios of decays with $\ell=\tau$ and $\ell=e,\mu$ are compared.
Deviations at the level of 3--4$\,\sigma$ are also present in \bTosll decays, which are mediated through a loop in the SM, when comparing the branching ratios for $\ell=e$ and $\ell=\mu$.

Such deviations can be interpreted theoretically in three steps.
Firstly, constraints on NP contributions to the short-distance Wilson coefficients of the Hamiltonian can be extracted from the experimental results using the effective Hamiltonian approach.
Secondly, simple NP models that provide the desired short-distance contributions via the exchange of one or two mediators can be built without contradicting the limits on direct production or indirect contribution to low-energy processes.
Finally, full-fledged NP models with a proper ultraviolet completion can be designed to match the previous models at lower energies.

In this context, it is of paramount importance to confirm or refute these hints of LU violation promptly.
Both the \belleTwo and \lhcb experiments will be in an ideal position to provide additional information by significantly reducing the uncertainties on the LU observables already studied and by measuring new observables that will further constrain NP models.
The present situation should thus evolve rapidly with the combined efforts of experimentalists and theorists, and has the potential to provide very exciting news in the coming years.

%%%%%%%%%%%%%%%%%%%%%%%%%%%%%%%%%%%%
% !TEX root = main.tex
%%%%%%%%%%%%%%%%%%%%%%%%%%%%%%%%%%%%

%\clearpage
\section*{Acknowledgements}

The authors warmly thank Yasmine Amhis, Martino Borsato, Juan J. Saborido and Javier Virto for carefully proofreading the article.
The authors also thank Martino Borsato for helping with the figures of the projected sensitivities.

SB acknowledges support from the UK STFC Research Council under the grant agreement No.~ST/N000463/1.
SDG acknowledges partial support from Contract FPA2014-61478-EXP and from the European Union Horizon 2020 research and innovation programme under grant agreements No.~690575, No.~674896 and No.~692194.
ARV acknowledges support from the Xunta de Galicia under the grant agreement No.~2016-PG003.
% ==============================================================
%\clearpage
\addcontentsline{toc}{section}{References}
\setboolean{inbibliography}{true}
\bibliographystyle{LHCb}
\bibliography{ms,LHCb-PAPER}
% ==============================================================
\end{document}